\documentclass[a4paper,11pt]{article}
\pdfoutput=1
\usepackage{jcappub}
\usepackage{amsmath}
\usepackage{amsfonts}
\usepackage{amssymb}
\usepackage{graphicx}%
\usepackage{amstext}
\usepackage{amsmath}
\usepackage{latexsym}
\usepackage{amsfonts}
\usepackage{amssymb}
\usepackage{color}
\usepackage{mathtools}
\usepackage{graphics}
\usepackage{graphicx}
\usepackage{slashed}
\usepackage{epsfig}
\usepackage[utf8]{inputenc}
\usepackage{amsthm}
\usepackage{appendix}
\usepackage{hyperref}
\usepackage{titlesec}
\def\simlt{\stackrel{<}{{}_\sim}}
\def\simgt{\stackrel{>}{{}_\sim}}

\begin{document}
\title{The Sommerfeld Enhancement in the Scotogenic Model with Large 
Electroweak 
Scalar Multiplets}
\author[a]{Talal Ahmed Chowdhury}\author[b]{\& Salah Nasri}
\affiliation[a]{Department of Physics, University of Dhaka, P.O. Box 1000, 
Dhaka, Bangladesh}
\affiliation[b]{Department of Physics, UAE University, P.O. Box
17551, Al-Ain, United Arab Emirates} \emailAdd{talal@du.ac.bd, 
snasri@uaeu.ac.ae}

\abstract{
We investigate the Sommerfeld enhancement (SE) in the generalized scotogenic model with large electroweak multiplets. We focus on scalar dark matter (DM) candidate of the model and compare DM annihilation cross sections to $WW$, $ZZ$, $\gamma\gamma$ and $\gamma Z$ at present day in the galactic halo for scalar doublet and its immediate generalization, the quartet in their respective viable regions of parameter space. We find that larger multiplet has sizable Sommerfeld enhanced annihilation cross section compared to the doublet and because of that it is more likely to be constrained by the current H.E.S.S. results and future CTA sensitivity limits.}
\maketitle

\section{Introduction}\label{intro}

Though considerable amount of astrophysical and cosmological studies suggest 
the 
Dark Matter (DM) to be an essential
component of our universe, its conclusive particle nature is yet to be 
determined. 
Therefore currently the extensive searches for the DM have been carried out 
using 
various direct and indirect detection methods. While the direct detection 
mainly 
looks for the DM-nucleus scattering in the detector, the indirect detection 
focuses on the search of atypical $e^{+}$, $\overline{p}$, $\gamma$ and $\nu$ 
signatures of extraterrestrial origin i.e. from DM-DM annihilation or decay in 
the universe, using ground and satellite based detectors. In recent years, the 
gamma-ray observation by Cherenkov telescope has provided stringent and 
robust constraints 
\cite{Bringmann:2012ez, Wood:2013taa, Buckley:2013bha, Cirelli:2015gux, Profumo:2016idl}
and it is also reaching the sensitivity level of DM annihilation cross 
sections 
to different final states of the Standard Model (SM) particles for the DM in 
$O(1-100)$ TeV mass range \cite{Mora:2015vhq, Doro:2012xx}. As a result, the present 
constraints and sensitivity limits, such as
the latest result of H.E.S.S. (High Energy Stereoscopic System) for searching DM annihilation towards the inner 
galactic halo \cite{::2016jja, Abdalla:2016olq} and the projected reach of CTA (Cherenkov
Telescope Array) \cite{Carr:2015hta, Conrad:2016jww, Lefranc:2016dgx, Lefranc:2016fgn}, can allow one 
to 
investigate the viability of a particular DM model with TeV mass with respect 
to 
current observation and upcoming experiments.
 
Why is the TeV mass-ranged DM in the focus of the indirect detection? 
When 
the DM with mass $m_{\text{DM}}\gg m_{W}$, is charged under the SM gauge group 
$SU(2)_{L}\times U(1)_{Y}$ and non-relativistic, its annihilation cross section 
and hence the indirect detection rate is affected by a non-perturbative 
correction known as the Sommerfeld enhancement (SE). The SE, first discovered 
in 
the study of very slow electron scattering \cite{sommerfeldref}, has gained much focus 
in the context
of DM phenomenology \cite{Hisano:2002fk, Hisano:2003ec, Hisano:2004ds, 
Hisano:2005ec, Hisano:2006nn, Cirelli:2007xd, 
Cirelli:2009uv, ArkaniHamed:2008qn, Lattanzi:2008qa, Pieri:2009zi, 
Iengo:2009ni, Cassel:2009wt, Slatyer:2009vg, Hambye:2009pw, Feng:2009hw, Feng:2010zp, Hryczuk:2010zi, Hryczuk:2011vi, Beneke:2012tg, Tulin:2013teo, Fan:2013faa, Cohen:2013ama, 
Ovanesyan:2014fwa, Baumgart:2014vma, Beneke:2014gja, Cirelli:2015bda, 
Garcia-Cely:2015dda, Aoki:2015nza} (and references therein).
In the case of non-relativistic DM 
with electroweak quantum numbers and TeV mass, the gauge boson exchange between two DM 
particles will induce long ranged attractive (repulsive) force which in turn 
modifies the incoming asymptotic plane waves 
and greatly enhances (suppresses) the annihilation cross section than the 
typical tree-level or loop-induced cross section. Thus, such enhancement can
substantially 
increases the detection prospects of the DM annihilation in the Cherenkov 
telescopes or in satellites.

The Scotogenic model \cite{Ma:2006km} is a well-motivated model which not only provides 
fermionic and/or scalar dark matter candidates (depending on region of parameter space) but also generates the neutrino mass 
radiatively at one loop 
\cite{Kubo:2006yx, Sierra:2008wj, Suematsu:2009ww, Adulpravitchai:2009gi, 
Toma:2013zsa, Klasen:2013jpa, Vicente:2014wga}. As there is no symmetry reason to prevent 
extending the scotogenic model with scalars and fermions of larger $SU(2)_{L}$ 
representations, in its generalized version, the scalar doublet and singlet 
fermion of the minimal model can be replaced by an even dimensional 
$(J,Y)=(n/2,1/2)$ electroweak scalar multiplet and corresponding odd 
dimensional 
(either $(J,Y)=(\frac{n-1}{2},0)$ or $(J,Y)=(\frac{n+1}{2},0)$) fermionic 
multiplets 
\cite{Ma:2008cu, Law:2013saa, Ren:2011mh, Restrepo:2013aga, Chowdhury:2015sla, Ahriche:2016cio, Ahriche:2016ixu}. For the 
generalized scotogenic model, SE is expected to be significantly large compared 
to the minimal one as more states of larger electroweak multiplets will 
contribute to the enhancement which is the main focus of this investigation. 
The 
SE analysis of the fermionic DM in the scotogenic model resembles that of 
well-studied fermionic minimal dark matter model
\cite{Cirelli:2005uq, Cirelli:2007xd, Cirelli:2009uv}. 
For this reason, we are not going to pursue fermionic 
DM in this study and instead we will focus on the case of scalar DM.  
The indirect detection prospects of the scalar DM of the minimal scotogenic model 
have been 
addressed in the light of H.E.S.S. results and upcoming CTA limits in 
\cite{Queiroz:2015utg, Garcia-Cely:2015khw}.
As the scalar doublet and higher multiplet
have the same set of parameters, we will carry out a comparative study of SE for 
the scalar doublet and its immediate generalization, the quartet 
$(J,Y)=(3/2,1/2)$ representation \cite{AbdusSalam:2013eya}
in their respective viable regions of parameter space and see how much SE 
increases for larger multiplets.

The article is organized as follows. In section \ref{inertscalarmodel}, we  present the model and set up the notation for subsequent analysis.
In section \ref{sommerfeldsec}, we briefly sketch the 2-particle states, potential and annihilation matrix and present the Schrodinger equation to calculate Sommerfeld enhancement for scalar multiplet in the generalized scotogenic model. In section \ref{resultdiscussionsection}, we present the Sommerfeld enhanced DM annihilation cross sections for the doublet and the quartet cases and discuss their implications in the light of H.E.S.S. results and CTA sensitivity limits. We conclude in section \ref{conclusion}. In appendix \ref{matrixelem}, we present the non-relativistic limit of scalar components, 2-particle effective action, the potential and $S$-wave annihilation matrix elements of the generalized scotogenic model.

\section{The Model}\label{inertscalarmodel}
The scalar sector of the generalized scotogenic model has been presented in 
\cite{AbdusSalam:2013eya, Chowdhury:2015sla}. Here we briefly present the potential and the
particle states for setting up our notations.

\subsection{The Scalar Potential and the Particle 
states}\label{potentialandparticlestates}
The general Higgs-scalar potential that involves the $SU(2)_{L}$ 
scalar multiplet
$\Delta$ with isospin, $J=n/2$ (n odd) and hyper-charge, $Y=1/2$, symmetric 
under 
a
$Z_{2}$, is as follows, 
\begin{align} 
  \label{potq}
V_0(H,\Delta)&=- \mu^2 H^\dagger H + M_0^2 \Delta^\dagger \Delta +
  \lambda_1 (H^\dagger H)^2 + \lambda_2 (\Delta^\dagger \Delta)^2
  +\lambda_3 |\Delta^\dagger T^a \Delta|^2\nonumber\\
  &+\alpha H^\dagger H \Delta^\dagger \Delta
  +\beta H^\dagger
  \tau^a H \Delta^\dagger T^a \Delta
 +\gamma\left[ (H^T\epsilon \tau^a H) (\Delta^T
    C T^a \Delta)^\dagger+h.c\right] 
\end{align}
Here, $\tau^a$ and $T^{a}$ are the $SU(2)$ generators in fundamental
and $\Delta$'s representation respectively. $C$ is an antisymmetric matrix
such that $C T^{a} C^{-1}=-T^{aT}$.
As $C$  is  antisymmetric,  it can only be defined for even 
dimensional space, i.e only for half-integer representation. If the
isospin of the representation is $J$ then $C$ is $(2J+1)\times (2J+1)$
dimensional matrix. The generators are defined so that they
satisfy $Tr[\tau^a\tau^b]=\frac{1}{2} \delta^{ab}$ for fundamental 
representation
and $Tr[\tau^a\tau^b]=D_{2}(\Delta) \delta^{ab}$ for $\Delta$'s representation
where $D_{2}(\Delta)$ is the Dynkin index of $\Delta$.
In addition, the
$\gamma$ term is only allowed for representations with
$(J,Y)=(\frac{n}{2},\frac{1}{2})$ and it is essential for the mass splitting 
between scalar and pseudoscalar
components in the neutral field of the scalar multiplet in a renormalizable 
way. 

Incidentally for complex odd dimensional ($J=n,\,Y\neq 0$), ($n=1,2,..$) scalar 
multiplets, $\gamma$ term doesn't occur in the $Z_{2}$ symmetric scalar 
potential Eq.(\ref{potq}) and no mass splitting takes place between $S$ and $A$ 
of the neutral component \cite{Hambye:2009pw}. But higher dimensional operators \cite{fabrizio-Goran} such as
\begin{equation}
O_{\delta}\sim \frac{c}{\Lambda^2}\Delta\Delta H^{*} H^{*} H^{*} H^{*}+h.c
\label{highdim}
\end{equation}
where $\Delta$ can be $(J=1,2,..,Y=1)$, will induce the mass splitting between 
the scalar and pseudoscalar components of neutral field $\Delta^{0}$ of the 
multiplet. However, incorporating complex odd dimensional scalar multiplets in 
the generalized scotogenic model requires non-minimal extension \cite{Lu:2016dbc}, therefore 
will not be pursued here.

The scalar representation $(J,Y)=(\frac{n}{2},\frac{1}{2})$ with component 
fields is expressed as
\begin{equation}
  {\bf{\Delta_{\frac{n}{2}}}}=\left(
    \Delta^{(\frac{n+1}{2})}_{\frac{n}{2}},
    \Delta^{(\frac{n-1}{2})}_{\frac{n-2}{2}},
    ..,
    \Delta^{(Q)}_{m},
    ..,
    \Delta^{(0)}_{-\frac{1}{2}}\equiv\frac{1}{\sqrt{2}}(S+i\, A),
    ..,
    \Delta^{(-Q)}_{-m-1},
    ..,
    \Delta^{(-\frac{n-1}{2})}_{-\frac{n}{2}}\right)^{T}
  \label{inertreps}
\end{equation}
Here the charge of the component field with $T_{3}=m$ is denoted by 
$Q=m+\frac{1}{2}$. 
Moreover each component of the multiplet is a complex quantity and 
$(\Delta^{(Q)}_{m})^{*}=
\Delta^{(-Q)}_{m}$.

The neutral scalar $S$ and pseudoscalar $A$ component
associated with $J=n/2$ (n odd) multiplet have masses as,
\begin{align}
m_{S}^2 &= 
M_{0}^2+\frac{1}{2}\left(\alpha+\frac{1}{4}\beta+p(-1)^{p+1}\gamma\right)
v_{0}^2\label{masseqscalar}\\
m_{A}^2&=M_{0}^2+\frac{1}{2}\left(\alpha+\frac{1}{4}\beta-p(-1)^{p+1}
\gamma\right)
v_{0}^2\label{masseqnpseudo}
\end{align}
where, $p=\frac{1}{2}(n+1)$ that comes from $2p\times 2p$ $C$ matrix. Moreover, 
because of $Z_{2}$ symmetry, one can switch between $S\rightarrow A$ or $\gamma 
\rightarrow -\gamma$. Here $v_{0}$ is the Higgs VEV.

Apart from the largest charged component of the multiplet which has $T_3=n/2$, 
the $\gamma$ term
mixes the components carrying the same amount of charge $|Q|$,
i.e between $\Delta^{(Q)}_{m}$ and $\Delta^{(Q)}_{-m-1}$.
Therefore, the mixing matrix between
components with charge $|Q|$ is, 
\begin{equation}
  \label{sc1}
  M^2_{Q}=\begin{pmatrix}
  m^2_{(m)}&\frac{\gamma 
v_{0}^2}{4}\sqrt{\left(\frac{n}{2}-m\right)\left(\frac{n}{2}+m+1\right)}\\
  \\
  \frac{\gamma 
v_{0}^2}{4}\sqrt{\left(\frac{n}{2}-m\right)\left(\frac{n}{2}+m+1\right)}& 
m^2_{(-m-1)}
  \end{pmatrix}
\end{equation}
where, $m_{(m)}^{2}$ is given by
\begin{equation}
  m_{(m)}^2=M_{0}^2+\frac{1}{2}\left(\alpha-\frac{1}{2}\beta\,m\right) v_{0}^{2}.
  \label{chargedmass1}
\end{equation}

So the mass eigenstates are,
\begin{align}
\tilde{\Delta}_{1}^{(Q)}&=\cos\theta_{Q}\,\Delta^{(Q)}_{m}+\sin\theta_{Q}\,
\Delta_{-m-1}^{(Q)}\nonumber\\
 \tilde{\Delta}_{2}^{(Q)}&=-\sin\theta_{Q}\,\Delta^{(Q)}_{m}+\cos\theta_{Q}\,
\Delta_{-m-1}^{(Q)}
 \label{egstate}
\end{align}
with
\begin{equation}
 \tan 2\theta_{Q}=\frac{2(M^{2}_{Q})_{12}}{(M^{2}_{Q})_{11}-(M^{2}_{Q})_{22}}
 \label{egstate1}
\end{equation}

Therefore the scalar multiplet can be written in terms of mass eigenstates in 
the following way
\begin{equation}
 \Delta=\begin{pmatrix}
         \Delta^{(\frac{n+1}{2})}\\
         ...\\
         
\Delta^{(Q)}_{m}=\tilde{\Delta}^{(Q)}_{1}\cos\theta_{Q}-\tilde{\Delta}^{(Q)}_{2}
\sin\theta_{Q}\\
         ...\\
         \Delta^{(0)}=\frac{1}{\sqrt{2}}(S+ i A)\\
         ...\\
         
\Delta^{(-Q)}_{-m-1}=\tilde{\Delta}^{(-Q)}_{1}\sin\theta_{Q}+\tilde{\Delta}^{
(-Q)}_{2}\sin\theta_{Q}\\
         ...\\
        \end{pmatrix}
\label{inertreps}
\end{equation}

\subsection{Scalar Dark Matter}\label{scalarDM}
Because of $Z_2$ symmetry, one can set either $S$ or $A$ to be the DM 
candidate. 
In this work, we set $S$ to be DM candidate. This consideration leads to 
$|\gamma|>|\beta|/2$ and 
the following mass hierarchy in the
components of the multiplet,
\begin{equation}
m_{S}<m_{\tilde{\Delta}_{1}^{+}}<m_{\tilde{\Delta}_{1}^{++}}<..<m_{\tilde{\Delta
}_{1}^{(Q)}}<..
 <m_{\Delta^{\left(\frac{n+1}{2}\right)}}<m_{\tilde{\Delta}_{2}^{+}}<..<m_{\tilde
{\Delta}_{2}^{(Q)}}<..
 <m_{A}
 \label{masshier}
\end{equation}

For the doublet and quartet, we use the following notations
\begin{equation}
D=\begin{pmatrix}
\Delta^{+}\\
\frac{1}{\sqrt{2}}(S+i A)
\end{pmatrix},\,\,
\Delta=\begin{pmatrix}
\Delta^{++}\\
\tilde{\Delta}^{+}_{1}\cos\theta-\tilde{\Delta}^{+}_{2}\sin\theta\\
\frac{1}{\sqrt{2}}(S+ i A)\\
\tilde{\Delta}_{1}^{-}\sin\theta+\tilde{\Delta}^{-}_{2}\cos\theta
\end{pmatrix}
\label{doubquartnotation}
\end{equation}

In addition, for doublet and quartet, the Higgs-DM cubic coupling which induces 
the scattering of DM with nucleus in direct detection experiments, are
\begin{equation}
\lambda^{d}_{S}=\alpha+\frac{1}{4}\beta-\gamma,\,\,\,\lambda^{q}_{S}
=\alpha+\frac{1}{4}-2|\gamma|
\label{dmhiggscoup}
\end{equation} 

The theoretical constraints imposed on the doublet and the quartet couplings 
are 
given in \cite{AbdusSalam:2013eya}.

\section{Sommerfeld Enhancement with large Electroweak Scalar 
Multiplets}\label{sommerfeldsec}

At present the dark matter is non-relativistic (NR) and has average velocity 
$v=220\,\text{km}s^{-1}$ in the Milky Way. In this NR limit, the exchange of 
massive $W$ and $Z$ bosons will induce Yukawa potential, $V_{Y}$ and massless 
$\gamma$ will induce Coulomb potential, $V_{C}$ between two incoming component 
states of the multiplet as shown in Fig. \ref{exchangedig} (upper and middle 
panels). Now if the range of the potential is larger than the characteristic Bohr 
radius of 2-particle state, i.e for Yukawa case, $1/m_{W}\simgt 1/\alpha 
m_{S}$, where $\alpha=g^{2}/4\pi$, the wavefunction of the incoming state is 
significantly modified inside the Yukawa potential. In other words, at NR 
limit, 
the ladder diagram as shown in Fig. \ref{exchangedig} (lowest), is enhanced by 
$\alpha m_{S}/m_{W}$ for each $W$ boson exchange because when the mass 
splitting 
is very small compared to the mass, the intermediate $SU(2)_{L}$ partner states 
are almost on-shell (at the threshold) and thus enhances the diagram\footnote{For an instructive presentation of various limits of momenta and masses in a Feynman diagram at the threshold region, please see \cite{Beneke:1997zp}.}. 

As the interaction related to annihilation or production process takes place in 
much shorter distance than the long range interaction responsible for the 
modification of the wavefunction, one can disentangle short distance physics from the long distance one. The modified 2-particle wavefunction (connected to long distance physics) is determined by solving 
the Schrodinger equation with an appropriate matrix-valued potential. The 
method 
of computing SE is well-studied and we have followed the prescription presented 
in \cite{Hisano:2004ds, Cirelli:2007xd, Beneke:2014gja} to compute the Sommerfeld enhanced 
S-wave DM DM annihilation rates $SS\rightarrow VV$ for the
large scalar multiplet of the generalized scotogenic model. In the following 
sections, we present the relevant parts required for the computation.

\subsection{The 2-particle states, potential and annihilation 
matrix}\label{nrlimit}

\subsubsection{The 2-particle states}\label{2particlestates}
The DM-DM 2-particle state, $|SS\rangle$, is charge neutral and CP-even state 
hence it only mixes with other $Q=0,\,\text{CP}=1$ 2-particle states. 
Therefore, the 2-particle state 
vector with only charge neutral and CP even component, is given by
\begin{align}
 |\Phi_{\Delta_{\frac{n}{2}}}\rangle&=\left(SS,\,AA,\,\Delta^{\left(\frac{n+1}{2}
\right)}\Delta^{\left(-\frac{n+1}{2}\right)},...,\,
 \tilde{\Delta}_{1}^{(Q)}\tilde{\Delta}_{1}^{(-Q)},\,\tilde{\Delta}_{2}^{(Q)}
\tilde{\Delta}_{2}^{(-Q)},...,\right.\\
 &\,
 \left.\tilde{\Delta}_{1}^{(Q)}\tilde{\Delta}_{2}^{(-Q)},\,
 \tilde{\Delta}_{2}^{(Q)}\tilde{\Delta}_{1}^{(-Q)},...,
 \tilde{\Delta}_{2}^{\left(\frac{n-1}{2}\right)}\tilde{\Delta}_{1}^{\left(-\frac{
n-1}{2}\right)}\right)^{T}
 \label{pstates}
\end{align}
Here, the ordering of the components in the vector is arbitrary. One can adopt 
different ordering for convenience.

For the doublet and quartet cases, we have
the 2-particle state vector as
\begin{align}
 \text{Doublet},\,\,|\Phi_{D}\rangle &=(SS,\,AA,\,\Delta^{+}\Delta^{-})^{T}\\
 \text{Quartet},\,\,|\Phi_{\Delta}\rangle &= 
(SS,\,AA,\,\Delta^{++}\Delta^{--},\,\tilde{\Delta}_{1}^{+}\tilde{\Delta}_{1}^{-}
,\,
 \tilde{\Delta}_{2}^{+}\tilde{\Delta}_{2}^{-},\,\tilde{\Delta}_{1}^{+}\tilde{
\Delta}_{2}^{-},
 \,\tilde{\Delta}_{2}^{+}\tilde{\Delta}_{1}^{-})^{T}
 \label{doubletquartetps}
\end{align}

\subsubsection{The Potential and Annihilation 
Matrix}\label{potentialandannihilation}
\begin{figure}[h!]
\vspace{-1cm}
 \centerline{\includegraphics[width=10cm]{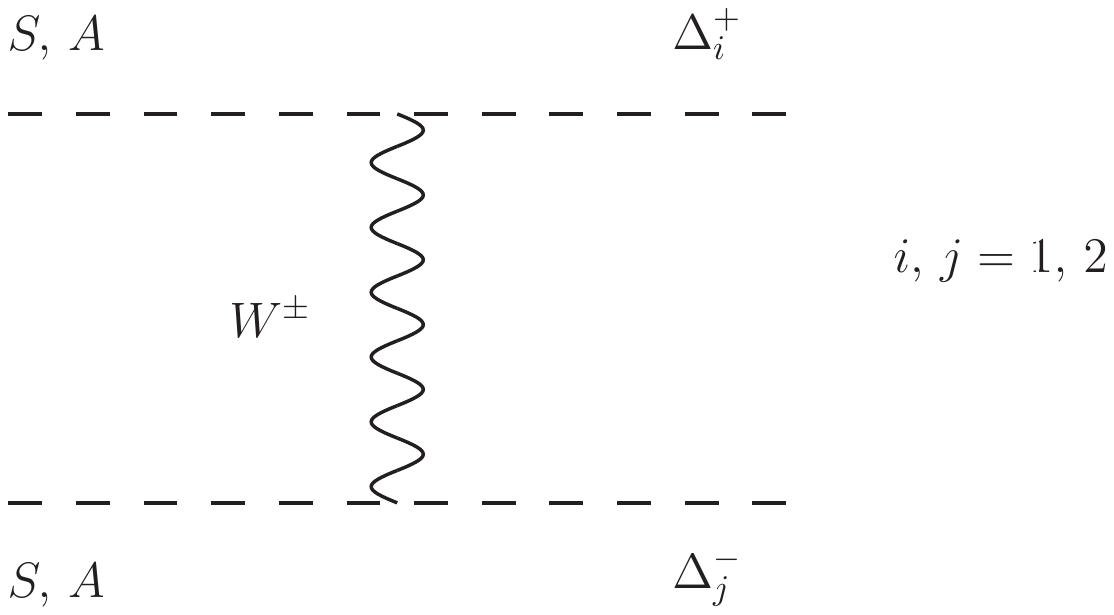}\hspace{-2cm}
 \includegraphics[width=10cm]{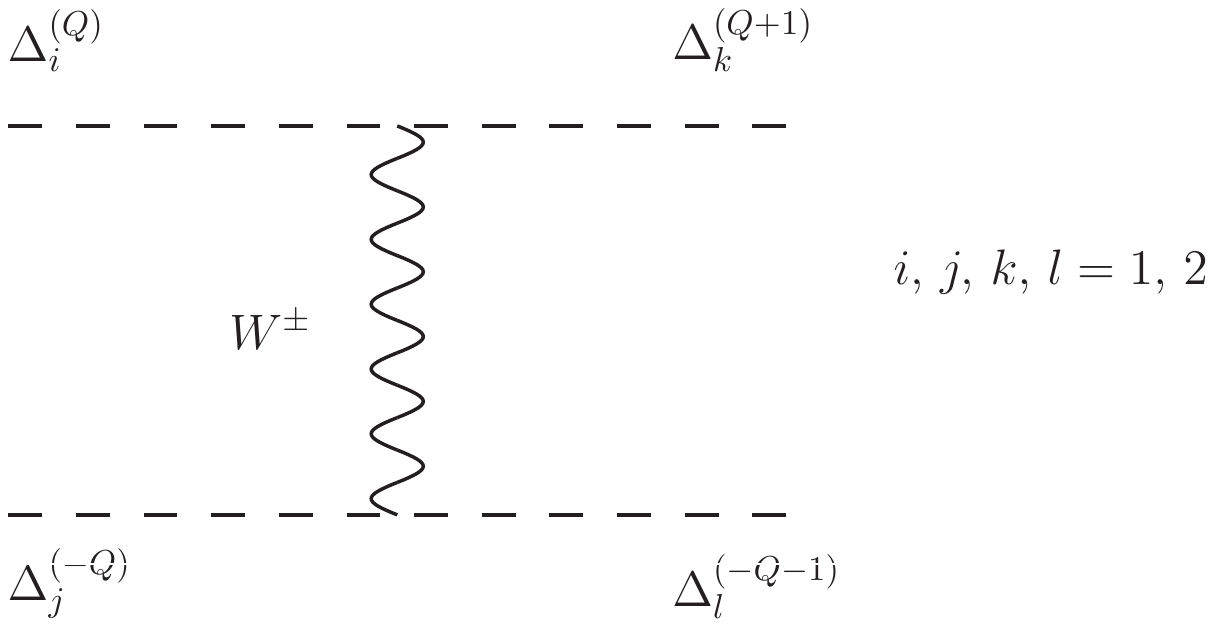}}\vspace{-10.5cm}
 \centerline{\includegraphics[width=10cm]{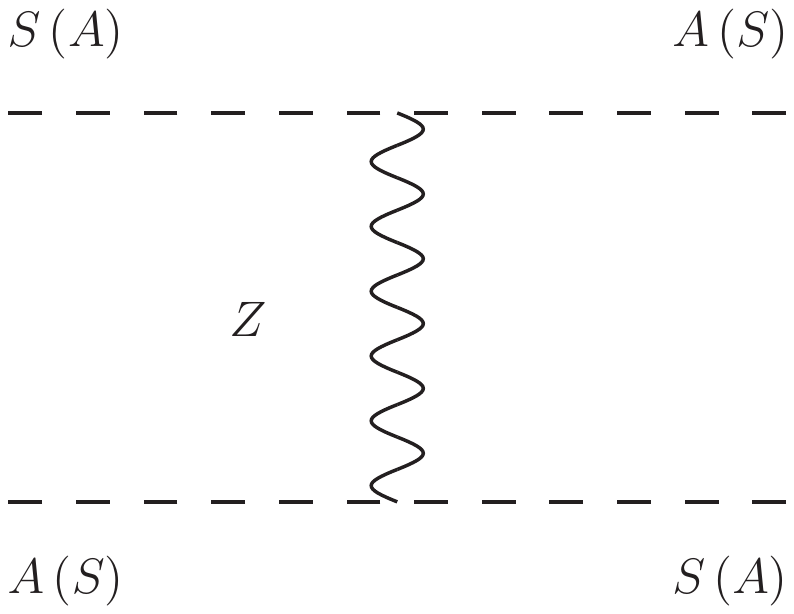}\hspace{-5.5cm}
 \includegraphics[width=10cm]{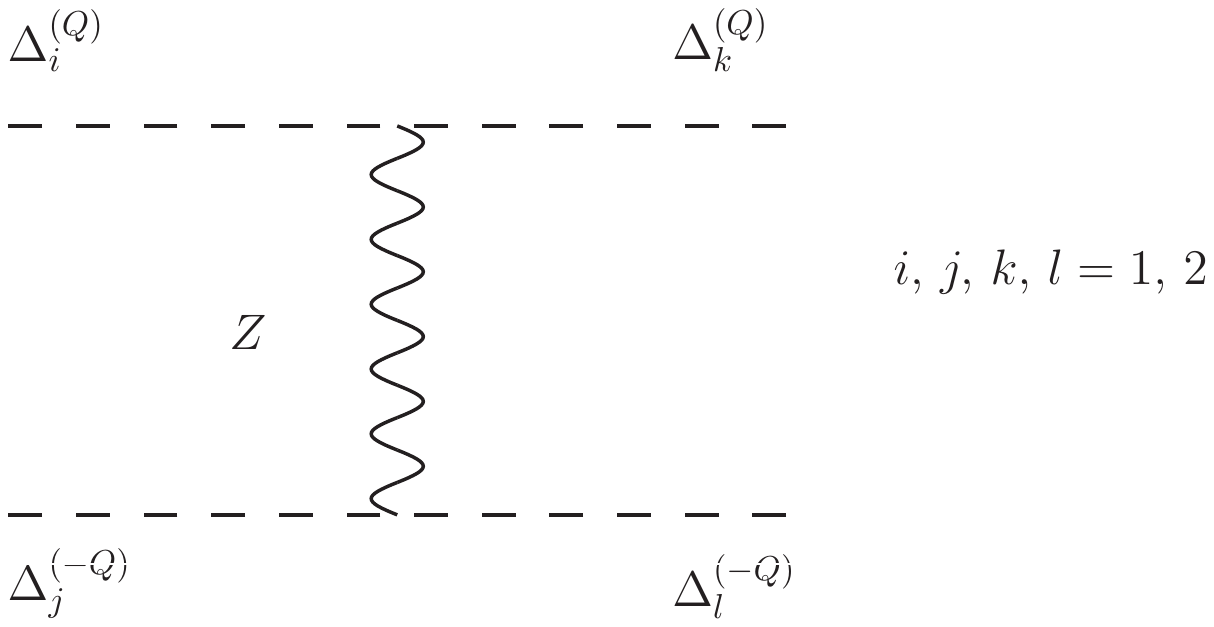}\hspace{-4cm}
 \includegraphics[width=10cm]{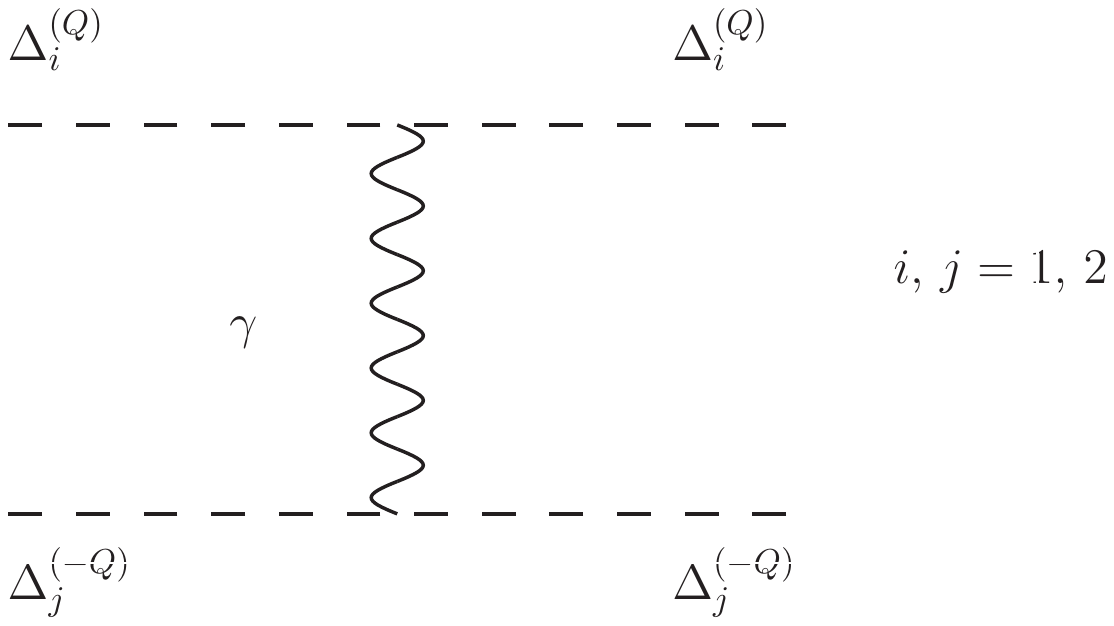}}\vspace{-11cm}
 \centerline{\includegraphics[width=12cm]{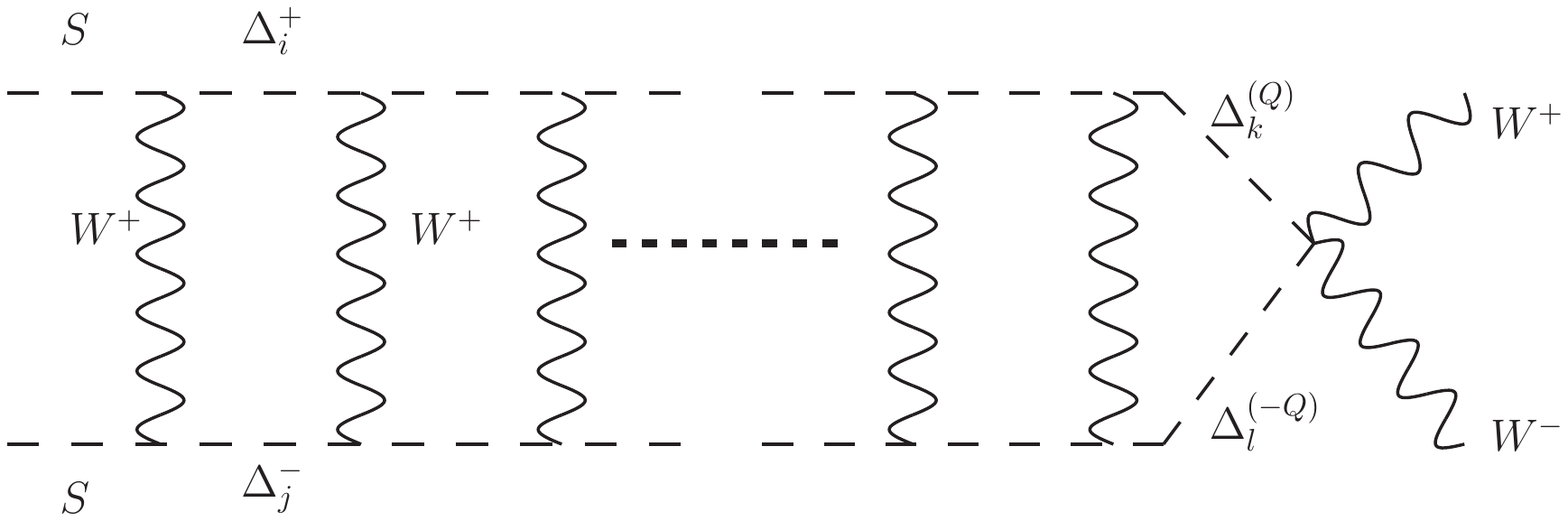}}
 \vspace{-11cm}
 \caption{In the NR limit, the Yukawa and Coulomb potential arising from the 
 exchange of $W$, $Z$ and $\gamma$ (first two rows) and at third row,
 an example of the ladder diagram arising
 from the multiple exchange of $W$ that will enhance $SS\rightarrow WW$ 
cross-section. 
 Similar ladder diagram will arise from multiple exchange of
 $Z$ and $\gamma$ and will also enhance $SS\rightarrow 
WW,\,ZZ,\,\gamma\gamma,\,Z\gamma$.}
 \label{exchangedig}
\end{figure}

The potential and the annihilation matrices arise from
the real and imaginary part of the 2-particle state Green's function \cite{Cirelli:2007xd}. 
Equivalently, by
integrating out the relativistic degrees of freedom, 
one can have effective potential and annihilation matrix
in the non-relativistic limit \cite{Hisano:2004ds, Camilo-thesis}. In the NR limit, the exchange of $W$ 
between charged currents $J^{\pm}_{W}$ and  $Z$ between neutral 
current $J_{Z}$ will lead to the Yukawa potential as shown in Fig. 
\ref{exchangedig}. If the relative distance between two DM 
particles 
in their center of Mass (C.M.),
frame is $\vec{r}$, 
the matrix element of this Yukawa potential will be of the form,
\begin{equation}
\langle ii' |V|jj'\rangle\equiv V_{ii',jj'}\sim \pm\frac{f_{ii',jj'}\alpha_{a} 
e^{-c_{a}m_{W}r}}{r}
\label{potential}
\end{equation}
Here, $|ii'\rangle$ and $|jj'\rangle$ denote the components of the 2-particle 
state vector given in Eq. (\ref{pstates}). 
Also, $f_{ii',jj'}$ is the 
the factor that involves the combination of group theoretical values associated 
with $|ii'\rangle$
and $|jj'\rangle$ 2-particle states and 
terms associated with the mixing between different mass eigenstates. Also,
$\alpha_{a}=\alpha$ and $c_{a}=1$ for $W$ boson exchange and 
$\alpha/\cos^{2}\theta_{W}$ and $
1/\cos\theta_{W}$ for $Z$ boson exchange respectively. In addition, the photon 
exchange
between $J_{A}$ currents leads to the Coulomb potential of the form 
$\frac{Q^2\alpha_{em}}{r}$.

The S-wave annihilation of 2-particle states into two gauge bosons, 
is the dominant channel in the non-relativistic limit. And it can be calculated 
from the imaginary or the absorptive part of the 2-particle propagator 
$ii'\rightarrow jj'$ as shown in Fig. \ref{annhildig} and is denoted by 
$\Gamma^{(f)}_{ii',jj'}$, where $f$ is the final state. In the appendix we 
have 
outlined the matrix elements of both potential and S-wave annihilation matrix 
elements related to the scotogenic model.
\begin{figure}[h!]
\vspace{-1cm}
\centerline{\includegraphics[width=10cm]{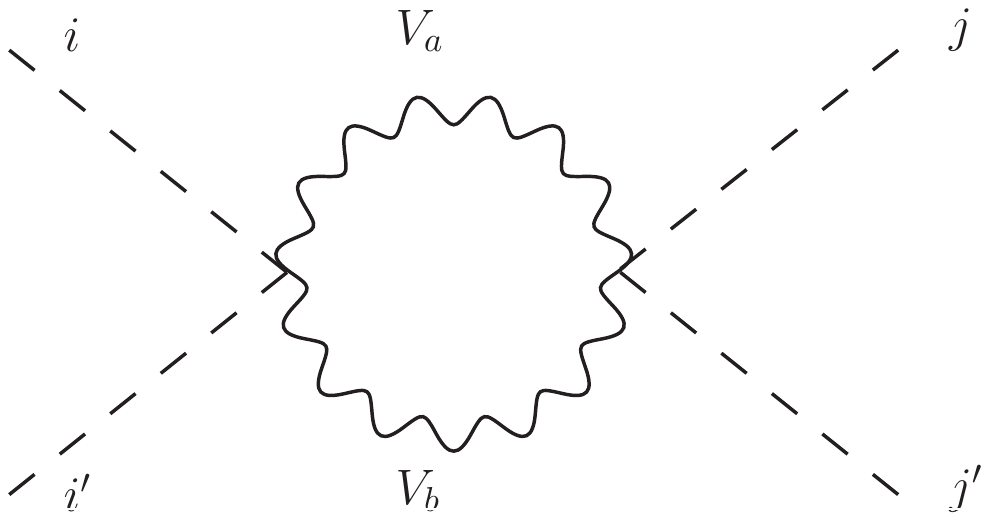}}
\vspace{-10cm}
\caption{The Feynman diagram related to the S-wave annihilation matrix 
elements. 
Here, $ii'$ and $jj'$ takes the 2-particle state components of 
$|\Phi_{\frac{n}{2}}\rangle$ and 
$V_{a}V_{b}=W^{+}W^{-},\,ZZ,\,\gamma\gamma,\,\gamma Z$.}
\label{annhildig}
\end{figure} 

\subsection{Sommerfeld enhanced annihilation cross section}\label{schrodingereq}
The radial Schrodinger matrix equation that determines the modified 
wavefunction 
in the presence of the effective potential $V$ is
\begin{equation}
 \frac{d^{2}\Psi_{jj',ii'}}{d 
r^2}+\left[\left((m_{S}v)^{2}-\frac{l(l+1)}{r^2}\right)\delta_{jj',kk'}- 
m_{S}V_{jj',kk'}(r)\right]
 \Psi_{kk',ii'}=0
 \label{schrod1}
\end{equation}
where, the kinetic energy of the incoming 2-particle state 
$|ii'=SS\rangle$ is, $E=m_{S}v^{2}$. The wavefunction $\Psi_{jj',ii'}$ gives 
the 
transition amplitude from $|ii'\rangle$ 
state to $|jj'\rangle$ state in the presence of $V$. Here $ii'$, $jj'$ and 
$kk'$ 
indices run over the components
of the 2-particle state vector, $|\Phi_{\Delta_{\frac{n}{2}}}\rangle$ given 
in 
Eq.(\ref{pstates}).

As our focus is on the S-wave annihilation, we set $l=0$ and have
\begin{equation}
 \frac{d^{2}\Psi_{jj',ii'}}{d r^2}+\left[k^{2}_{jj'}\delta_{jj',kk'}+
 m_{S}\left(\frac{f_{jj',kk'}\alpha_{a} 
e^{-n_{a}m_{W}r}}{r}+\frac{Q_{kk'}^{2}\alpha_{\text{em}}}{r}\delta_{jj',kk'}
\right)\right]
 \Psi_{kk',ii'}=0
 \label{schrod2} 
\end{equation}
Here, $k^{2}_{jj'}=m_{S}(m_{S}v^2-d_{jj'})$ is the momentum associated
with the 2-particle state, $|jj'\rangle$
and $d_{jj'}=m_{j}+m_{j'}-2m_{S}$ denotes the mass differences between DM and 
other states of
the multiplet. $Q_{kk'}$ is the electric charge associated with state 
$|kk'\rangle$. Also, $\alpha_{W}=\alpha$ and $n_{W}=1$ for W boson exchange and 
$\alpha_{Z}=
\alpha/\cos^{2}\theta_{W}$ and $n_{Z}=1/\cos\theta_{W}$ for Z boson exchange.

Now by using dimensionless variables defined as 
$x=\alpha m_{S} r$, $\epsilon_{\phi}=(m_{W}/m_{S})/\alpha$,
$\epsilon_{v}=(v/c)/\alpha$ and 
$\epsilon_{d_{ii'}}=\sqrt{d_{ii'}/m_{S}}/\alpha$, we re-write the
coupled radial Schrodinger equations as
\begin{equation}
\frac{d^{2}\Psi_{jj', ii'}}{dx^2}
+\left[\hat{k}_{jj'}^{2}\delta_{jj',kk'}+\frac{f_{jj',kk'}n_{a}^{2}e^{-n_{a}
\epsilon_{\phi}x}}{x}+\frac{Q_{kk'}^{2}\sin^{2}\theta_{W}}{x}\delta_{jj',kk'}
\right]\Psi_{kk',ii'}=0
\label{schrodeq}
\end{equation}
where the dimensionless momentum, 
$\hat{k}^{2}_{jj'}=\epsilon_{v}^2-\epsilon_{d_{jj'}}^{2}$.

At very large $x$, the solution of Eq.(\ref{schrodeq}) will become
$\Psi_{jj',ii'}\sim T_{jj',ii'}e^{i\hat{k}_{jj'}x}$
where $T$ is the $x$ independent amplitude of the wavefunction. As we have 
Yukawa potential in the diagonal and off-diagonal elements and Coulomb 
potential 
only in the diagonal elements of the potential, we would like to know which 
potential plays dominant role for the region of interest of the parameter 
space. 
It was pointed out in \cite{Cirelli:2007xd, Lattanzi:2008qa, Zhang:2013qza} that when $e_{v}\simlt 
e_{\phi}\simlt 1$ which is the case for today's DM velocity, $v\sim 10^{-3}$ 
and 
mass $m_{S}\in (1,\,30)$ TeV, the Yukawa potential will be more dominant than 
the Coulomb potential and will lead to resonances through the formation of 
(finite numbers of) zero-energy bound states. On the other hand, if 
$e_{\phi}\simlt e_{v}\simlt 1$, the Coulomb potential will be more important 
and 
lead to the formation of infinite number of (quasi-continuum) zero energy bound 
states and hence resonance behavior will be absent \cite{Zhang:2013qza}. Therefore as Yukawa 
potential is more dominant than Coulomb potential for our region of interest, 
in the subsequent numerical analysis, we focus on solving Eq.(\ref{schrodeq}) 
with Yukawa potential only. Finally using the optical theorem as in \cite{Hisano:2004ds}, the Sommerfeld enhanced
S-wave DM-DM annihilation cross section, $SS\rightarrow f$, is given by
\begin{equation}
\sigma_{SS\rightarrow f}=2(T.\Gamma^{(f)}.T^{\dagger})_{SS,SS}
\label{anncross}
\end{equation}
where the factor 2 appears as $|SS\rangle$ state has identical particles.

The Schrodinger equation in Eq.(\ref{schrodeq}) is solved using MATHEMATICA with modified Variable Phase Method described in \cite{martinazzo, Ershov:2011zz, Beneke:2014gja}. For highly degenerate mass spectrum, one can use the standard method described in 
\cite{Cohen:2013ama, Cirelli:2015bda, Garcia-Cely:2015dda} but when the mass-splitting becomes larger than the kinetic energy, which is the case for scotogenic model (section \ref{dmconstraintssec}), some components of the matrix solution become exponentially large while taking $x\rightarrow \infty$ and hence become numerically unstable but it can be alleviated with the above-mentioned method.

Moreover, Eq.(\ref{schrodingereq}) is written perturbatively at the zeroth order of absorptive part, $\Gamma$ as seen in Eq.(\ref{twofieldeffaction}). It has been shown in \cite{Blum:2016nrz}
that such expansion will lead to the violation of perturbative unitarity due to the formation of zero energy bound state at $v\sim 0$ and consistency requires the inclusion of the term $i\Gamma\delta^{(3)}(\vec{r})$ in the Schrodinger equation. 

\section{Result and Discussion}\label{resultdiscussionsection}
In this section we have presented the numerical results and discussed their 
implications on the large scalar multiplets. But before doing that, we have 
listed the DM constraints relevant for our study. Moreover, different scalar multiplets, for example the doublet and the quartet of the 
scotogenic model are controlled by same parameter set of the scalar potential 
$\{M,\,\alpha,\,\beta,\,\gamma\}$ but due to different mass spectrum, there 
are 
discernible phenomenological differences between smaller and larger multiplets. 
The collider constraints regarding these two multiplets is given in \cite{Chowdhury:2015sla}.

In this study our primary focus is on the indirect detection aspects of large 
scalar multiplet DM in the high mass regime ($m_{W}\ll m_{S}$) because at 
present the DM is non-relativistic
and has $O(\text{TeV})$ mass and therefore non-perturbative Sommerfeld 
enhancement 
takes place in the DM-DM annihilation cross sections to gauge or higgs boson 
pair. Such enhancement takes the
scalar DM of the generalized scotogenic model within the reach of H.E.S.S. \cite{Mora:2015vhq} and 
future CTA \cite{Doro:2012xx, Carr:2015hta}. We will see that how the current and future experimental limits 
put stricter constraints on the
larger scalar multiplets compared to the doublet case.

\subsection{DM constraints and allowed parameter space}\label{dmconstraintssec}

\paragraph{DM relic density} The relic density of the dark matter in the 
universe
is measured by Planck collaboration as 
$\Omega_{\text{DM}}h^{2}=0.1197\pm0.0022\,(68\%\,\text{C.L.})$ \cite{Ade:2015xua}. If scalar DM of 
the scotogenic model is the dominant component of the DM, this
relic density can be achieved either by thermal freeze-out or 
non-thermal process. For scalar DM, the thermal freeze-out processes are controlled by gauge and scalar interactions and proceed via the DM (co)annihilation into SM 
particles (for TeV scale DM, mostly into $WW$ and $ZZ$). It was shown in \cite{Hambye:2009pw} for doublet and \cite{Chowdhury:2015sla} for quartet that, certain bounds on mass splittings between the DM and other components of the scalar multiplet are to be satisfied so that scalar DM can have the correct relic density. One can also expect 
Sommerfeld enhancement of the (co)annihilation processes involved in thermal freeze-out. But as shown in \cite{Hisano:2006nn},
for the freeze-out temperature, $T_{F}$,
such that $m_{S}/T_{F}\sim 20$ (the typical freeze-out condition), the SE 
correction is not numerically significant and it only becomes important when 
$m_{S}/T_{F}\simgt 100$. Moreover it has been argued in \cite{Cohen:2013ama, Garcia-Cely:2015khw}
that the 
exclusion of SE in the thermal freeze out will modify the relic density at most 
by $30\%$.  

Apart from thermal freeze-out process, scalar DM in the scotogenic model can be produced non-thermally through the out-of-equilibrium decay of the fermion multiplet's components (for a quick review of non-thermal DM production please see \cite{Baer:2014eja}). For example, in the doublet case, singlet RH neutrino can decay into $S$ through $N_{1}\rightarrow S\,\nu$. For the quartet, components of triplet fermion produce $S$ through $F^{0}_{1}\rightarrow S\,\nu,\,F^{+}_{1}\rightarrow S\,l^{+}$. Here we consider the decay of lightest fermion multiplet, $N_{1}$ or $F_{1}$. Fermion triplet, unlike RH neutrino, will be kept in equilibrium by gauge interactions $F_{1}F_{1}\leftrightarrow VV$. In table \ref{table:1}, we have listed the decoupling temperature $T_{\text{dec}}$ below which the gauge interaction will be out of equilibrium for corresponding mass $m_{F_{1}}$. Also $\Gamma_{F_{1}}^{(\text{max})}$ is the maximum decay width of $F_{1}$ for $m_{F_{1}}$ so that inverse decay process will never be in equilibrium. By requiring $F$ to be decoupled from the thermal plasma, the corresponding decay width $ \Gamma_{F_{1}}$ and temperature $T_{D}$ are given in the following table for respective mass $m_{F_{1}}$.
\begin{table}[h!]
\centering
\begin{tabular}{|c|c|c|c|c|c|}
\hline
$m_{F_{1}}$ & $T_{\text{dec}}$ & $\Gamma_{F_{1}}^{(\text{max})}$ & $\Gamma_{F_{1}}$ & $T_{D}$\\
\hline
1 TeV & 31.25 GeV & $10^{-12}$ GeV & $10^{-16}$ GeV & 17 GeV\\
\hline
40 TeV & 1380 GeV & $10^{-9}$ GeV  & $5\times 10^{-13}$ GeV & 990 GeV\\
\hline
\end{tabular}
\caption{Decoupling temperature for gauge interaction $T_{\text{dec}}$, maximum decay width $\Gamma_{F_{1}}^{(\text{max})}$, decay width $\Gamma_{F_{1}}$ and temperature $T_{D}$ at decay for the respective masses $m_{F_{1}}$ of fermion triplet}
\label{table:1}
\end{table} 

Therefore in the case of $T_{D}\simlt T_{\text{dec}}$, non-thermal production will also contribute to the DM content of the universe. As a result, observed DM relic density is achievable for certain ranges of $\Gamma_{F_{1}}$ and $m_{S}$ and the bounds on mass splittings between DM and other scalar components coming from thermal freeze-out will be relaxed to some extent. For that reason,
we have focused on finding the role of the SE in present day DM annihilation rather than in the early universe. Besides the allowed range of mass splittings $|m_{i}-m_{S}|$ are set by the constraints on Electroweak Precision observables and DM direct detection bounds as we will see in subsequent paragraphs.

\paragraph{DM Direct Detection} 

In the scotogenic model, the elastic scattering of DM with nucleus is induced 
by 
the higgs exchange and thus controlled by the coupling $\lambda_{S}$ given in 
Eq.(\ref{dmhiggscoup}). The spin independent cross section is given by,
\begin{equation}
\sigma_{\text{SI}}=\frac{\lambda^{2}_{S}f^2}{4\pi}\frac{\mu^2 
m_{n}^2}{m^{4}_{h}m^{2}_{S}}
\label{directformula}
\end{equation}
Here, $\mu=m_n m_S/(m_n+m_S)$ is the DM-nucleon reduced mass. $f$
parameterizes the nuclear matrix element, $\sum_{u,d,s,c,b,t}\langle
n|m_q \bar{q}q|n\rangle\equiv f m_n \bar{n}{n}$ and from recent
lattice results \cite{Giedt:2009mr} $f=0.347131$.

\begin{figure}
\centerline{\includegraphics[width=8cm]{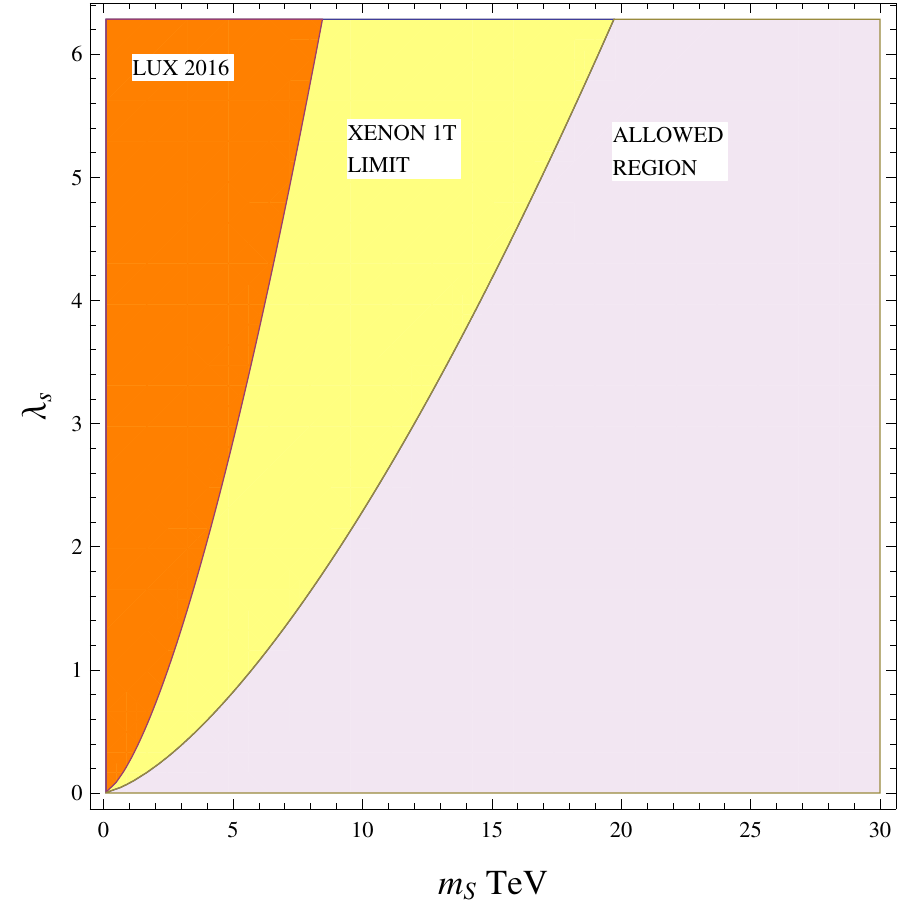}\hspace{0cm}
\includegraphics[width=8cm]{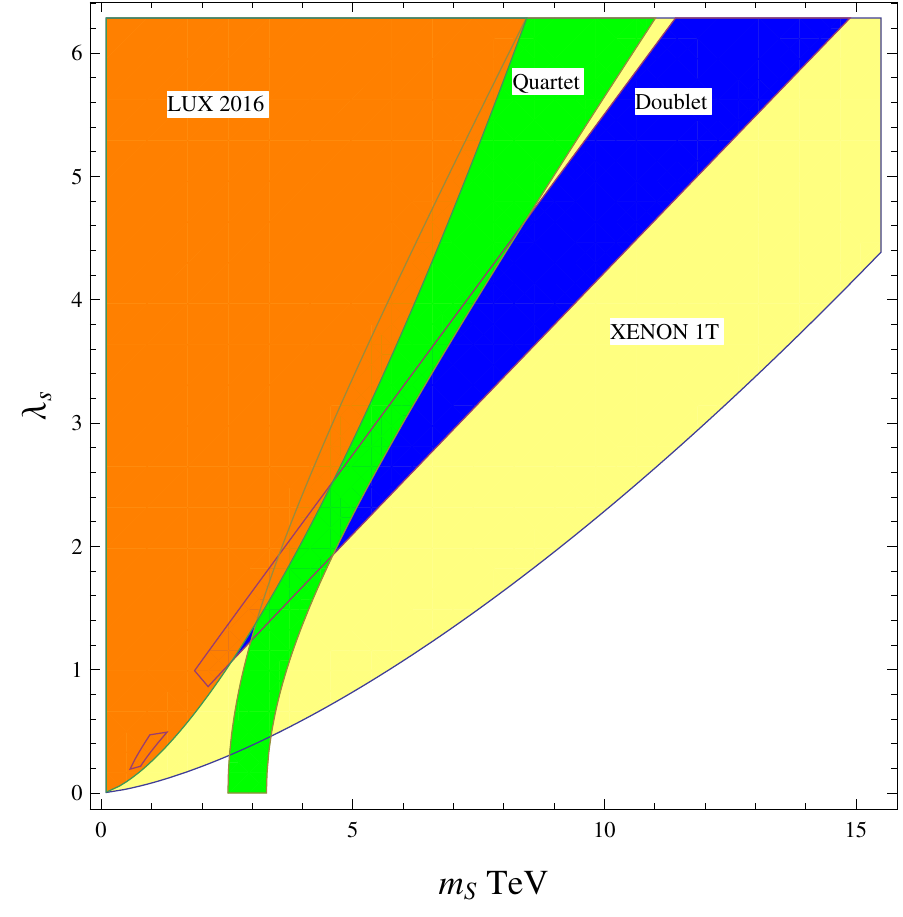}}
\caption{LUX(2016) exclusion limits and XENON 1T projected limits on 
$m_{S}-\lambda_{S}$ plane}
\label{directdig}
\end{figure}

The LUX 2016 \cite{Akerib:2016vxi} result has put limit on the $m_{S}-\lambda_{S}$ plane as 
shown in Fig. \ref{directdig} (left) and it can be seen that the direct 
detection experiments are reaching the sensitivity to probe dark matter in the 
high mass regime. Moreover the projected XENON 1T \cite{Aprile:2015uzo} can 
put 
stringent limit, if DM is not observed, on the $m_{S}-\lambda_{S}$ plane and will
reach the one loop corrected cross section of the order 
$O(10^{-48}-10^{-47}\text{cm}^{2})$ by $W$ and $Z$ bosons (as shown for the 
doublet in \cite{Klasen:2013btp}) even if $\lambda_{S}$ is tuned to be very 
small.

These direct detection limits also have important implications on the thermal 
freeze-out process in the scotogenic model. As we can see from Fig. 
\ref{directdig} (right) that LUX 2016 has already probed $1-5$ TeV and $1-7.5$ 
TeV region for doublet and quartet respectively. Here we have taken into 
account 
the $O(30\%)$ modification in the relic density for not considering SE 
correction in freeze-out. Finally, the entire thermal freeze-out region for 
both 
doublet and quartet is enclosed by the XENON1T sensitivity limit. 

\begin{figure}[h!]
\centerline{\includegraphics[width=8cm]{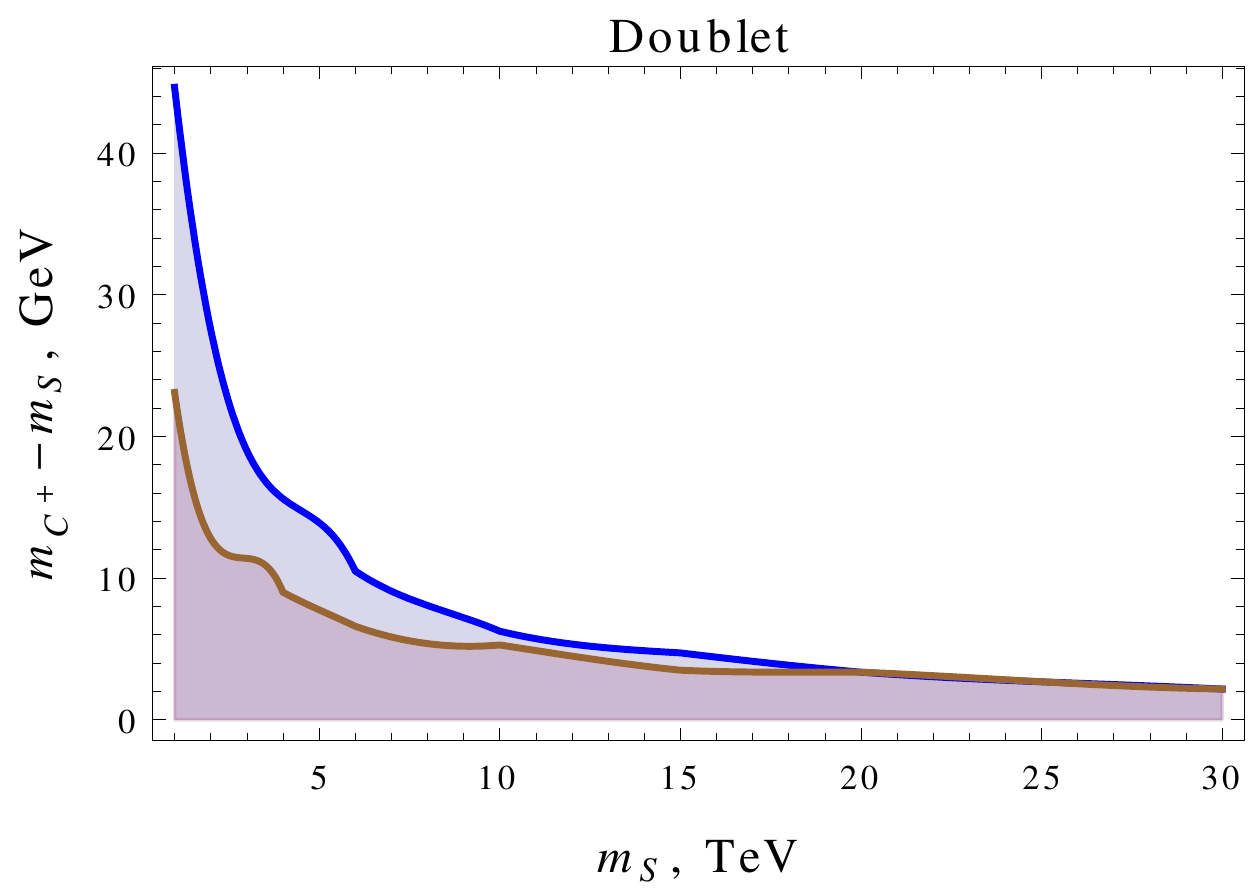}\hspace{0cm}
\includegraphics[width=8.1cm]{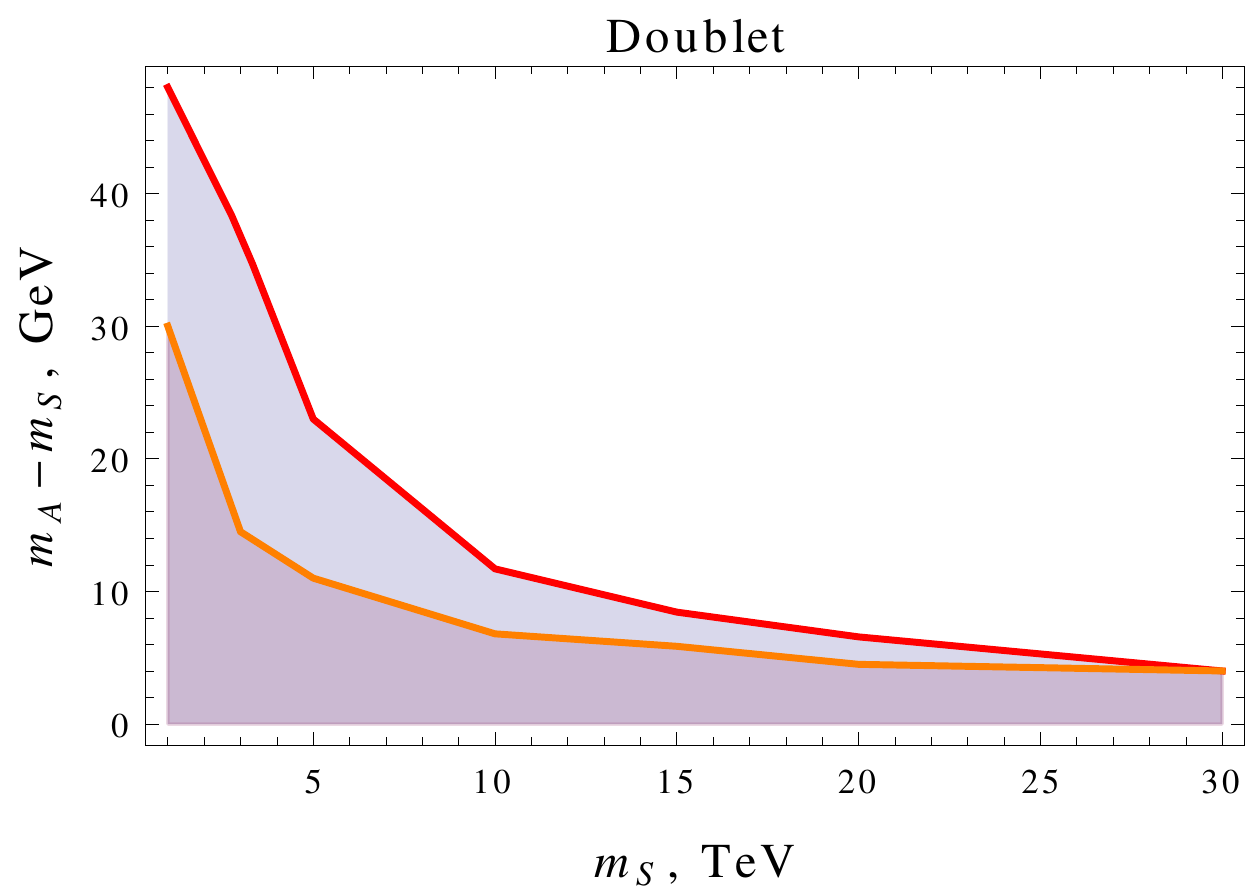}}\vspace{0cm}
\centerline{\includegraphics[width=8cm]{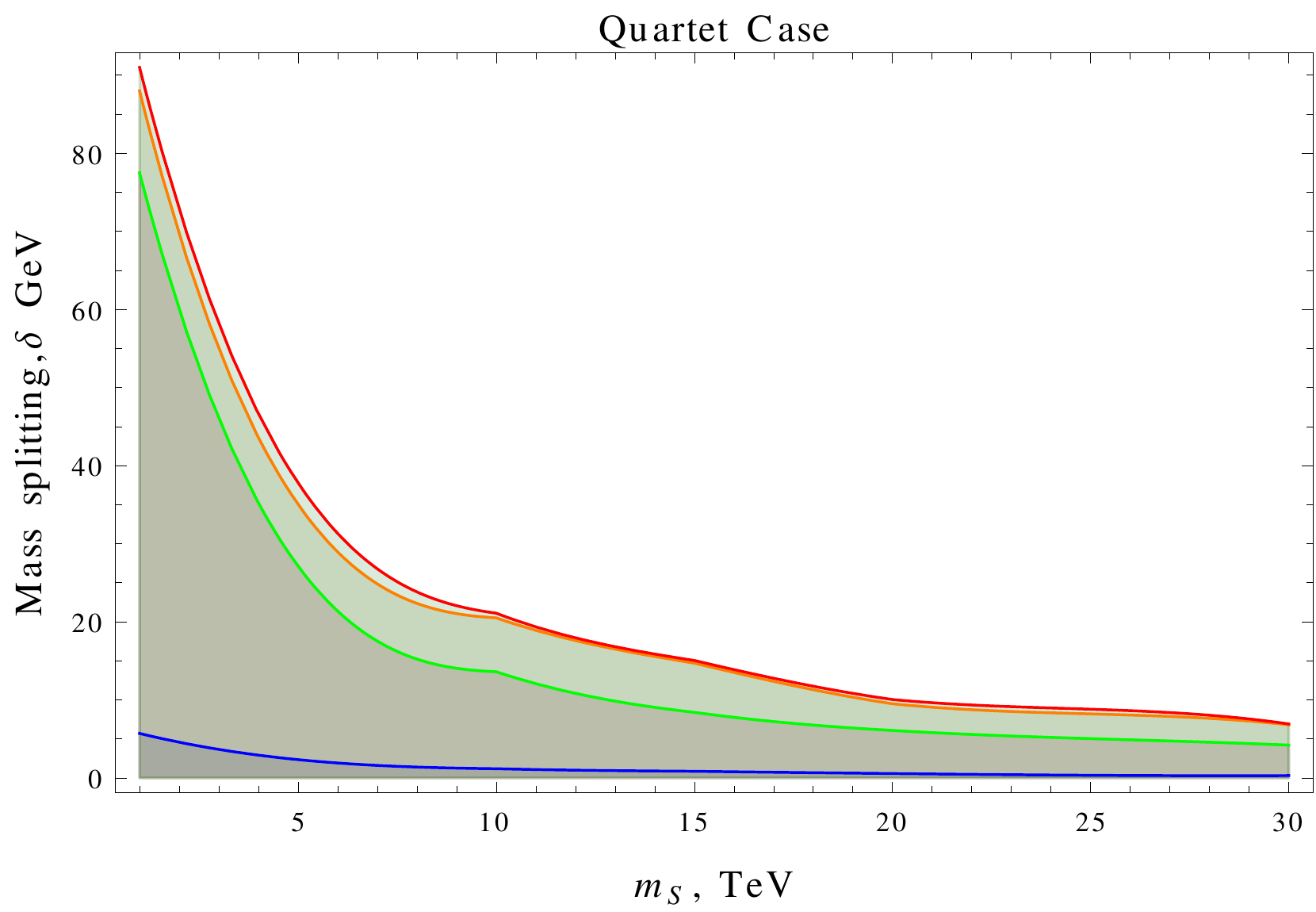}}
\caption{Allowed Mass splittings with DM mass, $m_{S}$ in the doublet and 
quartet cases. In the upper left fig. allowed maximum values of 
$\delta_{1}^{d}$ 
are marked by blue (LUX2016 result \cite{Akerib:2016vxi}) and brown (XENON1T \cite{Aprile:2015uzo}) lines. In 
the upper right fig. those of $\delta_{2}^{d}$ are marked by red (LUX2016) and 
orange (XENON1T) lines. In the lower fig. maximum values of $\delta^{q}_{1}$, 
$\delta^{q}_{2}$,$\delta^{q}_{3}$ and $\delta^{q}_{4}$, allowed by both LUX2016 
and XENON1T, are marked by blue, green, orange and red lines 
respectively.}\label{allowedmasssplit}
\end{figure}

\paragraph{Mass splitting in the scalar multiplet}\label{masssplitpara} Unlike 
the case of the minimal DM model where the mass splitting between the DM and 
the 
charged state of the multiplet
is induced radiatively after symmetry breaking and of the order 
$O(100\,\text{MeV})$ \cite{Cirelli:2005uq}, in the scotogenic model the mass 
splitting between the components of the scalar multiplet is set by the 
following 
terms in the potential,
\begin{equation}
V_{0}\supset \beta H^\dagger  \tau^a H \Delta^\dagger T^a \Delta +\gamma\left[ 
(H^T\epsilon \tau^a H) (\Delta^T C T^a \Delta)^\dagger+h.c\right]
\label{masssplitterms}
\end{equation}

The mass splittings among the components of the multiplet is constrained by 
Electroweak
Precision observables (EWPO) \cite{Kennedy:1988sn, Peskin:1991sw, Barbieri:2004qk, Olive:2016xmw}. They, in turn, put constraints on the allowed range for the 
couplings $\beta$ and $\gamma$. We use the results of the EWPO for the doublet 
\cite{Barbieri:2006dq, Martinez:2011ua, Melfo:2011ie, Chowdhury:2011ga} and the quartet \cite{AbdusSalam:2013eya} and determine the allowed 
splittings. The mass splittings are defined as
\begin{align}
\delta^{d}_{1}&=m_{\Delta^{+}}-m_{S}\,\,\,\text{and}\,\,\,\delta^{d}_{2}=m_{A}
-m_{S}\,\,\,\text{(doublet)}\nonumber\\ 
\delta^{q}_{1}&=m_{\Delta^{+}_{1}}-m_{S},\,\,\delta^{q}_{2}=m_{\Delta^{++}}-m_{S
},\,\,
\delta_{3}^{q}=m_{\Delta^{+}_{2}}-m_{S}\,\,\,\text{and}\,\,\,\delta^{q}_{4}=m_{A
}-m_{S}\,\,\,\text{(quartet)}
\label{masssplitdoubletquartet}
\end{align}

In addition, there is a lower bound on the mass splitting between scalar and 
pseudoscalar components in the scalar multiplet which arises due to the bounds 
on 
DM inelastic scattering with nuclei. As the $Z$ boson mediated
inelastic scattering with nuclei is of the order of 
$10^{-40}-10^{-39}\,\text{cm}^2$ and much
larger than the current direct detection limits, the mass splitting between $S$
and $A$ has to be large enough for this scattering to be kinematically 
forbidden 
and satisfy the
direct detection limits. If the velocity of DM is $v$, then the 
minimum mass splitting $\delta_{\text{min}}$ is
\begin{equation}
\delta_{\text{min}}=\frac{m_{S}M_{\text{nucleus}}v^{2}}{2(m_{S}+M_{\text
{nucleus}})}
 \label{masssplit1}
\end{equation}
For, $v\sim 10^{-3}$ and $M_{\text{nucleus}}\sim 130$ GeV, 
$\delta_{\text{min}}$
ranges from $\delta_{\text{min}}=57.5$ keV (for, $m_{S}=1$ TeV) to 
$\delta_{\text{min}}\sim 65$ keV (for, $m_{S}\gg O(\text{TeV})$). The allowed 
mass splittings for both the doublet and quartet cases are presented in Fig. 
\ref{allowedmasssplit}. In passing, from Eq.(\ref{highdim}) we see that 
$\delta_{\text{min}}$ for complex odd dimensional scalar multiplet with 
$m_{S}\sim 1$ TeV and $c\sim O(1)$ sets the cut off scale to be $\Lambda\sim 
O(200)$ TeV for above values of $v$ and $M_{\text{nucleus}}$.

Moreover, the direct detection limits put an upper bound on $|\lambda_{S}|$ and 
in turn constrain the allowed range of the couplings $\beta$ and $\gamma$ which translates into the allowed mass splittings for the scalar multiplets. As 
can be seen from Fig. \ref{allowedmasssplit} (upper panel), the LUX 2016 
results and XENON1T projected limits put additional constraints on the mass 
splittings apart from EWPO bounds on the doublet. On the other hand, for 
quartet, 
EWPO bounds are consistent with direct detection limits.

The mass splittings in the scalar multiplet have important impact on the 
Sommerfeld enhancement of the annihilation cross sections. If the mass 
splitting 
is much larger than the kinetic energy of the incoming DM particles, the almost
on-shell exchange of $SU(2)_{L}$ partner states in the ladder diagram 
which 
modifies the wavefunction and enhances the cross section, will be largely 
suppressed and for this reason, even if $\epsilon_{v}\simlt 1$ and 
$\epsilon_{\phi}\simlt 1$, the SE will be negligible for annihilation cross 
sections.

\subsection{Sommerfeld enhanced Cross sections}\label{SEcross-sectionresult}

In this section we present the Sommerfeld enhanced annihilation cross sections 
with DM mass to various final states that are relevant for indirect detection.

\subsubsection{$SS\rightarrow WW$} 

We have determined the cross sections for two cases. The 
first case is the (almost) degenerate limit, where the mass splittings among 
the 
components of the scalar multiplet, are set to their minimum values allowed by 
the constraints of section \ref{dmconstraintssec} and the second case is the 
maximal mass splitting limit where the mass splittings are set at their 
maximum 
allowed values. We have considered these two cases as the mass splitting is an 
important parameter for the SE in the scotogenic model. Also we have set the DM 
velocity $v=10^{-3}$, which is the scale of average velocity in the galactic halo. 

\begin{figure}[h!]
\centerline{\includegraphics[width=8.5cm]{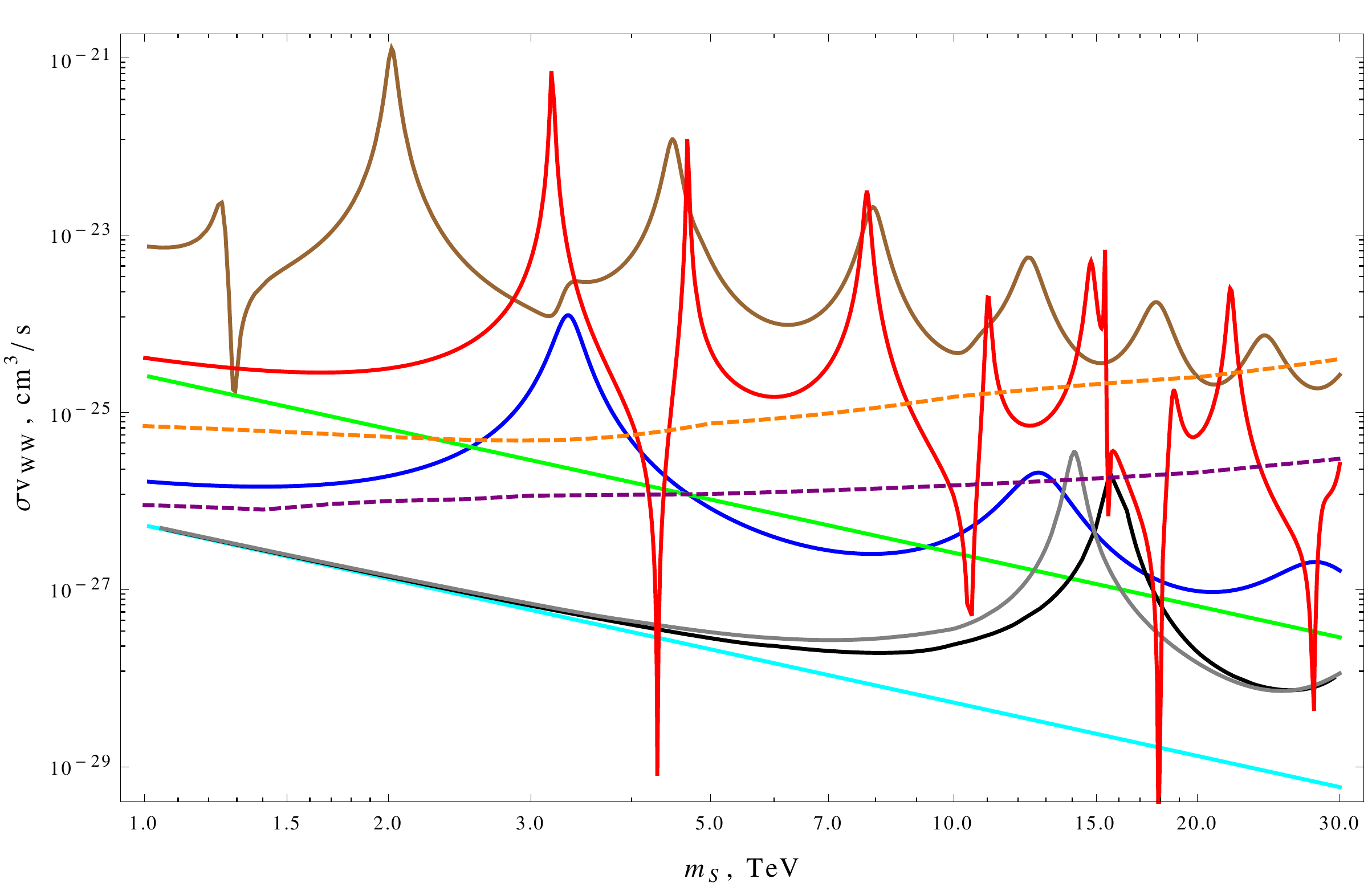}\hspace{0cm}
\includegraphics[width=8.5cm]{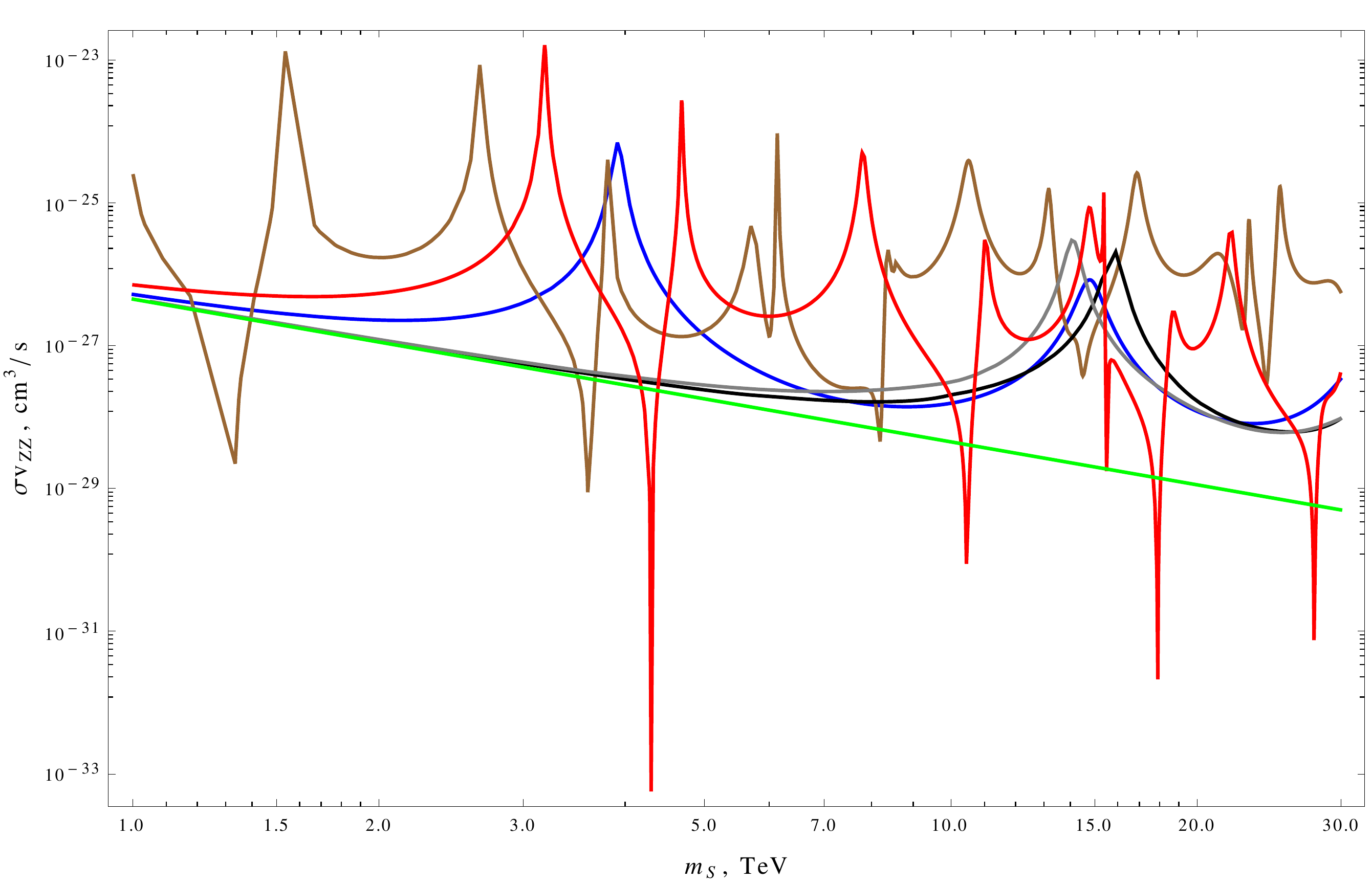}}
\caption{In the left fig. correlation between $\sigma v_{WW}$ and $m_{S}$. The 
blue (doublet) and brown (quartet) lines represent the annihilation cross 
section to $WW$ in the (almost) degenerate limit. The black (doublet, LUX2016), 
grey (doublet, XENON1T) and red (quartet) lines represent the cross section 
when 
mass splittings are taken as the maximum of allowed limit. The light blue 
(doublet) and green (quartet) lines are the tree-level annihilation cross 
sections. Moreover, the orange and purple dashed lines are H.E.S.S. \cite{::2016jja} and 
future CTA \cite{Carr:2015hta} limits respectively on $WW$ annihilation. Similarly in the right 
fig. correlation between $\sigma v_{ZZ}$ and $m_{S}$ for (almost) degenerate 
limit: blue (doublet) and brown (quartet) lines and for maximum limits of mass 
splittings: black (doublet, LUX2016), grey (doublet, XENON1T) and red (quartet) 
lines. The green line is the tree-level $\sigma v_{ZZ}$ for both doublet and 
quartet.}
\label{WWandZZse}
\end{figure}

In Fig. \ref{WWandZZse} (left) the blue and brown lines correspond to the 
doublet and quartet at their (almost) degenerate limit respectively. As we can see, the cross 
section in the quartet case is larger than that of the doublet. The 
annihilation 
matrix elements of $\Gamma^{(WW)}$ depends on the factor $j^2+j-m^2$ (appendix \ref{annihilationmatrix}) 
which increases with the isospin, $j$ of the multiplet. Moreover, the 
off-diagonal potential matrix elements involve the factor 
$V^{+2}_{j,m}=j^2+j-m-m^2$ which also depends on isospin. Therefore, in case of 
larger multiplets, relatively large Yukawa potential and annihilation matrix 
elements increase the Sommerfeld corrected annihilation cross section compared 
to the doublet case. The first resonance peak for the doublet and the quartet 
has occurred at $3.1$ TeV and $2$ TeV respectively. There is a dip at $1.3$ TeV 
in the quartet case though. The tree-level annihilation cross sections are 
marked as light blue (doublet) and green (quartet) lines.

Now we focus on the maximal mass splittings on the Sommerfeld enhanced cross 
sections. From Fig. \ref{WWandZZse} (left), we can see that, for doublet the 
resonance peak has been shifted to $14$ TeV (grey; for XENON1T mass splittings) 
and $16$ TeV (black; for LUX (2016) mass splittings). For $1-7$ TeV range, the 
Sommerfeld corrected cross section is numerically comparable to the tree level 
cross section in the case of doublet. On the other hand, for quartet (red) the 
resonance peak is shifted to $3$ TeV and the dip occurs at $4.2$ TeV. In this 
case, only for $1-2.5$ TeV range, the SE and tree-level cross sections are 
numerically comparable. In quartet case, from Fig. \ref{allowedmasssplit}, we 
see that for the mass splitting between $S$ and next to lightest component, 
$\tilde{\Delta}^{+}_{1}$, we have $3.3\simgt \epsilon_{\delta}\simgt 1.8$ and 
lead to comparable tree-level and SE cross sections. Again we have a small 
strip 
about $3.8$ TeV where again tree-level and SE cross sections are comparable. 

The orange and purple dashed lines in Fig. \ref{WWandZZse} (left) refer to the H.E.S.S. 
result and C.T.A sensitivity limits on DM annihilating into $WW$ which we have 
projected on appropriate plots without carrying out any comprehensive analysis in this work. 
The projected limits are to be used 
as a reference to compare Sommerfeld enhanced
cross sections of doublet and quartet. The complete analysis will be reported 
elsewhere. In section \ref{discussion} we have given qualitative explanation of 
resonance peaks and dips and their shifts towards smaller or larger mass.

\subsubsection{$SS\rightarrow ZZ$}\label{zzannsec}

Another important annihilation channel is the $SS\rightarrow ZZ$. In Fig. 
\ref{WWandZZse} (right), again the blue and brown lines represent the doublet 
and quartet ZZ annihilation in the (almost) degenerate limit. The first 
resonance peak for the doublet and quartet are $3.8$ and $1.6$ TeV respectively 
and the first dip occurs for quartet at $1.4$ TeV. In the maximal mass 
splitting 
limit, resonance peaks occur at 14 and 16 TeV respectively for the 
doublet: (grey: XENON1T mass splitting) and (black: LUX2016 mass 
splitting) respectively. For quartet (red), the first resonance and dip occur at 
3.1 and 4.2 TeV respectively. Again the mass range where the tree-level and SE 
cross sections are comparable, are $1-9$ TeV for doublet and $1-2.5$ TeV for 
quartet. Here, the maximum of $SS\rightarrow ZZ$ is at the order of 
$10^{-23}\,\text{cm}^{3}\text{s}^{-1}$ compared to 
$10^{-21}\,\text{cm}^{3}\text{s}^{-1}$ of $SS\rightarrow WW$ because in this 
case, the potential and annihilation matrix elements depend on the 
corresponding 
$T_{3}$ eigenvalues instead of $j$ itself like in $WW$ case.

\subsubsection{$SS\rightarrow \gamma \gamma\,\&\,\gamma Z$}\label{gammagammasec}
\begin{figure}[h!]
\centerline{\includegraphics[width=8.5cm]{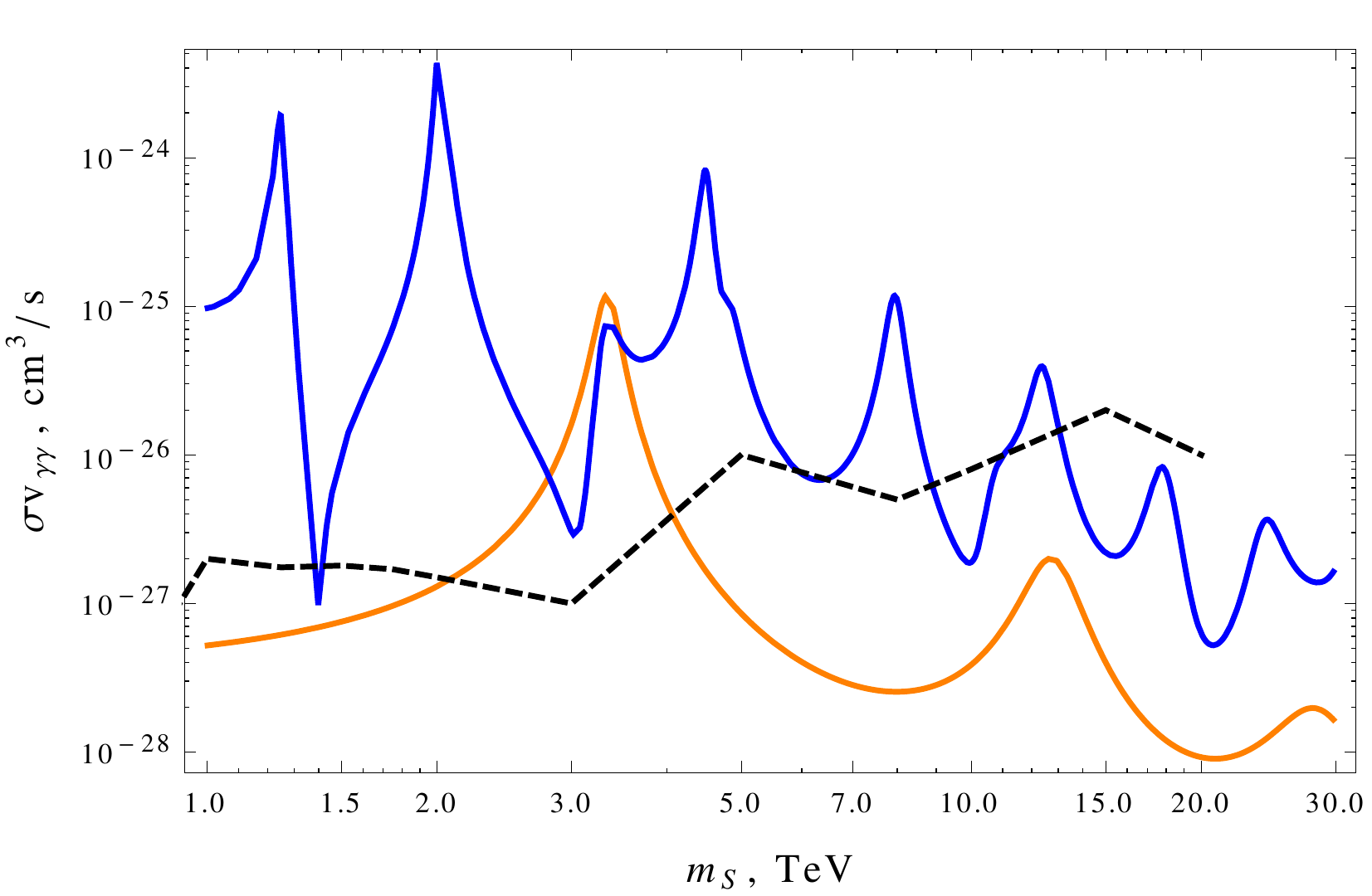}\hspace{0cm}
\includegraphics[width=8.5cm]{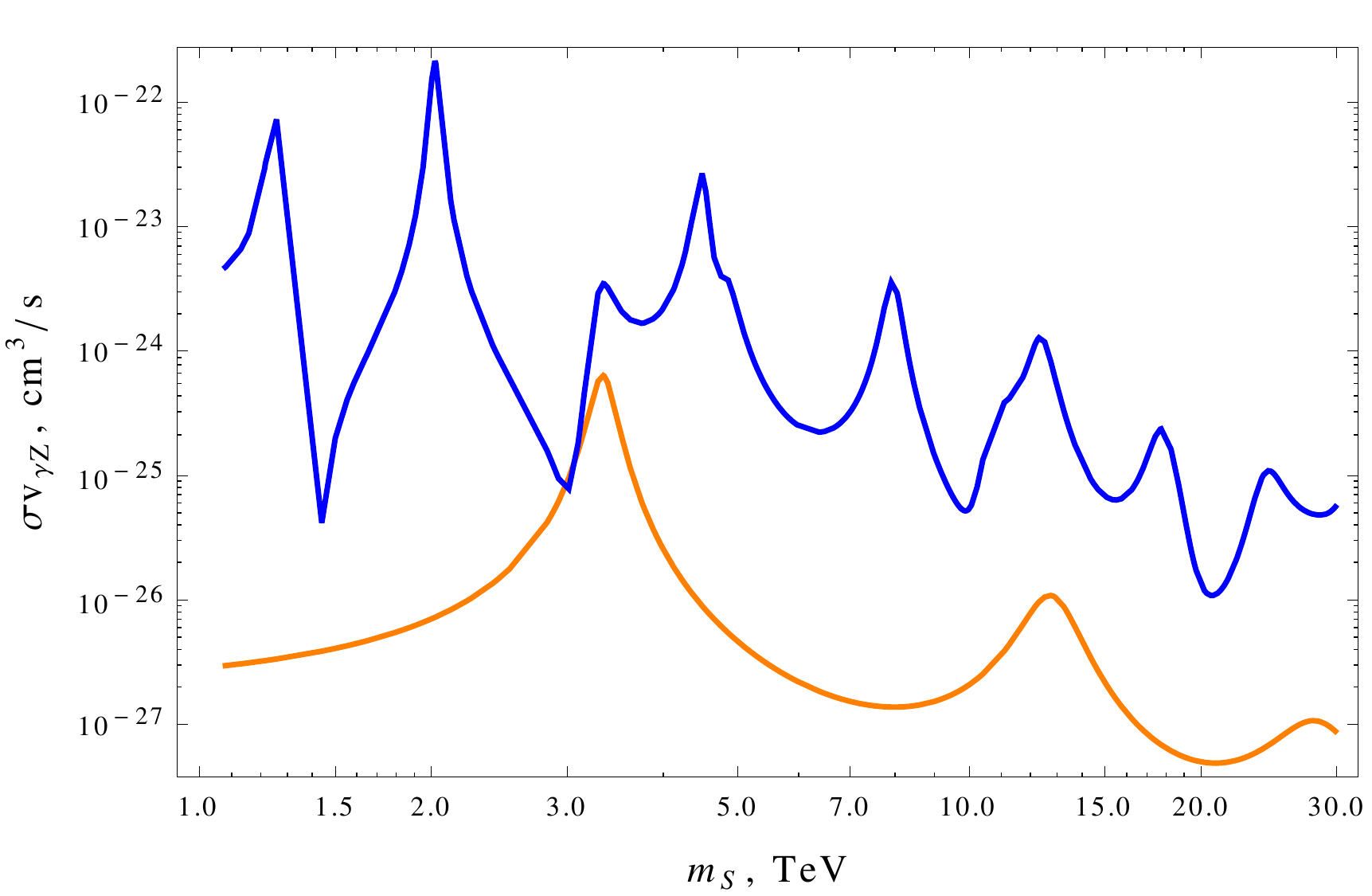}}
\caption{Correlation of $\sigma v_{\gamma\gamma}$ (left fig.) and $\sigma 
v_{\gamma Z}$  (right fig.) with $m_{S}$ for the doublet (orange line) and 
quartet (blue line) cases at the almost degenerate limits. Also the black 
dashed 
line in left fig is the H.E.S.S. limits on $\gamma\gamma$ annihilation \cite{Abdalla:2016olq}.}
\label{ggandgZcrossec}
\end{figure}

At almost degenerate limit, Fig. \ref{ggandgZcrossec} (left) and (right) 
describe $SS\rightarrow \gamma\gamma$ and $SS\rightarrow \gamma Z$ respectively 
for doublet (orange) and quartet (blue).
The occurrence of the resonances and dips in both processes are almost the same. The 
only difference is that for $SS\rightarrow \gamma\gamma$ the maximum peaks are 
of the order of $10^{-25}\,\text{cm}^{3}\text{s}^{-1}$ (doublet) and 
$10^{-24}\,\text{cm}^{3}\text{s}^{-1}$ (quartet). On the other hand, for 
$SS\rightarrow \gamma Z$, the maximum peaks are of the order 
$10^{-24}\,\text{cm}^{3}\text{s}^{-1}$ (doublet) and 
$10^{-22}\,\text{cm}^{3}\text{s}^{-1}$ (quartet). 

When the mass splittings are taken to be their maximum allowed limits, both 
$SS\rightarrow \gamma\gamma$ and $SS\rightarrow \gamma Z$ are greatly 
suppressed 
because large mass splittings don't allow the intermediate charged state to be 
on-shell (as the momentum becomes imaginary for $v\sim 10^{-3}$) in the ladder 
diagram and thus $T_{SS,\tilde{\Delta}^{(Q)}_{i}\tilde{\Delta}^{(-Q)}_{i}}$ 
factors that enter in the cross-section for $\SS\rightarrow \gamma\gamma$ are 
effectively zero. Similar arguments also hold for $SS\rightarrow \gamma Z$ 
case. 

Consequently, in such limit both processes can be induced by one-loop 
contribution of charged states. For the mass range of $1-30$ TeV, in case of 
$SS\rightarrow \gamma\gamma$, the cross sections are of the order, 
$10^{-28}-10^{-31}\,\text{cm}^{3}\text{s}^{-1}$ (both doublet and quartet) and 
in case of $SS\rightarrow \gamma Z$, they are of the order, 
$10^{-29}-10^{-32}\,\text{cm}^{3}\text{s}^{-1}$ (doublet) and 
$10^{-27}-10^{-30}\,\text{cm}^{3}\text{s}^{-1}$ (quartet).

\subsection{Discussion}\label{discussion}
In section \ref{SEcross-sectionresult} we have seen the shift of resonance 
peaks 
with increasing mass splitting and the occurrence of dips for the quartet case 
which was absent in the doublet case. Here we have used a two state system with 
finite square well potential to illustrate these features in a simple way.

The particle state is given by 
$|\chi\rangle=(\chi_{1},\chi_{2})$ and $\chi_{1}$ is considered to be DM state. 
If the 2-particle state vector is
$|\Psi\rangle=(\chi_{1}\chi_{1},\,\chi_{2}\chi_{2})$, the square-well 
potential in this basis is
\begin{equation}
 V=\begin{pmatrix}
    0 & -|V_{12}|\\
    -|V_{12}| & -|V_{22}|
   \end{pmatrix}\theta(L-r) +
   \begin{pmatrix}
    0 & 0\\
    0 & 2\delta
   \end{pmatrix}
\nonumber
\end{equation}
where, $L$ is the range of the potential. Moreover, the mass splitting is, 
$\delta=m_{\chi_{2}}-m_{\chi_{1}}$. The eigenvalues giving attractive and 
repulsive potential energy and the corresponding eigenstates are
\begin{equation}
 \lambda_{\pm}=\frac{1}{2}\left(-|V_{22}|+2\delta\pm 
\sqrt{(|V_{22}|-2\delta)^{2}+4 V_{12}^{2}}\right)\,\,
 \text{and}\,\,|\lambda_{\pm}\rangle=\begin{pmatrix}
	\frac{\lambda_{\mp}}{V_{12}\sqrt{1+\frac{\lambda_{\mp}^{2}}{V_{12}^{2}}}}\\
	\frac{1}{\sqrt{1+\frac{\lambda_{\mp}^{2}}{V_{12}^{2}}}}            
                   \end{pmatrix}
\label{2egvalue}
\end{equation}
The transition amplitude from $|\lambda_{a}\rangle$ state inside the potential 
well to $|i\rangle$ state outside the region of the potential, is given by 
\begin{equation}
T_{ai}=\frac{1}{\cos(k_{\text{in},a}L)-i\frac{k_{\text{out},i}}{k_{\text{in},a}}
\sin(k_{\text{in},a}
 L)}\label{tmatrix1}
\end{equation}
where, $k_{\text{in},a}=\sqrt{m_{\chi_{1}}(m_{\chi_{1}}v^{2}-\lambda_{a})}$ 
with 
$a=+,-$,
denotes the momentum inside the potential for $|\lambda_{\pm}\rangle$ states. 
And 
$k_{\text{out},1}=m_{\chi_{1}}v$ and 
$k_{\text{out},2}=\sqrt{m_{\chi_{1}}(m_{\chi_{1}}v^{2}-2\delta)}$
denote the momentum associated with $|\chi_{1}\chi_{1}\rangle$ and 
$|\chi_{2}\chi_{2}\rangle$
states respectively in the outside region of the potential.

\paragraph{1. $\mathbf{|V_{12}|\gg |V_{22}|,\,\delta}$ limit} In this limit, we 
have the eigenvalues,
\begin{equation}
 \lambda_{+}\sim |V_{12}|-\frac{1}{2}|V_{22}|+\delta,\,\,\,\lambda_{-}\sim 
-|V_{12}|-\frac{1}{2}|V_{22}|+\delta\label{sqegvalue1}
\end{equation}
and the eigenstates and the transformation matrix are,
\begin{equation}
 |\lambda_{+}\rangle=\begin{pmatrix}
                         c_{1+}\\
                         c_{2+}
                        \end{pmatrix},\,\,\,
 |\lambda_{-}\rangle=\begin{pmatrix}
                     c_{1-}\\
                     c_{2-}
                     \end{pmatrix},\,\,\,
 U_{ia}=\begin{pmatrix}
 c_{1+} & c_{1-}\\
 c_{2+} & c_{2-}
 \end{pmatrix}
 \label{sqegstate1}
\end{equation}

Here, the matrix elements $c_{ia}$ are,
\begin{align}
c_{1+} &\sim \frac{1}{\sqrt{2}}\left(-1-\frac{|V_{22}|}{4 
|V_{12}|}+\frac{\delta}{2 |V_{12}|}\right),\,\,\,c_{1-}\sim 
\frac{1}{\sqrt{2}}\left(1-\frac{|V_{22}|}{4 |V_{12}|}+\frac{\delta}{2 
|V_{12}|}\right)\nonumber\\
c_{2+} &\sim \frac{1}{\sqrt{2}}\left(1-\frac{|V_{22}|}{4 
|V_{12}|}+\frac{\delta}{2 |V_{12}|}\right),\,\,\,
c_{2-} \sim \frac{1}{\sqrt{2}}\left(+-\frac{|V_{22}|}{4 
|V_{12}|}-\frac{\delta}{2 |V_{12}|}\right)
\label{sqtrans1}
\end{align}
 
The transition amplitudes $T_{ai}$ in the limit, $v\rightarrow 0$\footnote{we take this limit to simplify the expressions further.}, are
\begin{align}
 T_{+i} &\sim 
\left[\cosh\left(\sqrt{m_{\chi_{1}}|V_{12}|}L\left(1+\frac{\delta}{2|V_{12}|}
-\frac{|V_{22}|}{4 |V_{12}|}\right)
 \right)+\sqrt{\frac{2\delta_{i}}{|V_{12}|}}\right.\nonumber\\
 &\hspace{5cm}\left.\sinh\left(\sqrt{m_{\chi_{1}}|V_{12}|}L\left(1+\frac{\delta}{
2|V_{12}|}-\frac{|V_{22}|}{4 |V_{12}|}\right)
 \right)\right]^{-1}\label{sqtrans11}\\
 T_{-i} &\sim 
\left[\cos\left(\sqrt{m_{\chi_{1}}|V_{12}|}L\left(1-\frac{\delta}{2|V_{12}|}
+\frac{|V_{22}|}{4 |V_{12}|}\right)
 \right)-\sqrt{\frac{2\delta_{i}}{|V_{12}|}}\right.\nonumber\\
 &\hspace{5cm}\left.\sin\left(\sqrt{m_{\chi_{1}}|V_{12}|}L\left(1-\frac{\delta}{
2|V_{12}|}+\frac{|V_{22}|}{4 |V_{12}|}\right)
 \right)\right]^{-1}\label{sqtrans12}
\end{align}
where, $\delta_{1}=0$ and $\delta_{2}=\delta$.

Now the transmission amplitude, $d_{ij}$ in the $|\chi_{i}\chi_{i}\rangle$ 
basis 
is,
\begin{equation}
 d_{ij}=U_{ia}T_{aj}\label{dmatsq}
\end{equation}
And if the annihilation matrix $\Gamma^{(VV)}$ is taken as,
\begin{equation}
 \Gamma^{(VV)}=\frac{\pi\alpha^2}{m_{\chi_{1}}^{2}}\begin{pmatrix}
         1 & 1\\
         1 & 1\\
        \end{pmatrix}
\label{testannmat}
\end{equation}
the Sommerfeld enhanced cross section for $\chi_{1}\chi_{1}\rightarrow VV$ is,
\begin{equation}
 \sigma_{11}=(d\Gamma d^{\dagger})_{11}=\frac{\pi\alpha^{2}}{m_{\chi_{1}}^{2}}
 \left(|d_{11}|^{2}+2\text{Re}(d_{11}^{*}d_{12})+|d_{12}|^{2}\right)\label{
sqanncross}
\end{equation}
where, $d_{11}=c_{1+}T_{+1}+c_{1-}T_{-1}$ and 
$d_{12}=c_{1+}T_{+2}+c_{1-}T_{-2}$. 

\begin{figure}[h!]
\centerline{\includegraphics[width=7cm]{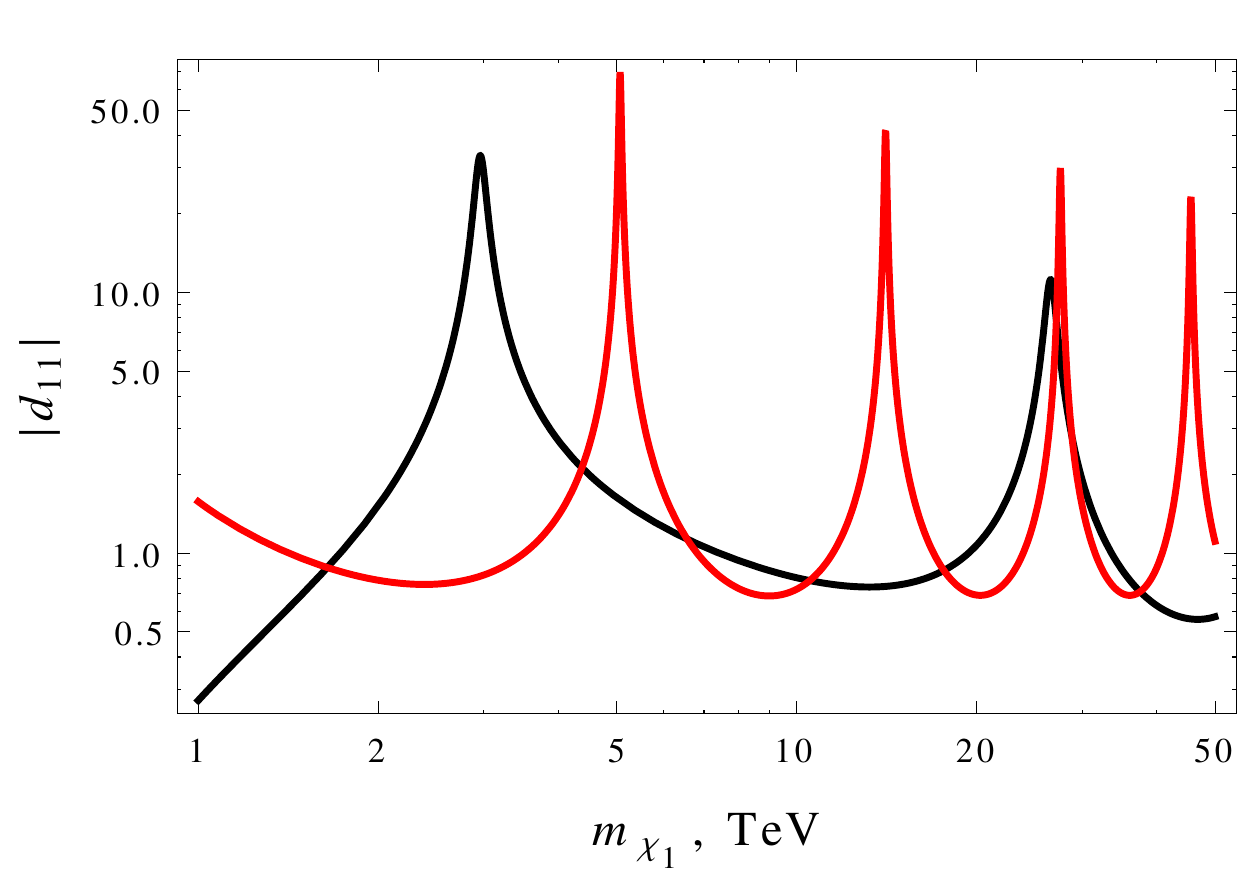}\hspace{0cm}
\includegraphics[width=7cm]{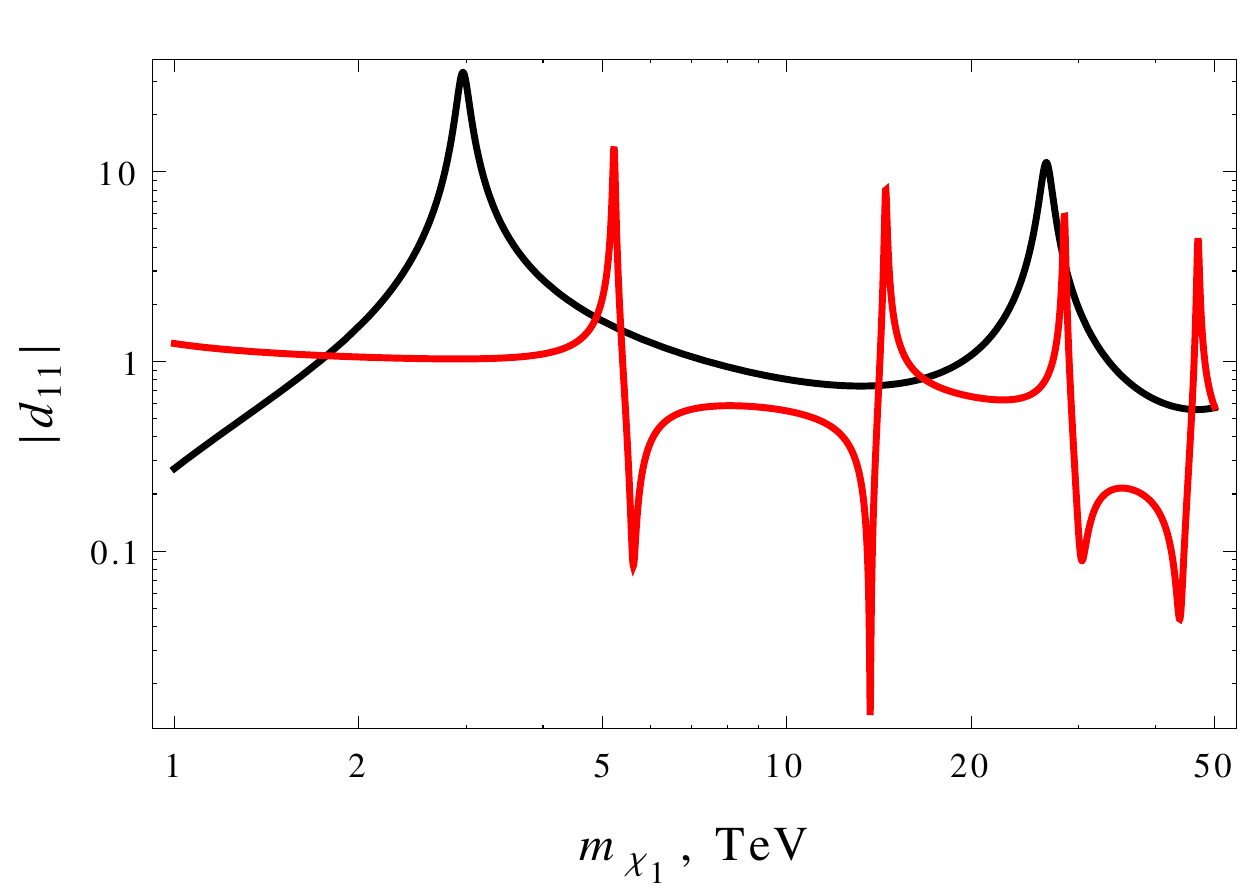}\hspace{0cm}}\vspace{0cm}
\centerline{\includegraphics[width=7cm]{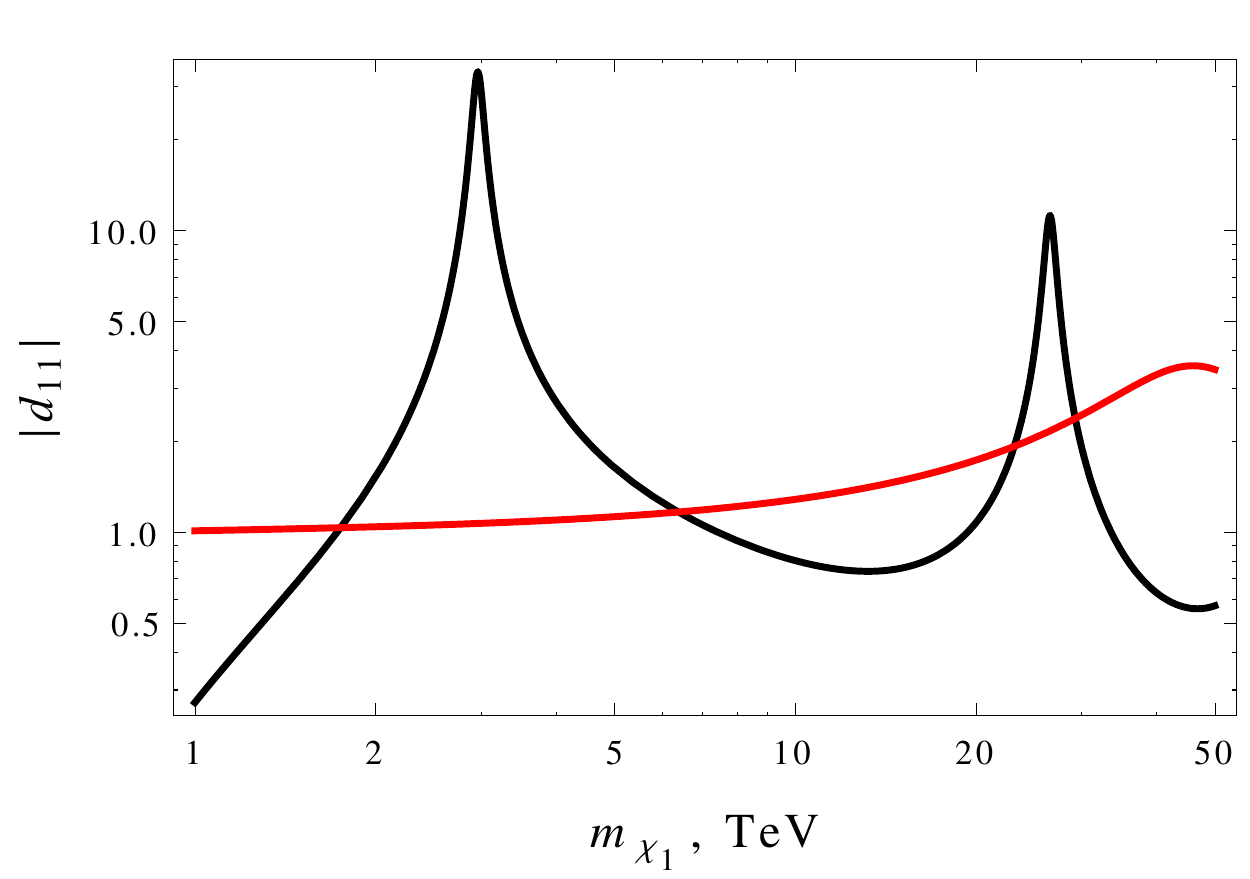}\hspace{0cm}
\includegraphics[width=7cm]{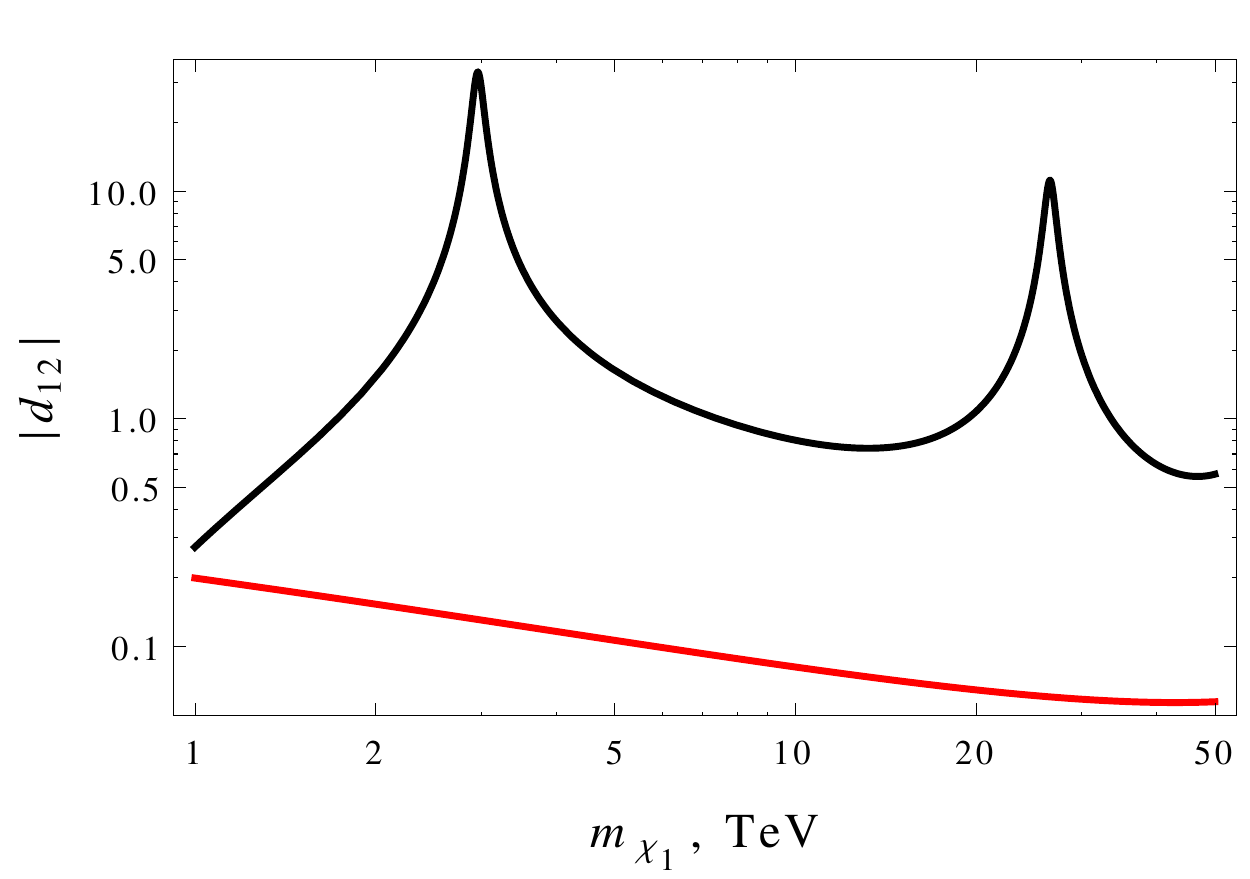}}
\caption{$|d_{11}|$ and $|d_{12}|$ with respect to $m_{\chi_{1}}$ in three 
cases 
and in all figures black line corresponds to $|V_{12}|=\sqrt{2}\alpha 
m_{W},\,|V_{22}|=\alpha m_{W},\,\delta=0$. Red line corresponds to in upper 
left 
fig. $|V_{12}|=10\alpha m_{W},\,|V_{22}|=\alpha m_{W},\,\delta=0$, upper right 
fig. $|V_{12}|=\sqrt{2}\alpha m_{W},\,|V_{22}|=10\alpha m_{W},\,\delta=0$. 
$|d_{12}|$ is similar so it is not shown for these two cases. In lower left and 
right fig. red line corresponds to $|d_{11}|$ and $|d_{12}|$ respectively for 
$|V_{12}|=\sqrt{2}\alpha m_{W},\,|V_{22}|=\alpha 
m_{W},\,\delta=50\,\text{GeV}$.}
\label{sqwellplots}
\end{figure}

Here, $T_{+1}$ and $T_{+2}$ are exponentially suppressed because of repulsive 
potential $\lambda_{+}$.
Moreover, the presence of small mass splitting $\delta$ reduces the 
attractive potential energy in $\lambda_{-}$. For simplicity, in the limit 
$\delta/|V_{12}|,\,|V_{22}|/|V_{12}|\ll 1$,
the resonance in $T_{-1}$ and $T_{-2}$ would occur in
\begin{equation}
 \sqrt{m_{\chi_{1}}|V_{12}|}L=\frac{(2n-1)\pi}{2},\,\,\,n=1,2,...\nonumber
\end{equation}
In addition, non-zero small $\delta\,(|V_{22}|)$ shifts the resonance to larger 
(smaller) $m_{\chi_{1}}$. Fig. \ref{sqwellplots} ( upper left) presents this 
behavior.

\paragraph{2. $\mathbf{|V_{22}|\gg |V_{12}|,\, \delta}$ limit} In this limit, 
the eigenvalues are
\begin{equation}
 \lambda_{+}\sim \frac{|V_{12}|^{2}}{2 
|V_{22}|}\left(1+\frac{2\delta}{|V_{22}|}\right),
 \,\,\,\lambda_{-}\sim -|V_{22}|-\frac{|V_{12}|^{2}}{|V_{22}|}+2\delta
 \label{sqegvalue2nd}
\end{equation} 
And the components $c_{ia}$ of the eigenvectors are,
\begin{align}
c_{+1}&\sim-1+O(|V_{12}|^{2}/|V_{22}|^{2})+O(4\delta^{2}/|V_{22}|^{2}),\,\,\,
c_{-1}\sim\frac{|V_{12}|}{|V_{22}|}+\frac{2 
|V_{12}|\delta}{|V_{22}|^{2}}\nonumber\\
c_{+2}&\sim\frac{|V_{12}|}{|V_{22}|}+\frac{2 
|V_{12}|\delta}{|V_{22}|^{2}},\,\,\,
c_{-2}\sim 
1+O(|V_{12}|^{2}/|V_{22}|^{2})+O(4\delta^{2}/|V_{22}|^{2})\label{sqtrans2}
\end{align}

For the second case, the amplitudes are,
\begin{align}
T_{+i}&\sim 
\left[\cosh\left(\sqrt{m_{\chi_{1}}\lambda_{+}}L\right)+\sqrt{\frac{2\delta_{i}}
{\lambda_{+}}}
\sinh(\sqrt{m_{\chi_{1}}\lambda_{+}}L)\right]^{-1}\label{sqtrans21}\\
 T_{-i}&\sim 
\left[\cos\left(\sqrt{m_{\chi_{1}}|V_{22}|}L\left(1-\frac{\delta}{|V_{22}|}
+\frac{|V_{12}|^{2}}{2 |V_{22}|^{2}}\right)\right)
 -\sqrt{\frac{2\delta_{i}}{|V_{22}|}}\right.\nonumber\\
 &\hspace{5cm}\left.\sin\left(\sqrt{m_{\chi_{1}}|V_{22}|}L
 \left(1-\frac{\delta}{|V_{22}|}+\frac{|V_{12}|^{2}}{2 
|V_{22}|^{2}}\right)\right)\right]^{-1}
 \label{sqtrans22}
\end{align}

Here, $T_{+i}$ will be suppressed with large $m_{\chi_{1}}$ and $T_{-i}$ will 
have resonances
for certain values of $m_{\chi_{1}}$. But compared with the first case 
Eq.(\ref{sqtrans11}), $T_{+i}$ in Eq.(\ref{sqtrans21}) will be less suppressed 
with $m_{\chi_{1}}$ because $\lambda_{+}$ for the second case 
Eq.(\ref{sqegvalue2nd}) is much smaller than that of the first case 
Eq.(\ref{sqegvalue1}). The $d_{11}$ and $d_{12}$ are
\begin{equation}
 d_{11}=c_{+1}T_{+1}+c_{-1}T_{-1},\,\,\,
 d_{12}=c_{+1}T_{+2}+c_{-1}T_{-2}
\end{equation}
As $|c_{+1}|\sim 1$ and $T_{+1}$ is not as suppressed as the first case, for 
small $m_{\chi_{1}}$, both terms
will contribute to the $|d_{11}|$ and cancellation between positive $|c_{\pm 
1}T_{\pm 1}|^2$ terms and the 
interference term $2\,\text{Re}(c^{*}_{+1}T^{*}_{+1}c_{-1}T_{-1})$ for 
particular values of $m_{\chi_{1}}$
leads to the resonance dips in $|d_{11}|$. For very large $m_{\chi_{1}}$, 
$T_{+1}$ will be infinitesimal, hence
$d_{11}\sim c_{-1}T_{-1}$ and no dips will occur in $|d_{11}|$. Similar 
reasoning also holds for dips in $|d_{12}|$. Fig. \ref{sqwellplots} (upper 
right) represents this behavior.

\paragraph{3. $\mathbf{\delta\gg |V_{12}|,|V_{22}|}$ limit} Finally, we 
consider 
the limit
$\delta\gg |V_{12}|,|V_{22}|$. In the zero energy limit,
as the mass difference is larger than the kinetic energy of DM state, 
the transition to $|\chi_{2}\chi_{2}\rangle$ is not kinematically possible 
and thus the cross section is not enhanced. 

In this limit, the eigenvalues
are,
\begin{equation}
\lambda_{+}\sim 
2\delta-|V_{22}|+\frac{|V_{12}|^{2}}{2\delta},\,\,\,\lambda_{-}\sim 
-\frac{|V_{12}|^{2}}{2\delta}
\label{egvalue3rd}
\end{equation}
Here, $\lambda_{+}\gg |\lambda_{-}|$. And the components of the eigenvectors are
\begin{align}
c_{1+} &\sim -\frac{|V_{12}|}{2\delta},\,\,\, c_{1-}\sim 1\nonumber\\
c_{2+} &\sim 1,\,\,\, c_{2-}\sim \frac{|V_{12}|}{2\delta}\label{sqtrans3}
\end{align}

In this limit, the amplitudes $T_{ai}$ are,
\begin{align}
 T_{+i}&\sim\left[\cosh\left(\sqrt{m_{\chi_{1}}\lambda_{+}}L\right)+\sqrt{\frac{
2\delta_{i}}{\lambda_{+}}}\sinh\left(\sqrt{m_{\chi_{1}}\lambda_{+}}
L\right)\right]^{-1}\label{sqtrans31}\\
T_{-i}&\sim\left[\cos\left(\sqrt{m_{\chi_{1}}|\lambda_{-}|}L\right)-\sqrt{\frac{
2\delta_{i}}{|\lambda_{-}|}}\sin\left(\sqrt{m_{\chi_{1}}|\lambda_{-}|}
L\right)\right]^{-1}\label{sqtrans32}
\end{align}
Here, $T_{+1}$ and $T_{+2}$ will be again exponentially suppressed as the first 
case. So
$d_{11}\sim c_{1-}T_{-1}$ and $d_{12}\sim c_{1-}T_{-2}$. In the $v\rightarrow 
0$ 
limit,
for $\delta\gg |V_{12}|$ such that, $|\lambda_{-}|\sim 0$, then $T_{-1}\sim 1$ 
and $T_{-2}$ scales as $T_{-2}\sim 1/\sqrt{m_{\chi_{1}}}$ and hence it is 
negligible for $O(\text{TeV})$ mass. So for very large mass splittings there 
won't be any enhancement in the cross section.

On the other hand, if $\delta\simgt |V_{12}|$ and $|\lambda_{-}|$ is small but 
non-zero, at certain large mass $m_{\chi_{1}}$ so that 
$\sqrt{m_{\chi_{1}}|\lambda_{-}|}L\sim \pi/2$, the resonance condition is 
fulfilled in $T_{-i}$ and we would again see resonance in the annihilation 
cross section which is non-observable in the smaller mass range. In Fig. 
\ref{sqwellplots} (lower left and right) we can see these behavior in $d_{11}$ 
and $d_{12}$ respectively.

These simplified results for the square-well can
be inferred for larger electroweak multiplets with short-ranged potential such 
as Yukawa potential. For larger representations, large group theoretic factors 
$f_{ab}$
associated with $SU(2)$ ladder operators enter as the coefficients of Yukawa 
potential, as in Eq.(\ref{schrod2}) and such large coefficients lead to large 
matrix elements $|V_{ab}|$. From the above three limits, we may encounter the 
following cases,

\begin{enumerate}
\item If the DM 2-particle state is denoted by $|1\rangle\equiv 
|\chi_{1}\chi_{1}\rangle$ state and $|a\neq 1\rangle$ as other charged and 
pseudoscalar 2-particle states $|\chi_{i}\chi_{j}\rangle$, any or more than 
one matrix elements $|V_{1a}|$ ($a=1,..,N$) of $N\times N$ potential matrix , 
satisfy $|V_{1a}|\gg |V_{bc}|,\delta_{a}$,
where, $b,c\neq 1$ and $\delta_{a}=m_{j}+m_{k}-2m_{1}$ for 
$|a=\chi_{j}\chi_{k}\rangle$ state, $|1\rangle$ will have larger overlapping 
with the attractive states $|\lambda_{-}\rangle$'s (as $c_{1-}\sim 1/\sqrt{2}$) 
and $d_{1a}$ will have resonances for certain $m_{\chi_{1}}$ and hence the 
annihilation cross section $\chi_{1}\chi_{1}\rightarrow f$ will be enhanced. 
Moreover, non-zero mass splitting shifts the resonance peaks to larger mass 
value. We have seen these features in the doublet case in Fig. \ref{WWandZZse} 
and \ref{ggandgZcrossec}.

\item In contrast, when $|V_{bc}|\gg |V_{1a}|,\,\delta_{a}$, where $b,c\neq 1$, 
$|1\rangle$ will have smaller overlapping with attractive states (as 
$c_{1-}\sim 
|V_{1a}|/|V_{bc}|$) and along with resonances, there will be resonance dips for 
certain $m_{\chi_{1}}$ values as seen for $2\times 2$ square-well potential 
case. We have seen the appearance of dips for the quartet case in Fig. 
\ref{WWandZZse} and \ref{ggandgZcrossec}.

\item Finally, for $\delta_{a}\gg |V_{bc}|$, where $|a\equiv 
\chi_{j}\chi_{k}\rangle$ and $j,k\neq 1$, the channel 
$\chi_{1}\chi_{1}\rightarrow \chi_{j}\chi_{k}$ will be kinematically closed and 
will
not contribute to the enhancement of the annihilation cross section. This was 
the case of $\gamma\gamma$ and $\gamma Z$ annihilation with large mass 
splittings.
\end{enumerate}

\section{Conclusion and outlook}\label{conclusion}

In this study we have demonstrated that the Sommerfeld enhancement of the DM annihilation cross section increases with the size of the multiplet but in the case of larger multiplet resonance dips or suppression for certain values of the DM mass appear along with resonances. In the multi-channel case, for larger multiplet, not only the number of potential matrix elements $V_{SS,jj'},\,jj'=SS,...,\tilde{\Delta}^{+}_{j}\tilde{\Delta}^{-}_{j'}$ is large than that of smaller multiplet but also the matrix elements are comparatively larger due to factors related to $SU(2)_{L}$ raising operator. As we have investigated the region where Yukawa potential has dominated over Coulomb potential, our case resembles the first case of square well example given in section \ref{discussion}.

In addition, the dips occur in the multi-channel because of large potential matrix elements $V_{ii',jj'},\,ii',jj'\neq SS$ which not only decrease the overlapping of the $|SS\rangle$ states on attractive eigenstates of the potential, which are responsible for the enhancement but also induce destructive interference in transition amplitudes $T_{SS,jj'},\,jj'=SS,...,\tilde{\Delta}^{+}_{j}\tilde{\Delta}^{-}_{j'}$ which enter into the cross section Eq.(\ref{anncross}). This case resembles the second case of square well example in section \ref{discussion}. Besides we can see that the dip can occur for short ranged potential in multi channel process, whereas it was attributed to Coulomb interaction in \cite{Chun:2012yt}.

Moreover, when mass splittings are taken as their maximum allowed limit, we can see from Fig. \ref{WWandZZse} (left) that for doublet, the tree-level and Sommerfeld enhanced cross sections are comparable for $1-7$ TeV mass range whereas only for $1-2.5$ TeV (outside the thermal DM region), one has comparable tree-level and SE cross-section of $SS\rightarrow WW$. Still there is a small strip about $3.8$ TeV, one again has the comparable tree-level and SE cross section. Apart from $10$, $17$ and $28$ TeV, SE cross section is larger than the tree-level cross section of $SS\rightarrow WW$. Comparatively larger SE is observed for the quartet, is due to the small mass splitting between DM and next to lightest single charged component allowed by EWPO, for which $\epsilon_{d_{\tilde{\Delta}^{+}_{1}\tilde{\Delta}^{-}_{1}}}\simlt 1$ except for few small mass ranges. Similar pattern is also seen in Fig. \ref{WWandZZse} (right) for $SS\rightarrow ZZ$ cross section.

Consequently large Sommerfeld enhanced DM annihilation cross section have important implications on the indirect detection. Without undertaking the comprehensive analysis for the scotogenic model, we have only projected the exclusion limits of H.E.S.S. \cite{::2016jja} and the future sensitivity limit of CTA \cite{Carr:2015hta} of annihilation cross section to $WW$ on our $SS\rightarrow WW$ cross section plot and exclusion limits of H.E.S.S. \cite{Abdalla:2016olq}
of $\gamma\gamma$ annihilation on our $SS\rightarrow \gamma\gamma$ plot so that we can see how far the experimental sensitivity has been reached to probe the indirect detection signals for the Scotogenic model with different sizes of scalar multiplets. It can be seen from Fig. \ref{WWandZZse} (left) that H.E.S.S. has already achieved the sensitivity to probe the entire $1-30$ TeV mass range for the quartet except $m_{S}\sim 27$ TeV for the (almost) degenerate limit and dips at certain mass values for the allowed maximum mass splitting. On the other hand, for the doublet, except for $2.5-4$ TeV for (almost) degenerate limit and almost all of $1-30$ TeV for allowed maximum limit, are below the H.E.S.S. limit. Future CTA sensitivity limit is improved by $O(10)$ compared to H.E.S.S. limits.

For $SS\rightarrow \gamma\gamma,\,\gamma Z$ cases, the Sommerfeld enhanced cross section is obtained only for (almost) degenerate limit because maximum allowed mass splitting suppress the $T_{SS,jj},\,jj=\tilde{\Delta}_{j}^{(Q)}\tilde{\Delta}^{(-Q)}_{j}$ factors and thus annihilation becomes negligible. For such case, the $\gamma\gamma$ and $\gamma Z$ annihilation proceed through one loop process via charged scalars exchange and has $10^{-32}-10^{-27}\,\text{cm}^{2}s^{-1}$ for doublet and quartet. From Fig. \ref{ggandgZcrossec} (left), we can see that H.E.S.S. limit can already probe $1-9$ (except for dip at 1.4 TeV) and $11.5-14$ TeV of their considered $1-20$ TeV mass range for the quartet whereas for the doublet only $2.1-4.1$ TeV out of $1-20$ TeV is within the reach of H.E.S.S.

From the present investigation we can infer that in scotogenic model, the Sommerfeld enhanced cross sections of higher scalar multiplet will be larger than the doublet. 
Although the projection of H.E.S.S. and future CTA sensitivity limits on $SS\rightarrow WW$ and $SS\rightarrow \gamma\gamma$ plots have already given us the order of magnitude prospects to detect the indirect signal of DM annihilation with larger multiplets, a detailed gamma-ray spectral analysis that includes specific features (as in the case for doublet \cite{Garcia-Cely:2013zga, Garcia-Cely:2015khw}) as well as the contribution from P-wave annihilation, specially in S-wave suppressed (occurance of dips) cases \cite{Das:2016ced}, are needed to be carried out for determining the viability of larger multiplets \cite{idmproject}. Nevertheless, the DM of higher scalar multiplet is more likely to be found in the current and future indirect detection experiments because of their large Sommerfeld enhanced annihilation cross sections.

\subsection*{Note Added}

During the preparation of this manuscript, \cite{Logan:2016ivc} has appeared which investigated the scalar DM relic density of $j=5/2$ and $j=7/2$ multiplets in the high mass regime but didn't include the prospect of detecting indirect signals in current H.E.S.S. or future CTA experiments.

\subsection*{Acknowledgements}

We would like to thank Xiaoyong Chu and Camilo Garcia-Cely for stimulating remarks and critical reading of the manuscript. T.A.C. is indebted to Fabrizio Nesti and Goran Senjanovi\'c for pointing out the role of higher dimensional operator in mass splittings of complex odd dimensional scalar multiplets and also would like to thank Mahbub Majumdar and Arshad Momen for discussion.

\begin{appendices}

\section{The Potential and Annihilation Matrix Elements}\label{matrixelem}
In this appendix we have made an inventory the NR limit of the component fields and 
the interaction part of the NR action that gives the potential and the S-wave 
annihilation matrix for the generalized
scotogenic model. Our calculation closely followed the methods given in \cite{Hisano:2004ds, Camilo-thesis}.

\subsection{NR Limit}\label{nrlimitsec}
In the NR limit, the component fields are
\begin{align}
S(A)(x) &=\frac{1}{\sqrt{2 m_{S(A)}}}\left(\zeta_{S(A)}(\vec{x},t)e^{-i 
m_{S(A)}t}+\zeta_{S(A)}^{*}(\vec{x},t)e^{-i m_{S(A)}t}\right)\nonumber\\
\tilde{\Delta}^{(Q)}_{i}(x)&=\frac{1}{\sqrt{2 
m_{\tilde{\Delta}^{(Q)}_{i}}}}\left(\xi_{\tilde{\Delta}^{(Q)}_{i}}(\vec{x},t)e^{
-i m_{\tilde{\Delta}^{(Q)}_{i}} 
t}+\eta_{\tilde{\Delta}^{(-Q)}_{i}}^{*}(\vec{x},t)e^{-i 
m_{\tilde{\Delta}^{(Q)}_{i}} t}\right)\label{nrcomponent}
\end{align}
Here, $\zeta_{S(A)}$ and $\zeta^{*}_{S(A)}$ are associated with the 
annihilation 
and creation of $S(A)$ fields. $\xi_{\tilde{\Delta}^{(Q)}_{i}}$ and $\eta^{*}_{ 
\tilde{\Delta}^{(-Q)}_{i}}$ corresponds to annihilation of 
$\tilde{\Delta}^{(Q)}_{i}$ and creation of $\tilde{\Delta}^{(-Q)}_{i}$ 
respectively and vice versa. At $O(\text{TeV})$ range, the allowed mass 
splittings between DM component and other components of the multiplet as 
mentioned in section \ref{allowedmasssplit}, is small compared to the mass 
itself. So in the subsequent analysis, we set, $m_{i}\sim m_{S}$ in 
Eq.(\ref{nrcomponent}) of the component fields. Moreover the annihilation and 
creation of 2-particle states $|SS\rangle$ or $|AA\rangle$ correspond to 
$\zeta_{i}\zeta_{i}$ and $\zeta^{*}_{i}\zeta^{*}_{i}$ respectively. Similarly 
those of the state $|\tilde{\Delta}^{(Q)}_{i}\tilde{\Delta}^{(-Q)}_{j}\rangle$ 
correspond to $\xi_{\tilde{\Delta}^{(Q)}_{i}}\eta_{\tilde{\Delta}^{(-Q)}_{j}}$ 
and $\xi^{*}_{\tilde{\Delta}^{(Q)}_{i}}\eta^{*}_{\tilde{\Delta}^{(-Q)}_{j}}$ 
respectively. 

In addition, at fixed time $x^{0}$, the corresponding the 2-particle fields, 
which are the components of the 2-particle field vector 
$\mathbf{\Phi}(x,\vec{r})$,  are defined as
\begin{align}
\Phi^{N}_{ii'}(x,\vec{r})&=\frac{1}{\sqrt{2}} 
\zeta_{i}(\vec{x}-\vec{r}/{2},x^{0})\zeta_{i}(\vec{x}+\vec{r}/{2},x^{0})\,\,\,i,
\,i'=S,\, A\nonumber\\
\Phi^{C}_{ii'}(x,\vec{r})&=\xi_{i}(\vec{x}-\vec{r}/{2},x^{0})\eta_{i}(\vec{x}
+\vec{r}/{2},x^{0})\,\,\,i=\tilde{\Delta}^{(Q)}_{i},\,i'=\tilde{\Delta}^{(-Q)}_{
i'}
\label{twostatefield}
\end{align}
where $x$ is the center of mass coordinate for 2-particle system and 
$\vec{r}$ 
represents the relative separation between them. Also $1/\sqrt{2}$ is due to 
identical particles. The NR effective action for 2-particle fields is now 
defined as
\begin{equation}
S_{\text{eff}}=\int d^{4}x d^{3}r 
\mathbf{\Phi}^{\dagger}(x,\vec{r})\left(i\partial_{x^{0}}+\frac{\nabla^{2}_{x}}{
4m_{S}}+\frac{\nabla^{2}_{r}}{m_{S}}-V(\vec{r})+2i\Gamma\delta^{(3)}(\vec{r}
)\right)\mathbf{\Phi}(x,\vec{r})
\label{twofieldeffaction}
\end{equation}

\subsection{The Potential Matrix elements}\label{potentialmatrixelements}
The potential matrix induced by the exchange of the gauge bosons can be derived 
from the gauge current interaction of the Lagrangian,
\begin{equation}
{\cal L}\supset 
J^{+}_{\mu}W^{+\mu}+J_{\mu}^{-}W^{\mu}+J^{Z}_{\mu}Z^{\mu}+J^{A}_{\mu}A^{\mu}
\label{gaugecurrent}
\end{equation}

After integrating out the light or relativistic degrees of freedom i.e gauge 
bosons, these terms induce the current current interactions in the effective 
action of the following form,
\begin{equation}
S_{\text{eff}}\supset -\int \frac{d^{4}xd^{3}y}{8\pi r}\left(
2J^{+}_{0}(x)J^{-}_{0}(x^{0},\vec{y})e^{-m_{W}r}+J^{Z}_{0}(x)J^{Z}_{0}(x^{0},
\vec{y})e^{-m_{Z}r}
+J^{A}_{0}(x)J^{A}_{0}(x^{0},\vec{y})\right)
\label{effecurrent}
\end{equation}
where $r=|\vec{x}-\vec{y}|$. Also for non-relativistic current, $J_{0}\gg J_{i}$.

\paragraph{Matrix elements due to Z and $\gamma$ exchange}
The part of the neutral current containing the charged component fields with 
$Q$ 
is
\begin{equation}
 J^{Z}_{0}\supset 
\frac{ig}{\cos\theta_{W}}\left(z_{11}\tilde{\Delta}^{(-Q)}_{1}\overleftrightarrow{
\partial_{0}}\tilde{\Delta}^{(Q)}_{1}+
 z_{22}\tilde{\Delta}^{(-Q)}_{2}\overleftrightarrow{\partial_{0}}\tilde{\Delta}^{(Q)}_{2}+
 z_{12}(\tilde{\Delta}^{(-Q)}_{1}\overleftrightarrow{\partial_{0}}\tilde{\Delta}^{(Q)}_{2}+
 \tilde{\Delta}^{(-Q)}_{2}\overleftrightarrow{\partial_{0}}\tilde{\Delta}^{(Q)}_{1})\right)
\end{equation}

Here the factors $z_{ij}$ are
\begin{align}
 z_{11}&=(m-Q\sin^{2}\theta_{W})\cos^{2}\theta_{Q}
 +(m+1-Q\sin^{2}\theta_{W})\sin^{2}\theta_{Q}\nonumber\\
 z_{22}&=(m-Q\sin^{2}\theta_{W})\sin^{2}\theta_{Q}
 +(m+1-Q\sin^{2}\theta_{W})\cos^{2}\theta_{Q}\nonumber\\
 z_{12}&=\cos\theta_{Q}\sin\theta_{Q}
\end{align}

The matrix elements of potential matrix induced by the exchange of Z boson and 
photon
are the following
\begin{align}
& V_{SS,SS}=V_{AA,AA}=0,\,\,\,
V_{SS,AA}=-\frac{\alpha}{4\cos^{2}\theta_{W}} \frac{e^{-m_{Z}}r}{r}\nonumber\\
&V_{\Delta^{(\frac{n+1}{2})}\Delta^{(-\frac{n+1}{2})},
\Delta^{(\frac{n+1}{2})}\Delta^{(-\frac{n+1}{2})}}=
-\frac{\alpha}{\cos^{2}\theta_{W}}
 \left(\frac{n}{2}-\frac{n+1}{2}\sin^{2}\theta_{W}\right)^2\frac{e^{-m_{Z}}r}{r}
-\frac{\alpha_{em}\left(\frac{n+1}{2}\right)^2}{r}\nonumber\\
& V_{\tilde{\Delta}^{(Q)}_{1}\tilde{\Delta}^{(-Q)}_{1},
\tilde{\Delta}^{(Q)}_{1}\tilde{\Delta}^{(-Q)}_{1}}=
-\frac{\alpha 
z_{11}^{2}}{\cos^{2}\theta_{W}}\frac{e^{-m_{Z}}r}{r}-\frac{\alpha_{em}Q^{2}}{r}
\nonumber\\
&V_{\tilde{\Delta}^{(Q)}_{2}\tilde{\Delta}^{(-Q)}_{2},
\tilde{\Delta}^{(Q)}_{2}\tilde{\Delta}^{(-Q)}_{2}}=
-\frac{\alpha 
z_{22}^{2}}{\cos^{2}\theta_{W}}\frac{e^{-m_{Z}}r}{r}-\frac{\alpha_{em}Q^{2}}{r}
\nonumber\\
&V_{\tilde{\Delta}^{(Q)}_{1}\tilde{\Delta}^{(-Q)}_{1},
\tilde{\Delta}^{(Q)}_{1}\tilde{\Delta}^{(-Q)}_{2}}=
-\frac{\alpha z_{11}z_{12}}{\cos^{2}\theta_{W}}\frac{e^{-m_{Z}}r}{r}
=V_{\tilde{\Delta}^{(Q)}_{1}\tilde{\Delta}^{(-Q)}_{1},
\tilde{\Delta}^{(Q)}_{2}\tilde{\Delta}^{(-Q)}_{1}}\nonumber\\
&V_{\tilde{\Delta}^{(Q)}_{2}\tilde{\Delta}^{(-Q)}_{2},
\tilde{\Delta}^{(Q)}_{1}\tilde{\Delta}^{(-Q)}_{2}}=
-\frac{\alpha z_{22}z_{12}}{\cos^{2}\theta_{W}}\frac{e^{-m_{Z}}r}{r}=
V_{\tilde{\Delta}^{(Q)}_{2}\tilde{\Delta}^{(-Q)}_{2},
\tilde{\Delta}^{(Q)}_{2}\tilde{\Delta}^{(-Q)}_{1}}\nonumber\\
&V_{\tilde{\Delta}^{(Q)}_{1}\tilde{\Delta}^{(-Q)}_{2},
\tilde{\Delta}^{(Q)}_{1}\tilde{\Delta}^{(-Q)}_{2}}=
-\frac{\alpha 
z_{11}z_{22}}{\cos^{2}\theta_{W}}\frac{e^{-m_{Z}}r}{r}-\frac{\alpha_{em}Q^{2}}{r
}\nonumber\\
&V_{\tilde{\Delta}^{(Q)}_{1}\tilde{\Delta}^{(-Q)}_{2},
\tilde{\Delta}^{(Q)}_{2}\tilde{\Delta}^{(-Q)}_{1}}=
-\frac{\alpha z_{12}^{2}}{\cos^{2}\theta_{W}}\frac{e^{-m_{Z}}r}{r}\nonumber
\end{align}
These matrix elements are read off from Eq.(\ref{effecurrent}) first by 
replacing the fields with their NR limits as given in Eq.(\ref{nrcomponent}) 
and 
then re-arranging them in the 2-particle fields $\Phi^{N}_{ii'}$ and 
$\Phi^{C}_{jj'}$ so to match the $V(\vec{r})$ term of the 2-particle 
effective 
action in Eq.(\ref{twofieldeffaction}).

\paragraph{Matrix elements induced by $W^{\pm}$ boson}\label{matrixW}
Now the matrix elements induced by the exchange of W boson are the following. 
The charged current involving the charged components of the multiplet is
\begin{equation}
 J^{+}_{\mu}\supset i 
g\left(\sum_{i}a_{ii}\tilde{\Delta}^{(-Q)}_{i}\overleftrightarrow{\partial_{\mu}}\tilde{\Delta}^{
(Q+1)}_{i}
 +\sum_{i\neq 
j}c_{ij}\tilde{\Delta}^{(-Q)}_{i}\overleftrightarrow{\partial_{\mu}}\tilde{\Delta}^{(Q+1)}_{j}
\right)\nonumber
\end{equation}
Here, $i,j=1,2$.

We denote $V^{+}_{j,m}=\sqrt{(j-m)(j+m+1)}$ and 
$V^{-}_{j,m}=\sqrt{(j+m)(j-m+1)}$. Also $V^{+}_{j,m}=V^{-}_{j,m+1}$. In the 
following we collect the
the relevant matrix element of the potential matrix induced by the exchange of 
the $W^{\pm}$ boson.

First let us consider the matrix elements induced by exchange of the W boson 
among the neutral pairs of charged components with the
largest charge, $Q=\frac{n+1}{2}$ $(m=\frac{n}{2})$
and charge, $Q-1=\frac{n-1}{2}$ $(m=\frac{n-2}{2})$ of the multiplet. They are

\begin{align}
V_{\Delta^{(\frac{n+1}{2})}\Delta^{(-\frac{n+1}{2})},\tilde{\Delta}^{(\frac{n-1}
{2})}_{1}\tilde{\Delta}^{(-\frac{n-1}{2})}_{1}} &=
-\frac{\alpha 
e^{-m_{W}r}}{r}(V^{+}_{\frac{n}{2},\frac{n-2}{2}})^2\cos^2\theta_{\frac{n-1}{2}}
\nonumber\\
 V_{\Delta^{(\frac{n+1}{2})}\Delta^{(-\frac{n+1}{2})},\tilde{\Delta}^{(\frac{n-1}
{2})}_{2}\tilde{\Delta}^{(-\frac{n-1}{2})}_{2}} &=
 -\frac{\alpha 
e^{-m_{W}r}}{r}(V^{+}_{\frac{n}{2},\frac{n-2}{2}})^2\sin^2\theta_{\frac{n-1}{2}}
\nonumber\\
  V_{\Delta^{(\frac{n+1}{2})}\Delta^{(-\frac{n+1}{2})},\tilde{\Delta}^{(\frac{n-1}
{2})}_{1}\tilde{\Delta}^{(-\frac{n-1}{2})}_{2}} &=
 -\frac{\alpha 
e^{-m_{W}r}}{r}(V^{+}_{\frac{n}{2},\frac{n-2}{2}})^2\sin\theta_{\frac{n-1}{2}}
\cos\theta_{\frac{n-1}{2}}\nonumber
\end{align}

Now we focus on the matrix elements induced by the exchange of W bosons
among the neutral pairs of charged states with charge $Q$ and $Q+1$ of the 
multiplet.
\begin{align}
 V_{\tilde{\Delta}^{(Q)}_{1}\tilde{\Delta}^{(-Q)}_{1},\tilde{\Delta}^{(Q+1)}_{1}
\tilde{\Delta}^{(-Q-1)}_{1}}
 &=-\frac{\alpha e^{-m_{W}r}}{r}a_{11}^2,\,\,\,
V_{\tilde{\Delta}^{(Q)}_{1}\tilde{\Delta}^{(-Q)}_{1},\tilde{\Delta}^{(Q+1)}_{2}
\tilde{\Delta}^{(-Q-1)}_{2}}
 =-\frac{\alpha e^{-m_{W}r}}{r}c_{12}^2\nonumber\\
V_{\tilde{\Delta}^{(Q)}_{1}\tilde{\Delta}^{(-Q)}_{1},\tilde{\Delta}^{(Q+1)}_{1}
\tilde{\Delta}^{(-Q-1)}_{2}}
&=V_{\tilde{\Delta}^{(Q)}_{1}\tilde{\Delta}^{(-Q)}_{1},\tilde{\Delta}^{(Q+1)}_{2
}\tilde{\Delta}^{(-Q-1)}_{1}}
 =-\frac{\alpha e^{-m_{W}r}}{r}a_{11}c_{12}\nonumber\\
V_{\tilde{\Delta}^{(Q)}_{1}\tilde{\Delta}^{(-Q)}_{2},\tilde{\Delta}^{(Q+1)}_{1}
\tilde{\Delta}^{(-Q-1)}_{1}}
&=V_{\tilde{\Delta}^{(Q)}_{2}\tilde{\Delta}^{(-Q)}_{1},\tilde{\Delta}^{(Q+1)}_{1
}\tilde{\Delta}^{(-Q-1)}_{1}}=
 -\frac{\alpha e^{-m_{W}r}}{r}a_{11}c_{21}\nonumber\\
V_{\tilde{\Delta}^{(Q)}_{1}\tilde{\Delta}^{(-Q)}_{2},\tilde{\Delta}^{(Q+1)}_{1}
\tilde{\Delta}^{(-Q-1)}_{2}}
&=V_{\tilde{\Delta}^{(Q)}_{2}\tilde{\Delta}^{(-Q)}_{1},\tilde{\Delta}^{(Q+1)}_{1
}\tilde{\Delta}^{(-Q-1)}_{2}}=
 -\frac{\alpha e^{-m_{W}r}}{r}a_{11}a_{22}\nonumber\\
V_{\tilde{\Delta}^{(Q)}_{1}\tilde{\Delta}^{(-Q)}_{2},\tilde{\Delta}^{(Q+1)}_{2}
\tilde{\Delta}^{(-Q-1)}_{1}}
&=V_{\tilde{\Delta}^{(Q)}_{2}\tilde{\Delta}^{(-Q)}_{1},\tilde{\Delta}^{(Q+1)}_{2
}\tilde{\Delta}^{(-Q-1)}_{1}}=
 -\frac{\alpha e^{-m_{W}r}}{r}c_{12}c_{21}\nonumber\\
V_{\tilde{\Delta}^{(Q)}_{1}\tilde{\Delta}^{(-Q)}_{2},\tilde{\Delta}^{(Q+1)}_{2}
\tilde{\Delta}^{(-Q-1)}_{2}}
&=V_{\tilde{\Delta}^{(Q)}_{2}\tilde{\Delta}^{(-Q)}_{1},\tilde{\Delta}^{(Q+1)}_{2
}\tilde{\Delta}^{(-Q-1)}_{2}}=
 -\frac{\alpha e^{-m_{W}r}}{r}c_{12}a_{22}\nonumber\\
V_{\tilde{\Delta}^{(Q)}_{2}\tilde{\Delta}^{(-Q)}_{2},\tilde{\Delta}^{(Q+1)}_{1}
\tilde{\Delta}^{(-Q-1)}_{2}}
&=V_{\tilde{\Delta}^{(Q)}_{2}\tilde{\Delta}^{(-Q)}_{2},\tilde{\Delta}^{(Q+1)}_{2
}\tilde{\Delta}^{(-Q-1)}_{1}}=
 -\frac{\alpha e^{-m_{W}r}}{r}c_{21}a_{22}\nonumber\\
V_{\tilde{\Delta}^{(Q)}_{2}\tilde{\Delta}^{(-Q)}_{2},\tilde{\Delta}^{(Q+1)}_{1}
\tilde{\Delta}^{(-Q-1)}_{1}}
 &=-\frac{\alpha e^{-m_{W}r}}{r}c_{21}^{2},\,\,\, 
V_{\tilde{\Delta}^{(Q)}_{2}\tilde{\Delta}^{(-Q)}_{2},\tilde{\Delta}^{(Q+1)}_{2}
\tilde{\Delta}^{(-Q-1)}_{2}}
 =-\frac{\alpha e^{-m_{W}r}}{r}a_{22}^2\nonumber
 \end{align}
 
Here the factors $a_{11}$, $a_{22}$, $c_{12}$ and $c_{21}$ are
\begin{align}
 a_{11}&=V^{+}_{\frac{n}{2},m}\cos\theta_{Q}\cos\theta_{Q+1}-
 V^{+}_{\frac{n}{2},-m-2}\sin\theta_{Q}\sin\theta_{Q+1}\nonumber\\
 a_{22}&=V^{+}_{\frac{n}{2},m}\sin\theta_{Q}\sin\theta_{Q+1}-
 V^{+}_{\frac{n}{2},-m-2}\cos\theta_{Q}\cos\theta_{Q+1}\nonumber\\
 c_{12}&=V^{+}_{\frac{n}{2},m}\cos\theta_{Q}\sin\theta_{Q+1}+
 V^{+}_{\frac{n}{2},-m-2}\sin\theta_{Q}\cos\theta_{Q+1}\nonumber\\
 c_{21}&=V^{+}_{\frac{n}{2},m}\sin\theta_{Q}\cos\theta_{Q+1}+
 V^{+}_{\frac{n}{2},-m-2}\cos\theta_{Q}\sin\theta_{Q+1}\nonumber
\end{align}

The charged current involving $S$ and $A$ is
\begin{equation}
J^{+}_{0}\supset\frac{ig}{2}(d_{S1}S\overleftrightarrow{\partial_{0}}\tilde{\Delta}^{+}_
{1}-
 d_{S2}S\overleftrightarrow{\partial_{0}}\tilde{\Delta}^{+}_{2}
 -id_{A1}A\overleftrightarrow{\partial_{0}}\tilde{\Delta}^{+}_{1}
 +id_{A2}A\overleftrightarrow{\partial_{0}}\tilde{\Delta}^{+}_{2})\nonumber
\end{equation}
Here the factors are
\begin{align}
d_{S1}&=V^{+}_{\frac{n}{2},-\frac{1}{2}}\cos\theta_{1}-V^{+}_{\frac{n}{2},-\frac
{3}{2}}\sin\theta_{1}\nonumber\\
d_{S2}&=V^{+}_{\frac{n}{2},-\frac{1}{2}}\sin\theta_{1}+V^{+}_{\frac{n}{2},-\frac
{3}{2}}\cos\theta_{1}\nonumber\\
d_{A1}&=V^{+}_{\frac{n}{2},-\frac{1}{2}}\cos\theta_{1}+V^{+}_{\frac{n}{2},-\frac
{3}{2}}\sin\theta_{1}\nonumber\\
d_{A2}&=V^{+}_{\frac{n}{2},-\frac{1}{2}}\sin\theta_{1}-V^{+}_{\frac{n}{2},-\frac
{3}{2}}\cos\theta_{1}\nonumber
\end{align}

The matrix elements between the $SS$, $AA$ states and single charged states 
induced by the W boson are
\begin{align}
V_{SS,\tilde{\Delta}^{+}_{1}\tilde{\Delta}^{-}_{1}}&=
-\frac{\alpha e^{-m_{W}r}}{\sqrt{2}r} d_{S1}^2,\,\,\,
V_{SS,\tilde{\Delta}^{+}_{2}\tilde{\Delta}^{-}_{2}}=
-\frac{\alpha e^{-m_{W}r}}{\sqrt{2}r} d_{S2}^2\nonumber\\
V_{SS,\tilde{\Delta}^{+}_{1}\tilde{\Delta}^{-}_{2}}&=
V_{SS,\tilde{\Delta}^{+}_{2}\tilde{\Delta}^{-}_{1}}
=-\frac{\alpha e^{-m_{W}r}}{\sqrt{2}r} d_{S1}d_{S2}\nonumber\\
V_{AA,\tilde{\Delta}^{+}_{1}\tilde{\Delta}^{-}_{1}}&=
-\frac{\alpha e^{-m_{W}r}}{\sqrt{2}r} d_{A1}^2,\,\,\,
V_{AA,\tilde{\Delta}^{+}_{2}\tilde{\Delta}^{-}_{2}}=
-\frac{\alpha e^{-m_{W}r}}{\sqrt{2}r} d_{A2}^2\nonumber\\
V_{AA,\tilde{\Delta}^{+}_{1}\tilde{\Delta}^{-}_{2}}&=
V_{AA,\tilde{\Delta}^{+}_{2}\tilde{\Delta}^{-}_{1}}
=-\frac{\alpha e^{-m_{W}r}}{\sqrt{2}r} d_{A1}d_{A2}\nonumber
\end{align}

\subsection{The Annihilation Matrix Elements}\label{annihilationmatrix}
In the non-relativistic limit, the $2\rightarrow 2$ S-wave annihilation channel 
is the most important channel. Therefore, we
focus on the S-wave annihilation matrix elements.
It can be derived from the gauge-scalar quartic couplings of the Lagrangian. 
The 
neutral sector and charged sector of quartic interactions are
\begin{align}
\text{Neutral:}&\,\,\,\,{\cal L}\supset 
T_{W^{+}W^{-}}W^{+}_{\mu}W^{-\mu}+T_{ZZ}Z_{\mu}Z^{\mu}+T_{AA}A_{\mu}A^{\mu}+
T_{ZA}Z_{\mu}A^{\mu}
\nonumber\\
\text{Charged:}&\,\,\,{\cal L}\supset 
T_{W^{+}W^{+}}W^{+}_{\mu}W^{+\mu}+T_{W^{+}Z}W^{+}_{\mu}Z^{\mu}+T_{W^{+}A}W^{+}_{
\mu}A^{\mu}
+\text{h.c.}
\label{gaugescalar}
\end{align}
with
\begin{align}
T_{W^{+}W^{-}}&=g^{2}\Delta^{\dagger}(T^{+}T^{-}+T^{-}T^{+})\Delta,\,\,\,
T_{AA}= e^2 \Delta^{\dagger}\hat{Q}^{2}\Delta\nonumber\\
T_{ZZ}&=g^2\cos^{2}\theta_{W}\Delta^{\dagger}(T^{3})^{2}\Delta+gg'\cos\theta_{W}
\sin\theta_{W}\Delta^{\dagger}T^{3}Y\Delta+
g'^{2}\sin^{2}\theta_{W}\Delta^{\dagger}Y^2\Delta\nonumber\\
T_{ZA}&= g^2\sin 
2\theta_{W}\Delta^{\dagger}(T^{3})^{2}\Delta+gg'\cos2\theta_{W}\Delta^{\dagger}
T^{3}Y\Delta-
g'^{2}\sin 2\theta_{W}\Delta^{\dagger}Y^2\Delta\nonumber\\
T_{W^{+}W^{+}}&=g^2\Delta^{\dagger}T^{+}T^{+}\Delta,\,\,\,T_{W^{+}Z}= 
g^2\cos\theta_{W}\Delta^{\dagger}T^{+}T^{3}\Delta,\,\,\,
T^{W^{+}A}= g^2\sin\theta_{W}\Delta^{\dagger}T^{+}T^{3}\Delta
\label{anncouplings}
\end{align}

The S-wave annihilation matrix elements $\Gamma^{VV}$ can be read from the 
terms 
in the effective action after integrating out the relativistic gauge bosons,
\begin{equation}
S_{\text{eff}}\supset \frac{i}{2\pi}\int d^{4}x d^{3}y 
\left(T^{\dagger}_{VV}(x)T_{VV}(x^{0},\vec{y})\right)_{ii',jj'}\delta^{(3)}(\vec
{x}-\vec{y})\equiv 2i \int d^{4}x d^{3}r 
\Gamma^{(VV)}_{ii',jj'}\delta^{(3)}(\vec{r})\Phi^{*}_{ii'}(x,\vec{r})
\Phi_{jj'}(x,\vec{r})\nonumber
\end{equation}
after arranging mass eigenstates $\tilde{\Delta}_{i}$'s into 2-particle 
fields 
$\Phi_{ii'}$ using their NR limits given in Eq.(\ref{nrcomponent}).

\paragraph{$\mathbf{\Gamma^{(WW)}_{ij,kl}}$}\label{annWW}
The matrix elements are,
\begin{equation}
\Gamma^{(WW)}_{SS,SS}=\frac{\pi\alpha^2}{2m_{S}^2}\left(j^2+j-\frac{1}{4}
\right)^2
 =\Gamma^{(WW)}_{SS,AA}=\Gamma^{(WW)}_{AA,AA}
 \end{equation}
Here, $j=n/2$ is the isospin of the multiplet.

\begin{equation}
\Gamma^{(WW)}_{\Delta^{(\frac{n+1}{2})}\Delta^{(-\frac{n+1}{2})},\Delta^{(\frac{
n+1}{2})}\Delta^{(-\frac{n+1}{2})}}=
 \frac{\pi\alpha^2}{m_{S}^2}\frac{n^2}{4}\nonumber
\end{equation}

\begin{equation}
\Gamma^{(WW)}_{\Delta^{(\frac{n+1}{2})}\Delta^{(-\frac{n+1}{2})},SS}=
\frac{\pi\alpha^2}{\sqrt{2}m_{S}^2}j\left(j^2+j-\frac{1}{4}\right)=
\Gamma^{(WW)}_{\Delta^{(\frac{n+1}{2})}\Delta^{(-\frac{n+1}{2})},AA}
\end{equation}

For convenience, let us define, $R_{j,m}=(j^2+j-m^2)$.
\begin{align}
&\Gamma^{(WW)}_{\Delta^{(\frac{n+1}{2})}\Delta^{(-\frac{n+1}{2})},\tilde{\Delta}
^{(Q)}_{1}\tilde{\Delta}^{(-Q)}_{1}}
=\frac{\pi\alpha^2}{m_{S}^2}j(R_{j,m}\cos^2\theta_{Q}+R_{j,-m-1}\sin^2\theta_{Q}
)\nonumber\\
&\Gamma^{(WW)}_{\Delta^{(\frac{n+1}{2})}\Delta^{(-\frac{n+1}{2})},\tilde{\Delta}
^{(Q)}_{2}\tilde{\Delta}^{(-Q)}_{2}}
=\frac{\pi\alpha^2}{m_{S}^2}j(R_{j,m}\sin^2\theta_{Q}+R_{j,-m-1}\cos^2\theta_{Q}
)\nonumber\\
&\Gamma^{(WW)}_{\Delta^{(\frac{n+1}{2})}\Delta^{(-\frac{n+1}{2})},\tilde{\Delta}^
{(Q)}_{1}\tilde{\Delta}^{(-Q)}_{2}}
=\frac{\pi\alpha^2}{m_{S}^2}j(R_{j,m}-R_{j,-m-1})\sin\theta_{Q}\cos\theta_{Q}
\nonumber\\
&\Gamma^{(WW)}_{SS,\tilde{\Delta}^{(Q)}_{1}\tilde{\Delta}^{(-Q)}_{1}}=\frac{
\pi\alpha^2}{\sqrt{2}m_{S}^2}R_{j,-\frac{1}{2}} 
(R_{j,m}\cos^2\theta_{Q}+R_{j,-m-1}\sin^2\theta_{Q})=\Gamma^{(WW)}_{AA,\tilde{
\Delta}^{(Q)}_{1}\tilde{\Delta}^{(-Q)}_{1}}\nonumber\\
&\Gamma^{(WW)}_{SS,\tilde{\Delta}^{(Q)}_{2}\tilde{\Delta}^{(-Q)}_{2}}=\frac{
\pi\alpha^2}{\sqrt{2}m_{S}^2}R_{j,-\frac{1}{2}}
(R_{j,m}\sin^2\theta_{Q}+R_{j,-m-1}\cos^2\theta_{Q})=\Gamma^{(WW)}_{AA,\tilde{
\Delta}^{(Q)}_{2}\tilde{\Delta}^{(-Q)}_{2}}\nonumber\\
&\Gamma^{(WW)}_{SS,\tilde{\Delta}^{(Q)}_{1}\tilde{\Delta}^{(-Q)}_{2}}=\frac{
\pi\alpha^2}{\sqrt{2}m_{S}^2}R_{j,-\frac{1}{2}}
(R_{j,m}-R_{j,-m-1})\sin\theta_{Q}\cos\theta_{Q}=\Gamma^{(WW)}_{AA,\tilde{\Delta
}^{(Q)}_{1}\tilde{\Delta}^{(-Q)}_{2}}\nonumber\\
&\Gamma^{(WW)}_{\tilde{\Delta}^{(Q)}_{1}\tilde{\Delta}^{(-Q)}_{1},\tilde{\Delta}
^{(Q)}_{1}\tilde{\Delta}^{(-Q)}_{1}}
=\frac{\pi\alpha^2}{m_{S}^2}(R_{j,m}\cos^2\theta_{Q}+R_{j,-m-1}\sin^2\theta_{Q}
)^2\nonumber\\
&\Gamma^{(WW)}_{\tilde{\Delta}^{(Q)}_{2}\tilde{\Delta}^{(-Q)}_{2},\tilde{\Delta}
^{(Q)}_{2}\tilde{\Delta}^{(-Q)}_{2}}
=\frac{\pi\alpha^2}{m_{S}^2}(R_{j,m}\sin^2\theta_{Q}+R_{j,-m-1}\cos^2\theta_{Q}
)^2\nonumber\\
&\Gamma^{(WW)}_{\tilde{\Delta}^{(Q)}_{1}\tilde{\Delta}^{(-Q)}_{1},\tilde{\Delta}
^{(Q)}_{2}\tilde{\Delta}^{(-Q)}_{2}}
=\frac{\pi\alpha^2}{m_{S}^2}(R_{j,m}\cos^2\theta_{Q}+R_{j,-m-1}\sin^2\theta_{Q}
)\nonumber\\
 &(R_{j,m}\sin^2\theta_{Q}+R_{j,-m-1}\cos^2\theta_{Q})\nonumber\\
&\Gamma^{(WW)}_{\tilde{\Delta}^{(Q)}_{1}\tilde{\Delta}^{(-Q)}_{2},\tilde{\Delta}
^{(Q)}_{1}\tilde{\Delta}^{(-Q)}_{2}}
=\frac{\pi\alpha^2}{m_{S}^2}(R_{j,m}-R_{j,-m-1})^2\sin^2\theta_{Q}\cos^2\theta_{
Q}\nonumber\\
&\Gamma^{(WW)}_{\tilde{\Delta}^{(Q)}_{1}\tilde{\Delta}^{(-Q)}_{1},\tilde{\Delta}
^{(Q)}_{1}\tilde{\Delta}^{(-Q)}_{2}}
=\frac{\pi\alpha^2}{m_{S}^2}(R_{j,m}\cos^2\theta_{Q}+R_{j,-m-1}\sin^2\theta_{Q}
)\nonumber\\
 &(R_{j,m}-R_{j,-m-1})\sin\theta_{Q}\cos\theta_{Q}\nonumber\\
&\Gamma^{(WW)}_{\tilde{\Delta}^{(Q)}_{2}\tilde{\Delta}^{(-Q)}_{2},\tilde{\Delta}
^{(Q)}_{1}\tilde{\Delta}^{(-Q)}_{2}}
=\frac{\pi\alpha^2}{m_{S}^2}(R_{j,m}\sin^2\theta_{Q}+R_{j,-m-1}\cos^2\theta_{Q}
)\nonumber\\
 &(R_{j,m}-R_{j,-m-1})\sin\theta_{Q}\cos\theta_{Q}
\end{align}

\paragraph{$\mathbf{\Gamma^{(ZZ)}_{ij,kl}}$} First we define, 
$V^{(z)}_{m}=m^2\cos^{2}\theta_{W}-m\sin^{2}\theta_{W}+\frac{1}{4}
\tan^{2}\theta_{W}\sin^2\theta_{W}$.
\begin{equation}
\Gamma^{(ZZ)}_{\Delta^{(\frac{n+1}{2})}\Delta^{(-\frac{n+1}{2})},\Delta^{(\frac{
n+1}{2})}\Delta^{(-\frac{n+1}{2})}}
 =\frac{2\pi\alpha^2}{m_{S}^2}(V^{(z)}_{\frac{n}{2}})^2\nonumber
\end{equation}

\begin{equation}
\Gamma^{(ZZ)}_{SS,SS}=\frac{\pi\alpha^2}{m_{S}^2}(V^{(z)}_{-\frac{1}{2}}
)^2=\Gamma^{(ZZ)}_{AA,AA}=\Gamma^{(ZZ)}_{SS,AA}
\end{equation}

\begin{align}
&\Gamma^{(ZZ)}_{\tilde{\Delta}^{(Q)}_{1}\tilde{\Delta}^{(-Q)}_{1},\tilde{\Delta}
^{(Q)}_{1}\tilde{\Delta}^{(-Q)}_{1}}=
\frac{2\pi\alpha^2}{m_{S}^2}(V^{(z)}_{m}\cos^2\theta_{Q}+V^{(z)}_{-m-1}
\sin^2\theta_{Q})^2\nonumber\\
&\Gamma^{(ZZ)}_{\tilde{\Delta}^{(Q)}_{2}\tilde{\Delta}^{(-Q)}_{2},\tilde{\Delta}
^{(Q)}_{2}\tilde{\Delta}^{(-Q)}_{2}}=
\frac{2\pi\alpha^2}{m_{S}^2}(V^{(z)}_{m}\sin^2\theta_{Q}+V^{(z)}_{-m-1}
\cos^2\theta_{Q})^2\nonumber\\
&\Gamma^{(ZZ)}_{\tilde{\Delta}^{(Q)}_{1}\tilde{\Delta}^{(-Q)}_{1},\tilde{\Delta}
^{(Q)}_{2}\tilde{\Delta}^{(-Q)}_{2}}=
\frac{2\pi\alpha^2}{m_{S}^2}(V^{(z)}_{m}\cos^2\theta_{Q}+V^{(z)}_{-m-1}
\sin^2\theta_{Q})\nonumber\\
 &(V^{(z)}_{m}\sin^2\theta_{Q}+V^{(z)}_{-m-1}\cos^2\theta_{Q})\nonumber\\
&\Gamma^{(ZZ)}_{\tilde{\Delta}^{(Q)}_{1}\tilde{\Delta}^{(-Q)}_{2},\tilde{\Delta}
^{(Q)}_{1}\tilde{\Delta}^{(-Q)}_{2}}=
\frac{2\pi\alpha^2}{m_{S}^2}(V^{(z)}_{m}-V^{(z)}_{-m-1})^{2}\sin^{2}\theta_{Q}
\cos^{2}\theta_{Q}\nonumber\\
&\Gamma^{(ZZ)}_{\tilde{\Delta}^{(Q)}_{1}\tilde{\Delta}^{(-Q)}_{1},\tilde{\Delta}
^{(Q)}_{1}\tilde{\Delta}^{(-Q)}_{1}}=
\frac{2\pi\alpha^2}{m_{S}^2}(V^{(z)}_{m}\cos^2\theta_{Q}+V^{(z)}_{-m-1}
\sin^2\theta_{Q})\nonumber\\
 &(V^{(z)}_{m}-V^{(z)}_{-m-1})\sin\theta_{Q}\cos\theta_{Q}\nonumber\\
&\Gamma^{(ZZ)}_{\tilde{\Delta}^{(Q)}_{1}\tilde{\Delta}^{(-Q)}_{1},\tilde{\Delta}
^{(Q)}_{1}\tilde{\Delta}^{(-Q)}_{2}}=
\frac{2\pi\alpha^2}{m_{S}^2}(V^{(z)}_{m}\sin^2\theta_{Q}+V^{(z)}_{-m-1}
\cos^2\theta_{Q})^2\nonumber\\
 &(V^{(z)}_{m}-V^{(z)}_{-m-1})\sin\theta_{Q}\cos\theta_{Q}\nonumber\\
&\Gamma^{(ZZ)}_{\Delta^{(\frac{n+1}{2})}\Delta^{(-\frac{n+1}{2})},\tilde{\Delta}
^{(Q)}_{1}\tilde{\Delta}^{(-Q)}_{1}}=
\frac{2\pi\alpha^2}{m_{S}^2}V^{(z)}_{\frac{n}{2}}(V^{(z)}_{m}\cos^2\theta_{Q}+V^
{(z)}_{-m-1}\sin^2\theta_{Q})\nonumber\\
&\Gamma^{(ZZ)}_{\Delta^{(\frac{n+1}{2})}\Delta^{(-\frac{n+1}{2})},\tilde{\Delta}
^{(Q)}_{2}\tilde{\Delta}^{(-Q)}_{2}}=
\frac{2\pi\alpha^2}{m_{S}^2}V^{(z)}_{\frac{n}{2}}(V^{(z)}_{m}\sin^2\theta_{Q}+V^
{(z)}_{-m-1}\cos^2\theta_{Q})\nonumber\\
&\Gamma^{(ZZ)}_{\Delta^{(\frac{n+1}{2})}\Delta^{(-\frac{n+1}{2})},\tilde{\Delta}
^{(Q)}_{1}\tilde{\Delta}^{(-Q)}_{1}}=
\frac{2\pi\alpha^2}{m_{S}^2}V^{(z)}_{\frac{n}{2}}(V^{(z)}_{m}-V^{(z)}_{-m-1}
)\sin\theta_{Q}\cos\theta_{Q}\nonumber\\
 &\Gamma^{(ZZ)}_{SS,\tilde{\Delta}^{(Q)}_{1}\tilde{\Delta}^{(-Q)}_{1}}=
\frac{\sqrt{2}\pi\alpha^2}{m_{S}^2}V^{(z)}_{-\frac{1}{2}}(V^{(z)}_{m}
\cos^2\theta_{Q}+V^{(z)}_{-m-1}\sin^2\theta_{Q})
=\Gamma^{(ZZ)}_{AA,\tilde{\Delta}^{(Q)}_{1}\tilde{\Delta}^{(-Q)}_{1}}\nonumber\\
  &\Gamma^{(ZZ)}_{SS,\tilde{\Delta}^{(Q)}_{2}\tilde{\Delta}^{(-Q)}_{2}}=
\frac{\sqrt{2}\pi\alpha^2}{m_{S}^2}V^{(z)}_{-\frac{1}{2}}(V^{(z)}_{m}
\sin^2\theta_{Q}+V^{(z)}_{-m-1}\cos^2\theta_{Q})
=\Gamma^{(ZZ)}_{AA,\tilde{\Delta}^{(Q)}_{2}\tilde{\Delta}^{(-Q)}_{2}}\nonumber\\
  &\Gamma^{(ZZ)}_{SS,\tilde{\Delta}^{(Q)}_{1}\tilde{\Delta}^{(-Q)}_{2}}=
\frac{\sqrt{2}\pi\alpha^2}{m_{S}^2}V^{(z)}_{-\frac{1}{2}}(V^{(z)}_{m}-V^{(z)}_{
-m-1})\sin\theta_{Q}\cos\theta_{Q}
 =\Gamma^{(ZZ)}_{AA,\tilde{\Delta}^{(Q)}_{1}\tilde{\Delta}^{(-Q)}_{2}}\nonumber
\end{align}

\paragraph{$\mathbf{\Gamma^{(\gamma\gamma)}_{ij,kl}}$}
\begin{align}
 &\Gamma^{(\gamma\gamma)}_{\Delta^{(\frac{n+1}{2})}\Delta^{(-\frac{n+1}{2})},
\Delta^{(\frac{n+1}{2})}\Delta^{(-\frac{n+1}{2})}}=
 \frac{2\pi\alpha_{em}^2}{m_{S}^2}(\frac{n+1}{2})^2\nonumber\\
&\Gamma^{(\gamma\gamma)}_{\tilde{\Delta}^{(Q)}_{1}\tilde{\Delta}^{(-Q)}_{1},
\tilde{\Delta}^{(Q)}_{1}\tilde{\Delta}^{(-Q)}_{1}}=
\frac{2\pi\alpha_{em}^2Q^2}{m_{S}^2}\nonumber=\Gamma^{(\gamma\gamma)}_{\tilde{
\Delta}^{(Q)}_{2}\tilde{\Delta}^{(-Q)}_{2},\tilde{\Delta}^{(Q)}_{2}\tilde{\Delta
}^{(-Q)}_{2}}
 \nonumber\\
&\Gamma^{(\gamma\gamma)}_{\Delta^{(\frac{n+1}{2})}\Delta^{(-\frac{n+1}{2})},
\tilde{\Delta}^{(Q)}_{1}\tilde{\Delta}^{(-Q)}_{1}}=
 \frac{2\pi\alpha_{em}^2}{m_{S}^2}\frac{(n+1)Q}{2}
\nonumber=\Gamma^{(\gamma\gamma)}_{\Delta^{(\frac{n+1}{2})}\Delta^{(-\frac{n+1}{
2})},\tilde{\Delta}^{(Q)}_{2}\tilde{\Delta}^{(-Q)}_{2}}
 \nonumber
\end{align}

\paragraph{$\mathbf{\Gamma^{(\gamma Z)}_{ij,kl}}$} Here we define, 
$V^{(\gamma z)}_{m}= \left(m+\frac{1}{2}\right)\left(m \sin 
2\theta_{W}-\sin^{2}\theta_{W} \tan\theta_{W}\right)$.
\begin{align}
& \Gamma^{(\gamma 
Z)}_{\Delta^{(\frac{n+1}{2})}\Delta^{(-\frac{n+1}{2})},\Delta^{(\frac{n+1}{2})}
\Delta^{(-\frac{n+1}{2})}}=
 \frac{\pi\alpha^2}{m_{S}^2}(V^{(\gamma z)}_{\frac{n}{2}})^2\nonumber\\
 &\Gamma^{(\gamma 
Z)}_{\tilde{\Delta}^{(Q)}_{1}\tilde{\Delta}^{(-Q)}_{1},\tilde{\Delta}^{(Q)}_{1}
\tilde{\Delta}^{(-Q)}_{1}}=
 \frac{\pi\alpha^2}{m_{S}^2}(V^{(\gamma z)}_{m}\cos^2\theta_{Q}+V^{(\gamma 
z)}_{-m-1}\sin^2\theta_{Q})^2\nonumber\\
 &\Gamma^{(\gamma 
Z)}_{\tilde{\Delta}^{(Q)}_{2}\tilde{\Delta}^{(-Q)}_{2},\tilde{\Delta}^{(Q)}_{2}
\tilde{\Delta}^{(-Q)}_{2}}=
 \frac{\pi\alpha^2}{m_{S}^2}(V^{(\gamma z)}_{m}\sin^2\theta_{Q}+V^{(\gamma 
z)}_{-m-1}\cos^2\theta_{Q})^2\nonumber\\
 &\Gamma^{(\gamma 
Z)}_{\tilde{\Delta}^{(Q)}_{1}\tilde{\Delta}^{(-Q)}_{2},\tilde{\Delta}^{(Q)}_{1}
\tilde{\Delta}^{(-Q)}_{2}}=
 \frac{\pi\alpha^2}{m_{S}^2}(V^{(\gamma z)}_{m}-V^{(\gamma 
z)}_{-m-1})^{2}\sin^2\theta_{Q}\cos^2\theta_{Q}\nonumber\\
 &\Gamma^{(\gamma 
Z)}_{\Delta^{(\frac{n+1}{2})}\Delta^{(-\frac{n+1}{2})},\tilde{\Delta}^{(Q)}_{1}
\tilde{\Delta}^{(-Q)}_{1}}=
 \frac{\pi\alpha^2}{m_{S}^2}V^{(\gamma z)}_{\frac{n}{2}}(V^{(\gamma 
z)}_{m}\cos^2\theta_{Q}+V^{(\gamma z)}_{-m-1}\sin^2\theta_{Q})\nonumber\\
 &\Gamma^{(\gamma 
Z)}_{\Delta^{(\frac{n+1}{2})}\Delta^{(-\frac{n+1}{2})},\tilde{\Delta}^{(Q)}_{2}
\tilde{\Delta}^{(-Q)}_{2}}=
 \frac{\pi\alpha^2}{m_{S}^2}V^{(\gamma z)}_{\frac{n}{2}}(V^{(\gamma 
z)}_{m}\sin^2\theta_{Q}+V^{(\gamma z)}_{-m-1}\cos^2\theta_{Q})\nonumber\\
 &\Gamma^{(\gamma 
Z)}_{\Delta^{(\frac{n+1}{2})}\Delta^{(-\frac{n+1}{2})},\tilde{\Delta}^{(Q)}_{1}
\tilde{\Delta}^{(-Q)}_{2}}=
 \frac{\pi\alpha^2}{m_{S}^2}V^{(\gamma z)}_{\frac{n}{2}}(V^{(\gamma 
z)}_{m}-V^{(\gamma z)}_{-m-1})\sin\theta_{Q}\cos\theta_{Q}\nonumber\\
  &\Gamma^{(\gamma 
Z)}_{\tilde{\Delta}^{(Q)}_{1}\tilde{\Delta}^{(-Q)}_{1},\tilde{\Delta}^{(Q)}_{1}
\tilde{\Delta}^{(-Q)}_{2}}=
 \frac{\pi\alpha^2}{m_{S}^2}(V^{(\gamma z)}_{m}\cos^2\theta_{Q}+V^{(\gamma 
z)}_{-m-1}\sin^2\theta_{Q})\nonumber\\
 &(V^{(\gamma z)}_{m}-V^{(\gamma 
z)}_{-m-1})\sin\theta_{Q}\cos\theta_{Q}\nonumber\\
 &\Gamma^{(\gamma 
Z)}_{\tilde{\Delta}^{(Q)}_{2}\tilde{\Delta}^{(-Q)}_{2},\tilde{\Delta}^{(Q)}_{1}
\tilde{\Delta}^{(-Q)}_{2}}=
 \frac{\pi\alpha^2}{m_{S}^2}(V^{(\gamma z)}_{m}\sin^2\theta_{Q}+V^{(\gamma 
z)}_{-m-1}\cos^2\theta_{Q})\nonumber\\
 &(V^{(\gamma z)}_{m}-V^{(\gamma 
z)}_{-m-1})\sin\theta_{Q}\cos\theta_{Q}\nonumber
 \end{align}

\end{appendices}


\begin{thebibliography}{99}

%\cite{Bringmann:2012ez}
\bibitem{Bringmann:2012ez} 
  T.~Bringmann and C.~Weniger,
  %``Gamma Ray Signals from Dark Matter: Concepts, Status and Prospects,''
  Phys.\ Dark Univ.\  {\bf 1}, 194 (2012)
  doi:10.1016/j.dark.2012.10.005
  [arXiv:1208.5481 [hep-ph]].
  %%CITATION = doi:10.1016/j.dark.2012.10.005;%%
  %150 citations counted in INSPIRE as of 12 Nov 2016

%\cite{Wood:2013taa}
\bibitem{Wood:2013taa} 
  M.~Wood, J.~Buckley, S.~Digel, S.~Funk, D.~Nieto and M.~A.~Sanchez-Conde,
  %``Prospects for Indirect Detection of Dark Matter with CTA,''
  arXiv:1305.0302 [astro-ph.HE].
  %%CITATION = ARXIV:1305.0302;%%
  %29 citations counted in INSPIRE as of 12 Nov 2016
  
  %\cite{Buckley:2013bha}
\bibitem{Buckley:2013bha} 
  J.~Buckley {\it et al.},
  %``Working Group Report: WIMP Dark Matter Indirect Detection,''
  arXiv:1310.7040 [astro-ph.HE].
  %%CITATION = ARXIV:1310.7040;%%
  %31 citations counted in INSPIRE as of 12 Nov 2016
  
%\cite{Cirelli:2015gux}
\bibitem{Cirelli:2015gux} 
  M.~Cirelli,
  %``Status of Indirect (and Direct) Dark Matter searches,''
  arXiv:1511.02031 [astro-ph.HE].
  %%CITATION = ARXIV:1511.02031;%%
  %9 citations counted in INSPIRE as of 12 Nov 2016
  
  %\cite{Profumo:2016idl}
\bibitem{Profumo:2016idl} 
  S.~Profumo, F.~S.~Queiroz and C.~E.~Yaguna,
  %``Extending Fermi-LAT and H.E.S.S. Limits on Gamma-ray Lines from Dark Matter Annihilation,''
  doi:10.1093/mnras/stw1600
  arXiv:1602.08501 [astro-ph.HE].
  %%CITATION = doi:10.1093/mnras/stw1600;%%
  %4 citations counted in INSPIRE as of 12 Nov 2016
  
  %\cite{Mora:2015vhq}
\bibitem{Mora:2015vhq} 
  K.~Morå [H.E.S.S. Collaboration],
  %``Dark Matter Searches with H.E.S.S,''
  arXiv:1512.00698 [astro-ph.HE].
  %%CITATION = ARXIV:1512.00698;%%
  %3 citations counted in INSPIRE as of 12 Nov 2016
  
  %\cite{::2016jja}
\bibitem{::2016jja} 
  H.~Abdallah {\it et al.} [HESS Collaboration],
  %``Search for dark matter annihilations towards the inner Galactic halo from 10 years of observations with H.E.S.S,''
  Phys.\ Rev.\ Lett.\  {\bf 117}, no. 11, 111301 (2016)
  doi:10.1103/PhysRevLett.117.111301
  [arXiv:1607.08142 [astro-ph.HE]].
  %%CITATION = doi:10.1103/PhysRevLett.117.111301;%%
  %6 citations counted in INSPIRE as of 12 Nov 2016
  
  %\cite{Abdalla:2016olq}
\bibitem{Abdalla:2016olq} 
  H.~Abdalla {\it et al.} [HESS Collaboration],
  %``H.E.S.S. Limits on Linelike Dark Matter Signatures in the 100 GeV to 2 TeV Energy Range Close to the Galactic Center,''
  Phys.\ Rev.\ Lett.\  {\bf 117}, no. 15, 151302 (2016)
  doi:10.1103/PhysRevLett.117.151302
  [arXiv:1609.08091 [astro-ph.HE]].
  %%CITATION = doi:10.1103/PhysRevLett.117.151302;%%

%\cite{Doro:2012xx}
\bibitem{Doro:2012xx} 
  M.~Doro {\it et al.} [CTA Consortium Collaboration],
  %``Dark Matter and Fundamental Physics with the Cherenkov Telescope Array,''
  Astropart.\ Phys.\  {\bf 43}, 189 (2013)
  doi:10.1016/j.astropartphys.2012.08.002
  [arXiv:1208.5356 [astro-ph.IM]].
  %%CITATION = doi:10.1016/j.astropartphys.2012.08.002;%%
  %101 citations counted in INSPIRE as of 12 Nov 2016 
  
  %\cite{Carr:2015hta}
\bibitem{Carr:2015hta} 
  J.~Carr {\it et al.} [CTA Collaboration],
  %``Prospects for Indirect Dark Matter Searches with the Cherenkov Telescope Array (CTA),''
  PoS ICRC {\bf 2015}, 1203 (2016)
  [arXiv:1508.06128 [astro-ph.HE]].
  %%CITATION = ARXIV:1508.06128;%%
  %20 citations counted in INSPIRE as of 12 Nov 2016
    
  %\cite{Conrad:2016jww}
\bibitem{Conrad:2016jww} 
  J.~Conrad,
  %``CTA in the Context of Searches for Particle Dark Matter - a glimpse,''
  arXiv:1610.03258 [astro-ph.HE].
  %%CITATION = ARXIV:1610.03258;%%
  
  %\cite{Lefranc:2016dgx}
\bibitem{Lefranc:2016dgx} 
  V.~Lefranc, G.~A.~Mamon and P.~Panci,
  %``Prospects for annihilating Dark Matter towards Milky Way's dwarf galaxies by the Cherenkov Telescope Array,''
  JCAP {\bf 1609}, no. 09, 021 (2016)
  doi:10.1088/1475-7516/2016/09/021
  [arXiv:1605.02793 [astro-ph.HE]].
  %%CITATION = doi:10.1088/1475-7516/2016/09/021;%%
  %2 citations counted in INSPIRE as of 27 Nov 2016
  
  %\cite{Lefranc:2016fgn}
\bibitem{Lefranc:2016fgn} 
  V.~Lefranc, E.~Moulin, P.~Panci, F.~Sala and J.~Silk,
  %``Dark Matter in $\gamma$ lines: Galactic Center vs dwarf galaxies,''
  JCAP {\bf 1609}, no. 09, 043 (2016)
  doi:10.1088/1475-7516/2016/09/043
  [arXiv:1608.00786 [astro-ph.HE]].
  %%CITATION = doi:10.1088/1475-7516/2016/09/043;%%
  %4 citations counted in INSPIRE as of 27 Nov 2016
  
  \bibitem{sommerfeldref}
  A.~Sommerfeld,
  Ann.\ Phys.\ {\bf 11}, 257 (1931)
  
%\cite{Hisano:2002fk}
\bibitem{Hisano:2002fk} 
  J.~Hisano, S.~Matsumoto and M.~M.~Nojiri,
  %``Unitarity and higher order corrections in neutralino dark matter annihilation into two photons,''
  Phys.\ Rev.\ D {\bf 67}, 075014 (2003)
  doi:10.1103/PhysRevD.67.075014
  [hep-ph/0212022].
  %%CITATION = doi:10.1103/PhysRevD.67.075014;%%
  %84 citations counted in INSPIRE as of 12 Nov 2016  
  
%\cite{Hisano:2003ec}
\bibitem{Hisano:2003ec} 
  J.~Hisano, S.~Matsumoto and M.~M.~Nojiri,
  %``Explosive dark matter annihilation,''
  Phys.\ Rev.\ Lett.\  {\bf 92}, 031303 (2004)
  [hep-ph/0307216].
  %%CITATION = HEP-PH/0307216;%%
  %263 citations counted in INSPIRE as of 02 Nov 2015
  
%\cite{Hisano:2004ds}
\bibitem{Hisano:2004ds} 
  J.~Hisano, S.~Matsumoto, M.~M.~Nojiri and O.~Saito,
  %``Non-perturbative effect on dark matter annihilation and gamma ray signature from galactic center,''
  Phys.\ Rev.\ D {\bf 71}, 063528 (2005)
  [hep-ph/0412403].
  %%CITATION = HEP-PH/0412403;%%
  %316 citations counted in INSPIRE as of 02 Nov 2015
  
%\cite{Hisano:2005ec}
\bibitem{Hisano:2005ec} 
  J.~Hisano, S.~Matsumoto, O.~Saito and M.~Senami,
  %``Heavy wino-like neutralino dark matter annihilation into antiparticles,''
  Phys.\ Rev.\ D {\bf 73}, 055004 (2006)
  [hep-ph/0511118].
  %%CITATION = HEP-PH/0511118;%%
  %114 citations counted in INSPIRE as of 02 Nov 2015
  
  %\cite{Hisano:2006nn}
\bibitem{Hisano:2006nn} 
  J.~Hisano, S.~Matsumoto, M.~Nagai, O.~Saito and M.~Senami,
  %``Non-perturbative effect on thermal relic abundance of dark matter,''
  Phys.\ Lett.\ B {\bf 646}, 34 (2007)
  doi:10.1016/j.physletb.2007.01.012
  [hep-ph/0610249].
  %%CITATION = doi:10.1016/j.physletb.2007.01.012;%%
  %245 citations counted in INSPIRE as of 12 Nov 2016
  
  
%\cite{Cirelli:2007xd}
\bibitem{Cirelli:2007xd} 
  M.~Cirelli, A.~Strumia and M.~Tamburini,
  %``Cosmology and Astrophysics of Minimal Dark Matter,''
  Nucl.\ Phys.\ B {\bf 787}, 152 (2007)
  [arXiv:0706.4071 [hep-ph]].
  %%CITATION = ARXIV:0706.4071;%%
  %261 citations counted in INSPIRE as of 04 Nov 2015
  
%\cite{Cirelli:2009uv}
\bibitem{Cirelli:2009uv} 
  M.~Cirelli and A.~Strumia,
  %``Minimal Dark Matter: Model and results,''
  New J.\ Phys.\  {\bf 11}, 105005 (2009)
  [arXiv:0903.3381 [hep-ph]].
  %%CITATION = ARXIV:0903.3381;%%
  %84 citations counted in INSPIRE as of 04 Nov 2015
  
%\cite{ArkaniHamed:2008qn}
\bibitem{ArkaniHamed:2008qn}
  N.~Arkani-Hamed, D.~P.~Finkbeiner, T.~R.~Slatyer and N.~Weiner,
  %``A Theory of Dark Matter,''
  Phys.\ Rev.\ D {\bf 79} (2009) 015014
  [arXiv:0810.0713 [hep-ph]].
  %%CITATION = ARXIV:0810.0713;%%
  %981 citations counted in INSPIRE as of 04 Nov 2015
  
  %\cite{Lattanzi:2008qa}
\bibitem{Lattanzi:2008qa} 
  M.~Lattanzi and J.~I.~Silk,
  %``Can the WIMP annihilation boost factor be boosted by the Sommerfeld enhancement?,''
  Phys.\ Rev.\ D {\bf 79}, 083523 (2009)
  doi:10.1103/PhysRevD.79.083523
  [arXiv:0812.0360 [astro-ph]].
  %%CITATION = doi:10.1103/PhysRevD.79.083523;%%
  %170 citations counted in INSPIRE as of 12 Nov 2016
  
  %\cite{Pieri:2009zi}
\bibitem{Pieri:2009zi} 
  L.~Pieri, M.~Lattanzi and J.~Silk,
  %``Constraining the Sommerfeld enhancement with Cherenkov telescope observations of dwarf galaxies,''
  Mon.\ Not.\ Roy.\ Astron.\ Soc.\  {\bf 399}, 2033 (2009)
  doi:10.1111/j.1365-2966.2009.15388.x
  [arXiv:0902.4330 [astro-ph.HE]].
  %%CITATION = doi:10.1111/j.1365-2966.2009.15388.x;%%
  %41 citations counted in INSPIRE as of 12 Nov 2016
  
%\cite{Iengo:2009ni}
\bibitem{Iengo:2009ni} 
  R.~Iengo,
  %``Sommerfeld enhancement: General results from field theory diagrams,''
  JHEP {\bf 0905}, 024 (2009)
  doi:10.1088/1126-6708/2009/05/024
  [arXiv:0902.0688 [hep-ph]].
  %%CITATION = doi:10.1088/1126-6708/2009/05/024;%%
  %79 citations counted in INSPIRE as of 12 Nov 2016
  
  %\cite{Cassel:2009wt}
\bibitem{Cassel:2009wt} 
  S.~Cassel,
  %``Sommerfeld factor for arbitrary partial wave processes,''
  J.\ Phys.\ G {\bf 37}, 105009 (2010)
  doi:10.1088/0954-3899/37/10/105009
  [arXiv:0903.5307 [hep-ph]].
  %%CITATION = doi:10.1088/0954-3899/37/10/105009;%%
  %87 citations counted in INSPIRE as of 12 Nov 2016
  
%\cite{Slatyer:2009vg}
\bibitem{Slatyer:2009vg}
  T.~R.~Slatyer,
  %``The Sommerfeld enhancement for dark matter with an excited state,''
  JCAP {\bf 1002} (2010) 028
  [arXiv:0910.5713 [hep-ph]].
  %%CITATION = ARXIV:0910.5713;%%
  %55 citations counted in INSPIRE as of 04 Nov 2015
  
  %\cite{Hambye:2009pw}
\bibitem{Hambye:2009pw}
  T.~Hambye, F.-S.~Ling, L.~Lopez Honorez and J.~Rocher,
  %``Scalar Multiplet Dark Matter,''
  JHEP {\bf 0907} (2009) 090
   [JHEP {\bf 1005} (2010) 066]
  [arXiv:0903.4010 [hep-ph]].
  %%CITATION = ARXIV:0903.4010;%%
  %100 citations counted in INSPIRE as of 26 May 2015
  
  %\cite{Feng:2009hw}
\bibitem{Feng:2009hw} 
  J.~L.~Feng, M.~Kaplinghat and H.~B.~Yu,
  %``Halo Shape and Relic Density Exclusions of Sommerfeld-Enhanced Dark Matter Explanations of Cosmic Ray Excesses,''
  Phys.\ Rev.\ Lett.\  {\bf 104}, 151301 (2010)
  doi:10.1103/PhysRevLett.104.151301
  [arXiv:0911.0422 [hep-ph]].
  %%CITATION = doi:10.1103/PhysRevLett.104.151301;%%
  %152 citations counted in INSPIRE as of 27 Nov 2016
  
  %\cite{Feng:2010zp}
\bibitem{Feng:2010zp} 
  J.~L.~Feng, M.~Kaplinghat and H.~B.~Yu,
  %``Sommerfeld Enhancements for Thermal Relic Dark Matter,''
  Phys.\ Rev.\ D {\bf 82}, 083525 (2010)
  doi:10.1103/PhysRevD.82.083525
  [arXiv:1005.4678 [hep-ph]].
  %%CITATION = doi:10.1103/PhysRevD.82.083525;%%
  %127 citations counted in INSPIRE as of 27 Nov 2016
    
  %\cite{Hryczuk:2010zi}
\bibitem{Hryczuk:2010zi} 
  A.~Hryczuk, R.~Iengo and P.~Ullio,
  %``Relic densities including Sommerfeld enhancements in the MSSM,''
  JHEP {\bf 1103}, 069 (2011)
  doi:10.1007/JHEP03(2011)069
  [arXiv:1010.2172 [hep-ph]].
  %%CITATION = doi:10.1007/JHEP03(2011)069;%%
  %66 citations counted in INSPIRE as of 12 Nov 2016
  
  %\cite{Hryczuk:2011vi}
\bibitem{Hryczuk:2011vi} 
  A.~Hryczuk and R.~Iengo,
  %``The one-loop and Sommerfeld electroweak corrections to the Wino dark matter annihilation,''
  JHEP {\bf 1201}, 163 (2012)
  Erratum: [JHEP {\bf 1206}, 137 (2012)]
  doi:10.1007/JHEP01(2012)163, 10.1007/JHEP06(2012)137
  [arXiv:1111.2916 [hep-ph]].
  %%CITATION = doi:10.1007/JHEP01(2012)163, 10.1007/JHEP06(2012)137;%%
  %47 citations counted in INSPIRE as of 12 Nov 2016
  
  %\cite{Beneke:2012tg}
\bibitem{Beneke:2012tg} 
  M.~Beneke, C.~Hellmann and P.~Ruiz-Femenia,
  %``Non-relativistic pair annihilation of nearly mass degenerate neutralinos and charginos I. General framework and S-wave annihilation,''
  JHEP {\bf 1303}, 148 (2013)
  Erratum: [JHEP {\bf 1310}, 224 (2013)]
  doi:10.1007/JHEP10(2013)224, 10.1007/JHEP03(2013)148
  [arXiv:1210.7928 [hep-ph]].
  %%CITATION = doi:10.1007/JHEP10(2013)224, 10.1007/JHEP03(2013)148;%%
  %18 citations counted in INSPIRE as of 12 Nov 2016
  
  %\cite{Tulin:2013teo}
\bibitem{Tulin:2013teo} 
  S.~Tulin, H.~B.~Yu and K.~M.~Zurek,
  %``Beyond Collisionless Dark Matter: Particle Physics Dynamics for Dark Matter Halo Structure,''
  Phys.\ Rev.\ D {\bf 87}, no. 11, 115007 (2013)
  doi:10.1103/PhysRevD.87.115007
  [arXiv:1302.3898 [hep-ph]].
  %%CITATION = doi:10.1103/PhysRevD.87.115007;%%
  %145 citations counted in INSPIRE as of 14 Nov 2016
  
%\cite{Fan:2013faa}
\bibitem{Fan:2013faa} 
  J.~Fan and M.~Reece,
  %``In Wino Veritas? Indirect Searches Shed Light on Neutralino Dark Matter,''
  JHEP {\bf 1310}, 124 (2013)
  [arXiv:1307.4400 [hep-ph]].
  %%CITATION = ARXIV:1307.4400;%%
  %95 citations counted in INSPIRE as of 04 Nov 2015
  
  
%\cite{Cohen:2013ama}
\bibitem{Cohen:2013ama} 
  T.~Cohen, M.~Lisanti, A.~Pierce and T.~R.~Slatyer,
  %``Wino Dark Matter Under Siege,''
  JCAP {\bf 1310}, 061 (2013)
  [arXiv:1307.4082].
  %%CITATION = ARXIV:1307.4082;%%
  %95 citations counted in INSPIRE as of 04 Nov 2015
  
  %\cite{Ovanesyan:2014fwa}
\bibitem{Ovanesyan:2014fwa} 
  G.~Ovanesyan, T.~R.~Slatyer and I.~W.~Stewart,
  %``Heavy Dark Matter Annihilation from Effective Field Theory,''
  Phys.\ Rev.\ Lett.\  {\bf 114}, no. 21, 211302 (2015)
  doi:10.1103/PhysRevLett.114.211302
  [arXiv:1409.8294 [hep-ph]].
  %%CITATION = doi:10.1103/PhysRevLett.114.211302;%%
  %27 citations counted in INSPIRE as of 12 Nov 2016
  
  %\cite{Baumgart:2014vma}
\bibitem{Baumgart:2014vma} 
  M.~Baumgart, I.~Z.~Rothstein and V.~Vaidya,
  %``Calculating the Annihilation Rate of Weakly Interacting Massive Particles,''
  Phys.\ Rev.\ Lett.\  {\bf 114}, 211301 (2015)
  doi:10.1103/PhysRevLett.114.211301
  [arXiv:1409.4415 [hep-ph]].
  %%CITATION = doi:10.1103/PhysRevLett.114.211301;%%
  %20 citations counted in INSPIRE as of 12 Nov 2016
  
  %\cite{Beneke:2014gja}
\bibitem{Beneke:2014gja} 
  M.~Beneke, C.~Hellmann and P.~Ruiz-Femenia,
  %``Non-relativistic pair annihilation of nearly mass degenerate neutralinos and charginos III. Computation of the Sommerfeld enhancements,''
  JHEP {\bf 1505}, 115 (2015)
  doi:10.1007/JHEP05(2015)115
  [arXiv:1411.6924 [hep-ph]].
  %%CITATION = doi:10.1007/JHEP05(2015)115;%%
  %14 citations counted in INSPIRE as of 12 Nov 2016
  
%\cite{Cirelli:2015bda}
\bibitem{Cirelli:2015bda} 
  M.~Cirelli, T.~Hambye, P.~Panci, F.~Sala and M.~Taoso,
  %``Gamma ray tests of Minimal Dark Matter,''
  JCAP {\bf 1510}, no. 10, 026 (2015)
  [arXiv:1507.05519 [hep-ph]].
  %%CITATION = ARXIV:1507.05519;%%
  %9 citations counted in INSPIRE as of 04 Nov 2015
  
%\cite{Garcia-Cely:2015dda}
\bibitem{Garcia-Cely:2015dda} 
  C.~Garcia-Cely, A.~Ibarra, A.~S.~Lamperstorfer and M.~H.~G.~Tytgat,
  %``Gamma-rays from Heavy Minimal Dark Matter,''
  JCAP {\bf 1510}, no. 10, 058 (2015)
  [arXiv:1507.05536 [hep-ph]].
  %%CITATION = ARXIV:1507.05536;%%
  %8 citations counted in INSPIRE as of 04 Nov 2015
  
%\cite{Aoki:2015nza}
\bibitem{Aoki:2015nza} 
  M.~Aoki, T.~Toma and A.~Vicente,
  %``Non-thermal Production of Minimal Dark Matter via Right-handed Neutrino Decay,''
  JCAP {\bf 1509}, 063 (2015)
  [arXiv:1507.01591 [hep-ph]].
  %%CITATION = ARXIV:1507.01591;%%
  %8 citations counted in INSPIRE as of 04 Nov 2015
  
  
  %\cite{Ma:2006km}
\bibitem{Ma:2006km} 
  E.~Ma,
  %``Verifiable radiative seesaw mechanism of neutrino mass and dark matter,''
  Phys.\ Rev.\ D {\bf 73}, 077301 (2006)
  [hep-ph/0601225].
  %%CITATION = HEP-PH/0601225;%%
  %463 citations counted in INSPIRE as of 21 Apr 2015
  
  %\cite{Kubo:2006yx}
\bibitem{Kubo:2006yx}
  J.~Kubo, E.~Ma and D.~Suematsu,
  %``Cold Dark Matter, Radiative Neutrino Mass, $\mu \to e\gamma$, and Neutrinoless Double Beta Decay,''
  Phys.\ Lett.\ B {\bf 642} (2006) 18
  [hep-ph/0604114].
  %%CITATION = HEP-PH/0604114;%%
  %108 citations counted in INSPIRE as of 21 Apr 2015
  
%\cite{Sierra:2008wj}
\bibitem{Sierra:2008wj}
  D.~Aristizabal Sierra, J.~Kubo, D.~Restrepo, D.~Suematsu and O.~Zapata,
  %``Radiative seesaw: Warm dark matter, collider and lepton flavour violating signals,''
  Phys.\ Rev.\ D {\bf 79} (2009) 013011
  [arXiv:0808.3340 [hep-ph]].
  %%CITATION = ARXIV:0808.3340;%%
  %58 citations counted in INSPIRE as of 21 Apr 2015
  
%\cite{Suematsu:2009ww}
\bibitem{Suematsu:2009ww}
  D.~Suematsu, T.~Toma and T.~Yoshida,
  %``Reconciliation of CDM abundance and mu ---> e gamma in a radiative seesaw model,''
  Phys.\ Rev.\ D {\bf 79} (2009) 093004
  [arXiv:0903.0287 [hep-ph]].
  %%CITATION = ARXIV:0903.0287;%%
  %45 citations counted in INSPIRE as of 21 Apr 2015
  
%\cite{Adulpravitchai:2009gi}
\bibitem{Adulpravitchai:2009gi}
  A.~Adulpravitchai, M.~Lindner and A.~Merle,
  %``Confronting Flavour Symmetries and extended Scalar Sectors with Lepton Flavour Violation Bounds,''
  Phys.\ Rev.\ D {\bf 80} (2009) 055031
  [arXiv:0907.2147 [hep-ph]].
  %%CITATION = ARXIV:0907.2147;%%
  %14 citations counted in INSPIRE as of 21 Apr 2015
 
%\cite{Toma:2013zsa}
\bibitem{Toma:2013zsa}
  T.~Toma and A.~Vicente,
  %``Lepton Flavor Violation in the Scotogenic Model,''
  JHEP {\bf 1401} (2014) 160
  [arXiv:1312.2840, arXiv:1312.2840 [hep-ph]].
  %%CITATION = ARXIV:1312.2840,;%%
  %11 citations counted in INSPIRE as of 21 Apr 2015
  
  %\cite{Klasen:2013jpa}
\bibitem{Klasen:2013jpa} 
  M.~Klasen, C.~E.~Yaguna, J.~D.~Ruiz-Alvarez, D.~Restrepo and O.~Zapata,
  %``Scalar dark matter and fermion coannihilations in the radiative seesaw model,''
  JCAP {\bf 1304}, 044 (2013)
  doi:10.1088/1475-7516/2013/04/044
  [arXiv:1302.5298 [hep-ph]].
  %%CITATION = doi:10.1088/1475-7516/2013/04/044;%%
  %18 citations counted in INSPIRE as of 12 Nov 2016
  
%\cite{Vicente:2014wga}
\bibitem{Vicente:2014wga}
  A.~Vicente and C.~E.~Yaguna,
  %``Probing the scotogenic model with lepton flavor violating processes,''
  JHEP {\bf 1502} (2015) 144
  [arXiv:1412.2545 [hep-ph]].
  %%CITATION = ARXIV:1412.2545;%%
  %1 citations counted in INSPIRE as of 21 Apr 2015
  
%\cite{Ma:2008cu}
\bibitem{Ma:2008cu}
  E.~Ma and D.~Suematsu,
  %``Fermion Triplet Dark Matter and Radiative Neutrino Mass,''
  Mod.\ Phys.\ Lett.\ A {\bf 24} (2009) 583
  [arXiv:0809.0942 [hep-ph]].
  %%CITATION = ARXIV:0809.0942;%%
  %51 citations counted in INSPIRE as of 27 May 2015
  
%\cite{Law:2013saa}
\bibitem{Law:2013saa}
  S.~S.~C.~Law and K.~L.~McDonald,
  %``A Class of Inert N-tuplet Models with Radiative Neutrino Mass and Dark Matter,''
  JHEP {\bf 1309} (2013) 092
  [arXiv:1305.6467 [hep-ph]].
  %%CITATION = ARXIV:1305.6467;%%
  %21 citations counted in INSPIRE as of 27 May 2015
  
%\cite{Ren:2011mh}
\bibitem{Ren:2011mh}
  B.~Ren, K.~Tsumura and X.~G.~He,
  %``A Higgs Quadruplet for Type III Seesaw and Implications for $\mu \to e\gamma$ and $\mu - e$ Conversion,''
  Phys.\ Rev.\ D {\bf 84} (2011) 073004
  [arXiv:1107.5879 [hep-ph]].
  %%CITATION = ARXIV:1107.5879;%%
  %14 citations counted in INSPIRE as of 04 juin 2015
  
  %\cite{Restrepo:2013aga}
\bibitem{Restrepo:2013aga} 
  D.~Restrepo, O.~Zapata and C.~E.~Yaguna,
  %``Models with radiative neutrino masses and viable dark matter candidates,''
  JHEP {\bf 1311}, 011 (2013)
  doi:10.1007/JHEP11(2013)011
  [arXiv:1308.3655 [hep-ph]].
  %%CITATION = doi:10.1007/JHEP11(2013)011;%%
  %32 citations counted in INSPIRE as of 12 Nov 2016
  
  %\cite{Chowdhury:2015sla}
\bibitem{Chowdhury:2015sla} 
  T.~A.~Chowdhury and S.~Nasri,
  %``Lepton Flavor Violation in the Inert Scalar Model with Higher Representations,''
  JHEP {\bf 1512}, 040 (2015)
  doi:10.1007/JHEP12(2015)040
  [arXiv:1506.00261 [hep-ph]].
  %%CITATION = doi:10.1007/JHEP12(2015)040;%%
  %5 citations counted in INSPIRE as of 12 Nov 2016
  
  %\cite{Ahriche:2016cio}
\bibitem{Ahriche:2016cio} 
  A.~Ahriche, K.~L.~McDonald and S.~Nasri,
  %``The Scale-Invariant Scotogenic Model,''
  JHEP {\bf 1606}, 182 (2016)
  doi:10.1007/JHEP06(2016)182
  [arXiv:1604.05569 [hep-ph]].
  %%CITATION = doi:10.1007/JHEP06(2016)182;%%
  %15 citations counted in INSPIRE as of 14 Nov 2016
  
  %\cite{Ahriche:2016ixu}
\bibitem{Ahriche:2016ixu} 
  A.~Ahriche, A.~Manning, K.~L.~McDonald and S.~Nasri,
  %``Scale-Invariant Models with One-Loop Neutrino Mass and Dark Matter Candidates,''
  Phys.\ Rev.\ D {\bf 94}, no. 5, 053005 (2016)
  doi:10.1103/PhysRevD.94.053005
  [arXiv:1604.05995 [hep-ph]].
  %%CITATION = doi:10.1103/PhysRevD.94.053005;%%
  %12 citations counted in INSPIRE as of 14 Nov 2016
  
  %\cite{Cirelli:2005uq}
\bibitem{Cirelli:2005uq}
  M.~Cirelli, N.~Fornengo and A.~Strumia,
  %``Minimal dark matter,''
  Nucl.\ Phys.\ B {\bf 753} (2006) 178
  [hep-ph/0512090].
  %%CITATION = HEP-PH/0512090;%%
  %326 citations counted in INSPIRE as of 21 Apr 2015
    
  %\cite{Queiroz:2015utg}
\bibitem{Queiroz:2015utg} 
  F.~S.~Queiroz and C.~E.~Yaguna,
  %``The CTA aims at the Inert Doublet Model,''
  JCAP {\bf 1602}, no. 02, 038 (2016)
  doi:10.1088/1475-7516/2016/02/038
  [arXiv:1511.05967 [hep-ph]].
  %%CITATION = doi:10.1088/1475-7516/2016/02/038;%%
  %12 citations counted in INSPIRE as of 12 Nov 2016
  
  %\cite{Garcia-Cely:2015khw}
\bibitem{Garcia-Cely:2015khw} 
  C.~Garcia-Cely, M.~Gustafsson and A.~Ibarra,
  %``Probing the Inert Doublet Dark Matter Model with Cherenkov Telescopes,''
  JCAP {\bf 1602}, no. 02, 043 (2016)
  doi:10.1088/1475-7516/2016/02/043
  [arXiv:1512.02801 [hep-ph]].
  %%CITATION = doi:10.1088/1475-7516/2016/02/043;%%
  %13 citations counted in INSPIRE as of 12 Nov 2016
  
  %\cite{AbdusSalam:2013eya}
\bibitem{AbdusSalam:2013eya} 
  S.~S.~AbdusSalam and T.~A.~Chowdhury,
  %``Scalar Representations in the Light of Electroweak Phase Transition and Cold Dark Matter Phenomenology,''
  JCAP {\bf 1405}, 026 (2014)
  doi:10.1088/1475-7516/2014/05/026
  [arXiv:1310.8152 [hep-ph]].
  %%CITATION = doi:10.1088/1475-7516/2014/05/026;%%
  %13 citations counted in INSPIRE as of 12 Nov 2016
  
  \bibitem{fabrizio-Goran}
  F.~Nesti and G.~Senjanovi\'c,
  Private communication,
  September, 2014
  
  %\cite{Lu:2016dbc}
\bibitem{Lu:2016dbc} 
  W.~B.~Lu and P.~H.~Gu,
  %``Mixed Inert Scalar Triplet Dark Matter, Radiative Neutrino Masses and Leptogenesis,''
  arXiv:1611.02106 [hep-ph].
  %%CITATION = ARXIV:1611.02106;%%
  
  %\cite{Beneke:1997zp}
\bibitem{Beneke:1997zp} 
  M.~Beneke and V.~A.~Smirnov,
  %``Asymptotic expansion of Feynman integrals near threshold,''
  Nucl.\ Phys.\ B {\bf 522}, 321 (1998)
  doi:10.1016/S0550-3213(98)00138-2
  [hep-ph/9711391].
  %%CITATION = doi:10.1016/S0550-3213(98)00138-2;%%
  %434 citations counted in INSPIRE as of 20 Nov 2016
  
  %\cite{Zhang:2013qza}
\bibitem{Zhang:2013qza} 
  Z.~Zhang,
  %``Multi-Sommerfeld enhancement in dark sector,''
  Phys.\ Lett.\ B {\bf 734}, 188 (2014)
  doi:10.1016/j.physletb.2014.05.054
  [arXiv:1307.2206 [hep-ph]].
  %%CITATION = doi:10.1016/j.physletb.2014.05.054;%%
  %2 citations counted in INSPIRE as of 12 Nov 2016
  
  \bibitem{martinazzo}
  R.~Martinazzo, E.~Bodo, F.~A.~Gianturco, 
  %"A modified Variable-Phase algorithm for multichannel scattering with long-range potentials,"
   Computer Physics Communications, {\bf 151}, 187 (2003). 
http://dx.doi.org/10.1016/S0010-4655(02)00737-3.
  
  %\cite{Ershov:2011zz}
\bibitem{Ershov:2011zz} 
  S.~N.~Ershov, J.~S.~Vaagen and M.~V.~Zhukov,
  %``Modified variable phase method for the solution of coupled radial Schrodinger equations,''
  Phys.\ Rev.\ C {\bf 84}, 064308 (2011).
  doi:10.1103/PhysRevC.84.064308
  %%CITATION = doi:10.1103/PhysRevC.84.064308;%%
  %7 citations counted in INSPIRE as of 12 Nov 2016
  
  %\cite{Blum:2016nrz}
\bibitem{Blum:2016nrz} 
  K.~Blum, R.~Sato and T.~R.~Slatyer,
  %``Self-consistent Calculation of the Sommerfeld Enhancement,''
  JCAP {\bf 1606}, no. 06, 021 (2016)
  doi:10.1088/1475-7516/2016/06/021
  [arXiv:1603.01383 [hep-ph]].
  %%CITATION = doi:10.1088/1475-7516/2016/06/021;%%
  %4 citations counted in INSPIRE as of 12 Nov 2016
  
  
  %\cite{Ade:2015xua}
\bibitem{Ade:2015xua} 
  P.~A.~R.~Ade {\it et al.} [Planck Collaboration],
  %``Planck 2015 results. XIII. Cosmological parameters,''
  Astron.\ Astrophys.\  {\bf 594}, A13 (2016)
  doi:10.1051/0004-6361/201525830
  [arXiv:1502.01589 [astro-ph.CO]].
  %%CITATION = doi:10.1051/0004-6361/201525830;%%
  %2397 citations counted in INSPIRE as of 12 Nov 2016
  
  %\cite{Baer:2014eja}
\bibitem{Baer:2014eja} 
  H.~Baer, K.~Y.~Choi, J.~E.~Kim and L.~Roszkowski,
  %``Dark matter production in the early Universe: beyond the thermal WIMP paradigm,''
  Phys.\ Rept.\  {\bf 555}, 1 (2015)
  doi:10.1016/j.physrep.2014.10.002
  [arXiv:1407.0017 [hep-ph]].
  %%CITATION = doi:10.1016/j.physrep.2014.10.002;%%
  %93 citations counted in INSPIRE as of 20 Nov 2016
  
  %\cite{Giedt:2009mr}
\bibitem{Giedt:2009mr} 
  J.~Giedt, A.~W.~Thomas and R.~D.~Young,
  %{\it ``Dark matter, the CMSSM and lattice QCD,''}
  Phys.\ Rev.\ Lett.\  {\bf 103}, 201802 (2009)
  [arXiv:0907.4177 [hep-ph]].
  %%CITATION = ARXIV:0907.4177;%%
  %108 citations counted in INSPIRE as of 21 Mar 2013
  
  %\cite{Akerib:2016vxi}
\bibitem{Akerib:2016vxi} 
  D.~S.~Akerib {\it et al.},
  %``Results from a search for dark matter in the complete LUX exposure,''
  arXiv:1608.07648 [astro-ph.CO].
  %%CITATION = ARXIV:1608.07648;%%
  %61 citations counted in INSPIRE as of 12 Nov 2016
  
  %\cite{Aprile:2015uzo}
\bibitem{Aprile:2015uzo} 
  E.~Aprile {\it et al.} [XENON Collaboration],
  %``Physics reach of the XENON1T dark matter experiment,''
  JCAP {\bf 1604}, no. 04, 027 (2016)
  doi:10.1088/1475-7516/2016/04/027
  [arXiv:1512.07501 [physics.ins-det]].
  %%CITATION = doi:10.1088/1475-7516/2016/04/027;%%
  %96 citations counted in INSPIRE as of 12 Nov 2016
  
  %\cite{Klasen:2013btp}
\bibitem{Klasen:2013btp} 
  M.~Klasen, C.~E.~Yaguna and J.~D.~Ruiz-Alvarez,
  %``Electroweak corrections to the direct detection cross section of inert higgs dark matter,''
  Phys.\ Rev.\ D {\bf 87}, 075025 (2013)
  doi:10.1103/PhysRevD.87.075025
  [arXiv:1302.1657 [hep-ph]].
  %%CITATION = doi:10.1103/PhysRevD.87.075025;%%
  %30 citations counted in INSPIRE as of 12 Nov 2016
  
  %\cite{Kennedy:1988sn}
\bibitem{Kennedy:1988sn} 
  D.~C.~Kennedy and B.~W.~Lynn,
  %``Electroweak Radiative Corrections with an Effective Lagrangian: Four Fermion Processes,''
  Nucl.\ Phys.\ B {\bf 322}, 1 (1989).
  doi:10.1016/0550-3213(89)90483-5
  %%CITATION = doi:10.1016/0550-3213(89)90483-5;%%
  %418 citations counted in INSPIRE as of 12 Nov 2016
  
  %\cite{Peskin:1991sw}
\bibitem{Peskin:1991sw} 
  M.~E.~Peskin and T.~Takeuchi,
  %``Estimation of oblique electroweak corrections,''
  Phys.\ Rev.\ D {\bf 46}, 381 (1992).
  doi:10.1103/PhysRevD.46.381
  %%CITATION = doi:10.1103/PhysRevD.46.381;%%
  %1760 citations counted in INSPIRE as of 12 Nov 2016
  
  %\cite{Barbieri:2004qk}
\bibitem{Barbieri:2004qk} 
  R.~Barbieri, A.~Pomarol, R.~Rattazzi and A.~Strumia,
  %``Electroweak symmetry breaking after LEP-1 and LEP-2,''
  Nucl.\ Phys.\ B {\bf 703}, 127 (2004)
  doi:10.1016/j.nuclphysb.2004.10.014
  [hep-ph/0405040].
  %%CITATION = doi:10.1016/j.nuclphysb.2004.10.014;%%
  %458 citations counted in INSPIRE as of 12 Nov 2016
  
  %\cite{Olive:2016xmw}
\bibitem{Olive:2016xmw} 
  C.~Patrignani {\it et al.} [Particle Data Group Collaboration],
  %``Review of Particle Physics,''
  Chin.\ Phys.\ C {\bf 40}, no. 10, 100001 (2016).
  doi:10.1088/1674-1137/40/10/100001
  %%CITATION = doi:10.1088/1674-1137/40/10/100001;%%
  %52 citations counted in INSPIRE as of 13 Nov 2016  
  
  %\cite{Barbieri:2006dq}
\bibitem{Barbieri:2006dq} 
  R.~Barbieri, L.~J.~Hall and V.~S.~Rychkov,
  %``Improved naturalness with a heavy Higgs: An Alternative road to LHC physics,''
  Phys.\ Rev.\ D {\bf 74}, 015007 (2006)
  doi:10.1103/PhysRevD.74.015007
  [hep-ph/0603188].
  %%CITATION = doi:10.1103/PhysRevD.74.015007;%%
  %575 citations counted in INSPIRE as of 12 Nov 2016
  
  %\cite{Martinez:2011ua}
\bibitem{Martinez:2011ua}
  H.~Martinez, A.~Melfo, F.~Nesti, G.~Senjanovi\'c,
  %``Three Extra Mirror or Sequential Families: A Case for Heavy Higgs and Inert Doublet,''
  Phys.\ Rev.\ Lett.\  {\bf 106}, 191802 (2011).
  [arXiv:1101.3796 [hep-ph]]. 

%\cite{Melfo:2011ie}
\bibitem{Melfo:2011ie}
  A.~Melfo, M.~Nemev\v sek, F.~Nesti, G.~Senjanovi\' c, Y.~Zhang,
  %``Inert Doublet Dark Matter and Mirror/Extra Families after Xenon100,''
  Phys.\ Rev.\  {\bf D84 } (2011)  034009.
  [arXiv:1105.4611 [hep-ph]].
  	
%\cite{Chowdhury:2011ga}
\bibitem{Chowdhury:2011ga} 
  T.~A.~Chowdhury, M.~Nemev\v sek, G.~Senjanovi\'c and Y.~Zhang,
  %``Dark Matter as the Trigger of Strong Electroweak Phase Transition,''
  JCAP {\bf 1202}, 029 (2012)
  [arXiv:1110.5334 [hep-ph]].
  %%CITATION = ARXIV:1110.5334;%%
  
  %\cite{Chun:2012yt}
\bibitem{Chun:2012yt} 
  E.~J.~Chun, J.~C.~Park and S.~Scopel,
  %``Non-perturbative Effect and PAMELA Limit on Electro-Weak Dark Matter,''
  JCAP {\bf 1212}, 022 (2012)
  doi:10.1088/1475-7516/2012/12/022
  [arXiv:1210.6104 [astro-ph.CO]].
  %%CITATION = doi:10.1088/1475-7516/2012/12/022;%%
  %13 citations counted in INSPIRE as of 12 Nov 2016
  
  \bibitem{Camilo-thesis}
  C.~A.~Garcia-Cely,
  {\it Dark Matter Phenomenology in Scalar Extensions of the Standard Model}
  PhD thesis, Technische Universit\"at M\"unchen (TUM), (2014)
  
  %\cite{Garcia-Cely:2013zga}
\bibitem{Garcia-Cely:2013zga} 
  C.~Garcia-Cely and A.~Ibarra,
  %``Novel Gamma-ray Spectral Features in the Inert Doublet Model,''
  JCAP {\bf 1309}, 025 (2013)
  doi:10.1088/1475-7516/2013/09/025
  [arXiv:1306.4681 [hep-ph]].
  %%CITATION = doi:10.1088/1475-7516/2013/09/025;%%
  %22 citations counted in INSPIRE as of 14 Nov 2016
  
  %\cite{Das:2016ced}
\bibitem{Das:2016ced} 
  A.~Das and B.~Dasgupta,
  %``A Selection Rule for Enhanced Dark Matter Annihilation,''
  arXiv:1611.04606 [hep-ph].
  %%CITATION = ARXIV:1611.04606;%%
  
  \bibitem{idmproject}
  T.~A.~Chowdhury and S.~Nasri,
  In preparation.
  
  %\cite{Logan:2016ivc}
\bibitem{Logan:2016ivc} 
  H.~E.~Logan and T.~Pilkington,
  %``Large scalar multiplet dark matter in the high-mass region,''
  arXiv:1610.08835 [hep-ph].
  %%CITATION = ARXIV:1610.08835;%%
  
  
  
  
 
  
 
  
  
  
  







\end{thebibliography}
\end{document}